# Study of transport process and discharge structure of inductively coupled electronegative plasmas via fluid model and analytic theory collaboration

[a]Shu-Xia Zhao[1] and An-Qi Tang[2]

1. *Key Laboratory of Material Modification by Laser, Ion, and Electron Beams (Ministry of Education), School of Physics, Dalian University of Technology, 116024, Dalian, China*
2. *China Key System & Integrated Circuit Co., Ltd, 214026, Wuxi, China*

[a] *Correspondence: zhaonie@dlut.edu.cn*

## Abstract

The discharge structure of inductively coupled plasma is studied via fluid simulation and analytic theory collaboration. At low pressure, the discharge is stratified by the double layer, which is modelled as dipole moment. The parabolic profile is formed in the discharge core when recombination loss is negligible and both the electron and anions are the Boltzmann balanced. At increasing the pressure, the main characteristics, *i.e.*, parabolic, elliptic and flat-topped profile, are experienced, predicted by the simulation and analytics. Self-coagulation is accompanied at all considered pressures. It is more a chemistry process and provides new means of constricting plasma. At its influence, electron density deviates from the Boltzmann equilibrium. For satisfying the neutrality of bulk plasma, the ambi-polar self-coagulation mechanism is suggested. At high pressure, the self-coagulation-to-coil scheme causes the mass point behavior in the plasma. Minor cations re-self-coagulate at certain conditions and the correlation with the celestial bodies' formation is hypothesized.

## I. Introduction

Structure of glow gaseous discharge is important for people to understand the plasma generated, since it exhibits the physics and chemistry processes of plasma transport, *i.e.*, a window for observing the plasma inside. It is difficult to study the electronegative plasma because of the interfere of anion in ambi-polar diffusion and chemical bulk loss of plasma species, which thereby complicates the transport equations. At reasonable approximations such as the Boltzmann statistics of electron and anion, these complex transport equations are simplified, and accordingly important analytic solutions are found for illuminating the electronegative discharge structure [1-5]. Main discharge characteristics are specified, for instance the heavy ions density profiles at respectively the models of parabola, ellipse, and flat-top, as well as the stratification of discharge. At these models, the electron density is low and its spatial variation is negligible. There are in total two types of discharge stratification phenomena. The normal one represents the bulk discharge structure that is divided into electronegative core (predominantly consisting of cation and anion) and electropositive halo (consisting of cation and electron). These two parts are either continually joined or discontinuously linked by double layer. One more stratification is used to indicate that the density distributions of different species, *i.e.*, the electron, cation and anion, in the bulk is stratified, which is more happened at relatively small electronegativities [6]. The parabolic and flat-topped ions profiles discovered via the analytic theories are validated the most with the experimental observation. It is noticed that the parabolic ion density profile matches well with the experiments while the flat-top profile is not [7-10]. Experimentally, a bumped ion density peak is appeared at the end of the flat-top profile, which cannot be explained by the present analytic theory.

We recently reported in Ref. [11] a *delta* distribution of anion density in an $Ar/O_2$ inductively coupled plasma. It is discovered based on a two-dimensional fluid model simulation. When examining the temporal variation of this simulation, it is found that the combination of free diffusion and negative source term (*i.e.*, recombination loss of anion) plays important role for the anion, $O^-$, to self-coagulate to certain spatial location. Form the point of high mathematic view, these transport components consist of one quasi-Helmholtz equation, whose analytic solution can be a delta distribution, spatially independent. This new analytic work given by us unveils the self-coagulation behavior naturally happened inside the plasma. This transport scheme is more determined by the chemistry process, against the conventional ambi-polar diffusion transport which arises from the mass and polarity distinctions (physics). At the influence of chemistry, the continuous medium flows along with the direction of density gradient, *i.e.*, anti- free diffusion. A seemingly not understandable phenomenon. It in fact belongs to the self-organizations of dissipative structure, for it satisfies the main features, *e.g.*, nonlinear evolution and descending entropy [12-14]. More details about the self-coagulation can be found in the Refs. [11, 15]. This above self-coagulation is related to the anion dynamic. How is the cation coagulated for arriving at the plasma quasi-neutrality, at high electronegativity? At the assistance of electronegative internal sheath, when the negative chemical source is not satisfied. It is called as the ambi-polar self-coagulation. All the new discoveries of us involving the self-coagulation interpret well the experimentally observed density peak of ions. Herein, ions refer to both the cation and anion.

It is originally thought that the self-coagulation analytic theory just belongs to the less electronegative plasmas. It is nevertheless found it exists in the general electronegative plasma sources. In the present work, the discharge structure of $Ar/SF_6$ inductively coupled plasma is focused via a fluid simulation [11, 16]. The fluid simulation of $Ar/SF_6$ plasma satisfies well the predictions of the present and early analytic theories, qualitatively in accord to the experiments as well. In such a high electronegative discharge, the self-coagulation is found to happen within all charged species, including the light electrons. At its influence, the species deviates from the Boltzmann balance when chemistry process dominates over the physics. Besides, the essence of well-known double layer is investigated. Presently, self-consistent simulations, *i.e.*, fluid, particle or hybrid, are very useful but not focused on analyzing discharge structure. We feel it is a waste of computational resource since lot amount of interesting profiles simulated are ignored [17-21]. Hence, one more goal of this article is to trigger more interest of the community onto the basic discharge structure studies. It is emphasized that the two-dimensional fluid simulations of reactive inductively coupled plasmas are not a new topic and lots of relevant papers have been published [17-20, 22-26]. However, all these articles are not focused on the analysis of discharge structure and are neither related to the previous useful analytic works. It is the innovation embedded in our present conventional fluid simulations.

## II. Methodology

The formula of fluid model used is described in this section. It includes the mass, momentum and energy equations of plasma species, together with the Poisson and Maxwellian equations. The reactor used consists of discharge chamber, dielectric window and matching box. In the chamber, the wafer is seated at the bottom. The dimension and configuration of reactor can be found in Refs. [11,16]. The $Ar/SF_6$ gas-phase chemistry and surface kinetics are given in Tables.1 and 2.

Table 1. Chemical reaction set considered in the model

| No. | Reaction | Rate coefficient[a] | Threshold (eV) | Ref. |
|---|---|---|---|---|
| | | Elastic collisions | | |
| 1 | $e+Ar \rightarrow e+Ar$ | Cross Section | 0 | [27] |
| 2 | $e+SF_6 \rightarrow e+SF_6$ | Cross Section | 0 | [27] |
| 3 | $e+F_2 \rightarrow e+F_2$ | Cross Section | 0 | [27] |
| 4 | $e+F \rightarrow e+F$ | Cross Section | 0 | [27] |
| | | Excitation and deexcitation reactions | | |
| 5 | $e+Ar \rightarrow e+Ars$ | Cross Section | 11.6 | [27] |
| 6 | $e+Ars \rightarrow e+Ar$ | Cross Section | -11.6 | [27] |
| | | Ionization reactions | | |
| 7 | $e+Ar \rightarrow 2e+Ar^+$ | Cross Section | 15.76 | [27] |
| 8 | $e+Ars \rightarrow 2e+Ar^+$ | Cross Section | 4.43 | [27] |
| 9 | $e+SF_6 \rightarrow SF_5^+ + F + 2e$ | $1.2\times10^{-7}\exp(-18.1/T_e)$ | 16 | [28,29] |
| 10 | $e+SF_6 \rightarrow SF_4^+ + 2F + 2e$ | $8.4\times10^{-9}\exp(-19.9/T_e)$ | 20 | [28,29] |
| 11 | $e+SF_6 \rightarrow SF_3^+ + 3F + 2e$ | $3.2\times10^{-8}\exp(-20.7/T_e)$ | 20.5 | [28,29] |
| 12 | $e+SF_6 \rightarrow SF_2^+ + F_2 + 2F + 2e$ | $7.6\times10^{-9}\exp(-24.4/T_e)$ | 28 | [28,29] |
| 13 | $e+SF_6 \rightarrow SF^+ + F_2 + 3F + 2e$ | $1.2\times10^{-8}\exp(-26.0/T_e)$ | 37.5 | [28,29] |
| 14 | $e+SF_6 \rightarrow F^+ + SF_4 + F + 2e$ | $1.2\times10^{-8}\exp(-31.7/T_e)$ | 29 | [28,29] |
| 15 | $e+SF_6 \rightarrow S^+ + 4F + F_2 + 2e$ | $1.4\times10^{-8}\exp(-39.9/T_e)$ | 18 | [28,29] |
| 16 | $e+SF_5 \rightarrow SF_5^+ + 2e$ | $1.0\times10^{-7}\exp(-17.8/T_e)$ | 11 | [28,29] |
| 17 | $e+SF_5 \rightarrow SF_4^+ + F + 2e$ | $9.4\times10^{-8}\exp(-22.8/T_e)$ | 15 | [28,29] |
| 18 | $e+SF_4 \rightarrow SF_4^+ + 2e$ | $4.77\times10^{-8}\exp(-16.35/T_e)$ | 13 | [28,29] |
| 19 | $e+SF_4 \rightarrow SF_3^+ + F + 2e$ | $5.31\times10^{-8}\exp(-17.67/T_e)$ | 14.5 | [28,29] |
| 20 | $e+SF_3 \rightarrow SF_3^+ + 2e$ | $1.0\times10^{-7}\exp(-18.9/T_e)$ | 11 | [28,29] |
| 21 | $e+F \rightarrow F^+ + 2e$ | $1.3\times10^{-8}\exp(-16.5/T_e)$ | 15 | [28,29] |
| 22 | $e+S \rightarrow S^+ + 2e$ | $1.6\times10^{-7}\exp(-13.3/T_e)$ | 10 | [28,29] |
| 23 | $e+F_2 \rightarrow F_2^+ + 2e$ | $1.37\times10^{-8}\exp(-20.7/T_e)$ | 15.69 | [28,29] |
| | | Attachment and dissociative attachment reactions | | |
| 24 | $e + SF_6 \rightarrow SF_6^-$ | Cross Section | 0 | [27] |
| 25 | $e + SF_6 \rightarrow SF_5^- + F$ | Cross Section | 0.1 | [27] |
| 26 | $e + SF_6 \rightarrow SF_4^- + 2F$ | Cross Section | 5.4 | [27] |
| 27 | $e + SF_6 \rightarrow SF_3^- + 3F$ | Cross Section | 11.2 | [27] |
| 28 | $e + SF_6 \rightarrow SF_2^- + 4F$ | Cross Section | 12 | [27] |
| 29 | $e + SF_6 \rightarrow F^- + SF_5$ | Cross Section | 2.9 | [27] |
| 30 | $e + SF_6 \rightarrow F_2^- + SF_4$ | Cross Section | 5.4 | [27] |
| 31 | $e + F_2 \rightarrow F^- + F$ | Cross Section | 0 | [27] |
| | | Dissociation reactions | | |
| 32 | $e + SF_6 \rightarrow SF_5 + F + e$ | $1.5\times10^{-7}\exp(-8.1/T_e)$ | 9.6 | [28,29] |
| 33 | $e + SF_6 \rightarrow SF_4 + 2F + e$ | $9.0\times10^{-9}\exp(-13.4/T_e)$ | 12.4 | [28,29] |
| 34 | $e + SF_6 \rightarrow SF_3 + 3F + e$ | $2.5\times10^{-8}\exp(-33.5/T_e)$ | 16 | [28,29] |
| 35 | $e + SF_6 \rightarrow SF_2 + F_2 + 2F + e$ | $2.3\times10^{-8}\exp(-33.9/T_e)$ | 18.6 | [28,29] |
| 36 | $e + SF_6 \rightarrow SF + F_2 + 3F + e$ | $1.5\times10^{-9}\exp(-26.0/T_e)$ | 22.7 | [28,29] |
| 37 | $e + SF_5 \rightarrow SF_4 + F + e$ | $1.5\times10^{-7}\exp(-9.0/T_e)$ | 5 | [28,29] |
| 38 | $e + SF_4 \rightarrow SF_3 + F + e$ | $6.2\times10^{-8}\exp(-9.0/T_e)$ | 8.5 | [28,29] |
| 39 | $e + SF_3 \rightarrow SF_2 + F + e$ | $8.6\times10^{-8}\exp(-9.0/T_e)$ | 5 | [28,29] |
| 40 | $e + SF_2 \rightarrow SF + F + e$ | $4.5\times10^{-8}\exp(-9.0/T_e)$ | 8 | [28,29] |
| 41 | $e + SF \rightarrow S + F + e$ | $6.2\times10^{-8}\exp(-9.0/T_e)$ | 7.9 | [28,29] |
| 42 | $e + F_2 \rightarrow 2F + e$ | $1.2\times10^{-8}\exp(-5.8/T_e)$ | 1.6 | [28,29] |
| | | Neutral / neutral recombination reactions | | |
| 43 | $S + F \rightarrow SF$ | $2\times10^{-16}$ | 0 | [28,29] |
| 44 | $SF + F \rightarrow SF_2$ | $2.9\times10^{-14}$ | 0 | [28,29] |
| 45 | $SF_2 + F \rightarrow SF_3$ | $2.6\times10^{-12}$ | 0 | [28,29] |
| 46 | $SF_3 + F \rightarrow SF_4$ | $1.6\times10^{-11}$ | 0 | [28,29] |
| 47 | $SF_4 + F \rightarrow SF_5$ | $1.7\times10^{-11}$ | 0 | [28,29] |
| 48 | $SF_5 + F \rightarrow SF_6$ | $1.0\times10^{-11}$ | 0 | [28,29] |
| 49 | $SF_3 + SF_3 \rightarrow SF_2 + SF_4$ | $2.5\times10^{-11}$ | 0 | [28,29] |
| 50 | $SF_5 + SF_5 \rightarrow SF_4 + SF_6$ | $2.5\times10^{-11}$ | 0 | [28,29] |
| 51 | $SF + SF \rightarrow S + SF_2$ | $2.5\times10^{-11}$ | 0 | [28,29] |
| 52 | $SF_x + F_2 \rightarrow SF_{x+1} + F$ [b] | $7.0\times10^{-15}$ | 0 | [28,29] |
| | Ion / ion recombination reactions | | | |
| 53 | $X^+ + Y^- \rightarrow X + Y$ [c] | $5.0\times10^{-9}$ | 0 | [28,29] |
| | Detachment reactions | | | |
| 54 | $Z + Y^- \rightarrow Z + Y + e$ [d] | $5.27\times10^{-14}$ | 0 | [28,29] |
| | Other reactions | | | |
| 55 | $Ars + Ars \rightarrow e + Ar + Ar^+$ | $6.2\times10^{-10}$ | 0 | [28,29] |
| 56 | $Ars + Ar \rightarrow Ar + Ar$ | $3.0\times10^{-15}$ | 0 | [28,29] |
| 57 | $Ar^+ + SF_6 \rightarrow SF_5^+ + F + Ar$ | $9.0\times10^{-10}$ | 0 | [28,29] |
| 58 | $SF_5^+ + SF_6 \rightarrow SF_3^+ + SF_6 + F_2$ | $6.0\times10^{-12}$ | 0 | [28,29] |

[a]The unit of the rate coefficient is $cm^3s^{-1}$.

[b] $x$ stands for the number 1-5.

[c] X = $SF_5$、$SF_4$、$SF_3$、$SF_2$、SF、F、S or $F_2$ and Y = $SF_6$、$SF_5$、$SF_4$、$SF_3$、$SF_2$、F or $F_2$.

[d] Z = $SF_6$、$SF_5$、$SF_4$、$SF_3$、$SF_2$、SF、F、S or $F_2$ and Y = $SF_6$、$SF_5$、$SF_4$、$SF_3$、$SF_2$、F or $F_2$.

Table 2. Surface reaction set considered in the model

| No. | Surface reaction | Sticking coefficient | Ref. |
|---|---|---|---|
| 1 | $SF_x^+ + wall \rightarrow SF_x$; $x = 1-5$ | 1 | [28,29] |
| 2 | $F^+ + wall \rightarrow F$ | 1 | [28,29] |
| 3 | $F_2^+ + wall \rightarrow F_2$ | 1 | [28,29] |
| 4 | $S^+ + wall \rightarrow S$ | 1 | [28,29] |
| 5 | F+wall $\rightarrow 1/2\,F_2$ | 0.02 | [30] |
| 6 | $Ar^+$+wall $\rightarrow$ Ar | 1 | [28,29] |
| 7 | Ars+wall $\rightarrow$ Ar | 1 | [28,29] |

## (2.1) Electron equations

The equations of electron density and energy are given as

$$\begin{aligned} &\frac{\partial n_e}{\partial t} + \nabla \cdot \mathbf{\Gamma}_e = R_e - (\mathbf{u} \cdot \nabla) n_e, \\ &\frac{\partial n_\varepsilon}{\partial t} + \nabla \cdot \mathbf{\Gamma}_\varepsilon + \mathbf{E} \cdot \mathbf{\Gamma}_e = S_{en} - (\mathbf{u} \cdot \nabla) n_\varepsilon + P_{ohm}. \end{aligned} \tag{2.1}$$

The fluxes of electrons density and energy can be described as shown in Eq. (2.2)

$$\begin{aligned} \mathbf{\Gamma}_e &= -(\boldsymbol{\mu}_e \cdot \mathbf{E}) n_e - \mathbf{D}_e \cdot \nabla n_e, \\ \mathbf{\Gamma}_\varepsilon &= -(\boldsymbol{\mu}_{en} \cdot \mathbf{E}) n_\varepsilon - \mathbf{D}_\varepsilon \cdot \nabla n_\varepsilon. \end{aligned} \tag{2.2}$$

Here, $n_e$ and $n_\varepsilon$ are the number density and energy density of electrons, respectively. $\boldsymbol{\mu}_e$ and $\boldsymbol{\mu}_{en}$ are the electron mobility and electron energy mobility, respectively. $\mathbf{D}_e$ and $\mathbf{D}_\varepsilon$ are electron diffusivity and electron energy diffusivity, respectively. The relations among the above mass and energy transport coefficients are $\mathbf{D}_e = \boldsymbol{\mu}_e T_e$, $\mathbf{D}_\varepsilon = \boldsymbol{\mu}_\varepsilon T_e$, and $\boldsymbol{\mu}_\varepsilon = \frac{5}{3}\boldsymbol{\mu}_e$.

$R_e$ and $R_\varepsilon$ in Eq. (2.1) are the respective source terms of number density and energy density of electrons. Their expressions are stated in Eq. (2.3)

$$\begin{aligned} R_e &= \sum_{j=1}^{M} l_j k_j \prod_{m=1}^{P} n_m^{\,v_{jm}}, \\ R_\varepsilon &= \sum_{j=1}^{M} \varepsilon_j l_j k_j \prod_{m=1}^{P} n_m^{\,v_{jm}}. \end{aligned} \tag{2.3}$$

Here, $l_j$ is the number of electrons created or lost per collision of reactions that will either generate or deplete electrons. *M* is the number of such reactions. $k_j$ is the rate coefficient of reaction *j*, and these values can be found in Table 1. $n_m$ is the number density of reactant *m* of reaction *j*. $v$ is the stoichiometric coefficient of the reaction and *P* is the number of reactants. $\varepsilon_j$ is the energy loss per electron-impact reaction.

In Eq. (2.1), $(\mathbf{u} \cdot \Delta) n_e$ and $(\mathbf{u} \cdot \Delta) n_\varepsilon$ represent the transports of electron mass and mean energy by background gas advection, where the velocity field $\mathbf{u}$ is advective velocity of background gas and equals to zero since gas advection is not considered in the present work. The electron thermal velocity is higher than the advection, thus the direct influence of advection on electrons should be not significant. $P_{ohm}$ is the power density that is deposited through the Ohm's heating mechanism,

$$P_{ohm} = \frac{1}{2}\mathrm{Re}(\sigma|E_\theta|^2), \tag{2.4}$$

where $\sigma$ is the electron conductivity and $E_\theta$ is the azimuthal field that is given in Sec. 2.4.

Without the secondary electron emissions, the boundary conditions for the above equation are

$$\begin{aligned}\mathbf{n}\cdot\Gamma_e &= \frac{1-r_e}{1+r}(\frac{1}{2}v_{e,th}n_e),\\ \mathbf{n}\cdot\Gamma_\varepsilon &= \frac{1-r_e}{1+r_e}(\frac{5}{6}v_{e,th}n_e),\end{aligned} \tag{2.5}$$

where $v_{e,th}$ is the thermal velocity of electrons and $r_e$ is the reflection coefficient in the reactor wall, which is set to 0.2 in this model.

## 2.2 Heavy species equations

Heavy species that consists of *k= 1,2,...,Q* species are supposed as a reacting flow that consists of Q species, i.e., all plasma species except electrons. The equations for the mass transport of heavy species are

$$\rho\frac{\partial w_k}{\partial t} + \rho(\mathbf{u}\cdot\nabla)w_k = \nabla\cdot\mathbf{j}_k + R_k, \tag{2.6}$$

where, $\rho$ is the total mass density of heavy species, $w_k$ is the mass fraction of species *k*, and $\mathbf{u}$ is the velocity of background gas advection. $\mathbf{j}_k$ is the diffusive and drift flux of species *k* and expressed as

$$\begin{aligned}\mathbf{j}_k &= \rho w_k \mathbf{V}_k,\\ \mathbf{V}_k &= D_{k,m}\nabla\ln(w_k) - z_k\mu_m\mathbf{E},\end{aligned} \tag{2.7}$$

where $\mathbf{V}_k$ is the velocity of species *k*, $z_k$ is elementary charge number that species *k* carries, $\mu_m$ is the mobility, and **E** is the electrostatic field, calculated from the Poisson's equation. Note that the multi-component mass diffusion is taken into account and $D_{k,m}$ is the mixture averaged diffusion coefficient and expressed as

$$D_{k,m} = \frac{1-w_k}{\sum_{j\neq k}^{Q} x_j / D_{kj}}, \tag{2.8}$$

where $x_j$ is the number fraction of species *j* and $D_{k,j}$ is the Chapman-Enskog binary diffusion coefficient . The source term of Eq. (2.6), $R_k$ , is expressed as

$$R_k = M_k \sum_{j=1}^{N} l_{k,j} r_j, \tag{2.9}$$

where $M_k$ is molecular weight, $r_j$ is the rate of reaction $j$ that creates/consumes species $k$, $N$ is the reaction number, and $l_{k,j}$ is the particle number of species $k$ created or lost per collision of reaction $j$. The reaction rate , $r_j$, is expressed as

$$r_j = k_j \prod_{m=1}^{S} c_m^{\ \nu_{jm}}. \tag{2.10}$$

In this formula, $k$ is the rate coefficient, $S$ is the number of reactants, $\nu$ is the stoichiometric coefficients, and $c$ is the molar concentration of reactants. In sum, the heavy species equation, Eq. (2.6), sequentially describes the inertia term, advection, diffusion, drift and chemical kinetic of heavy species. Note that only $Q$-$1$ equations are used, since the mass fraction of feedstock gas is given by the mass constraint, $\omega_{Ar} = 1 - \sum_{k}^{Q-1} \omega_k$ .

The total mass density of heavy species, $\rho$, is obtained from ideal gas law,

$$\rho = \frac{P}{kT} \cdot \frac{M}{N_A}, \tag{2.11}$$

where $k$ is Boltzmann's constant, $T$ is the gas temperature equal to 300K, $P$ is the fixed gas pressure, 20mTorr, and $N_A$ is the Avogadro's constant. $M$ is the mole averaged molecular weight, which is given by

$$\frac{1}{M} = \sum_{k=1}^{Q} \frac{w_k}{M_k}. \tag{2.12}$$

The mean molecular weight $M$ is generally not a constant, since it is a function of mass fractions and molecular weight of various species, and the mass fractions, calculated by Eq. (2.6), are spatial and temporal dependent. Nevertheless, in inert gas discharges, all heavy species has the same mass and the mean molecular weight can be considered constant due to the fact that the sum of all species mass fractions is equal to 1. Accordingly, the total mass density is constant and the inertia term of Eq. (2.6) can be rewritten as

$$\frac{\partial(\rho w_k)}{\partial t} = \frac{\partial(\rho_k)}{\partial t} = \frac{\partial(M_k c_k)}{\partial t} = M_k \frac{\partial c_k}{\partial t} , \tag{2.13}$$

where $\rho_k$ is the mass density of species $k$, $M_k$ is the molecule weight, and $c_k$ is the molar density. The total flux boundary conditions that includes diffusion, drift and advection components, is used at the chamber wall, i.e., $\mathbf{\Gamma}_k = -\mathbf{n} \cdot \rho \omega_k (\mathbf{V}_k + \mathbf{u})$ .

## 2.3 Electromagnetic equations

To describe the electromagnetic field in the reactor, the Maxwell's equations are used to express the Ampere's law

$$(j\omega\sigma - \omega^2\varepsilon_0\varepsilon_r)\mathbf{A} + \nabla\times(\mu_0^{-1}\mu_r^{-1}\mathbf{B}) = \mathbf{J}_a, \tag{2.14}$$

where $j$ is the imaginary unit and $\omega$ is the angular frequency of the electric source. $\varepsilon_0$ and $\varepsilon_r$ are vacuum and relative permittivity, respectively. $\mu_0$ and $\mu_r$ are vacuum and relative permeability, respectively. $\mathbf{A}$ is the magnetic vector potential and $\mathbf{J}_a$ is the applied external current. $\sigma$ is the electron conductivity, calculated from the cold plasma approximation,

$$\sigma = \frac{n_e q^2}{m_e(v_e + j\varpi)}, \tag{2.15}$$

where $n_e$, $m_e$ and $q$ are the electron density, mass, and charge. $v_e$ is the collision frequency of electrons with neutrals. The magnetic insulation, $\mathbf{n}\times\mathbf{A} = 0$, are used at the boundaries.

### 2.4 Electrostatic equations

The Poisson's equation is used to calculate electrostatic field,

$$\begin{aligned} \mathbf{E} &= -\nabla V, \\ \nabla\cdot\mathbf{D} &= \rho_V, \end{aligned} \tag{2.16}$$

where $\rho_V$ is the charge density of the space. Zero potential boundary conditions are used at chamber walls and dielectric window surface.

~~~~~~~~~~~~~~~~~~~~~~~~~~~~~~~~~~~~~~~~~~~~~~~~~~~~~~~~~~~~~~~~~~~~~~~~~~~~~~~~

Catalogue of section III:

(III.1) Ar/$O_2$ plasma
(III.1.A) $\delta$-type anion density of fluid model and self-coagulation theory
(a) Steady state structure
(b) Temporal dynamics
(III.1.B) Space plasma characteristic
(III.2) Ar/$SF_6$ plasma
(III.2.A) Space-stratification phenomenon
(a) Stratification, anion Boltzmann relation and parabola theory
(b) Double layer and discontinuity theory
(c) Spontaneous self-coagulation of anions
(III.2.B) Self-coagulation-to-coil scheme of ions
(a) Pre-condition of self-coagulation
(b) Self-coagulation stage of anions
(c) Collapse of ambi-polar potential and electron transport change
(d) Mass point behavior and theory of post-self-coagulation stage
(e) Re-self-coagulation of minor cations and its astronomy significance
(f) Advective ambi-polar self-coagulation
(III.2.C) Pressure effect at high electronegativity
(a) Evolution of ions density profile beside for the self-coagulation
(b) Ellipse and flat-top theories
(c) Electron density profile at anion (heart) and its own (peripherical) self-coagulations
(d) Electron deviates the Boltzmann relation

~~~~~~~~~~~~~~~~~~~~~~~~~~~~~~~~~~~~~~~~~~~~~~~~~~~~~~~~~~~~~~~~~~~~~~~~~~~~~~~~

# III Results and discussion

## (III.1) Ar/$O_2$ plasma

### (III.1.A) $\delta$-type anion density of fluid model and self-coagulation theory

Variable definition declaration: in this section, for following the convention, without specific stress, the term, *ions*, is used to represent the positive ions. Anions are defined as negative ions.

In this section, the delta type anion density given the fluid model is given. Accordingly, the self-coagulation theory that produces the delta type distribution is described. More details about the model and theory can be found in our previous publication.

#### (a) Steady state structure

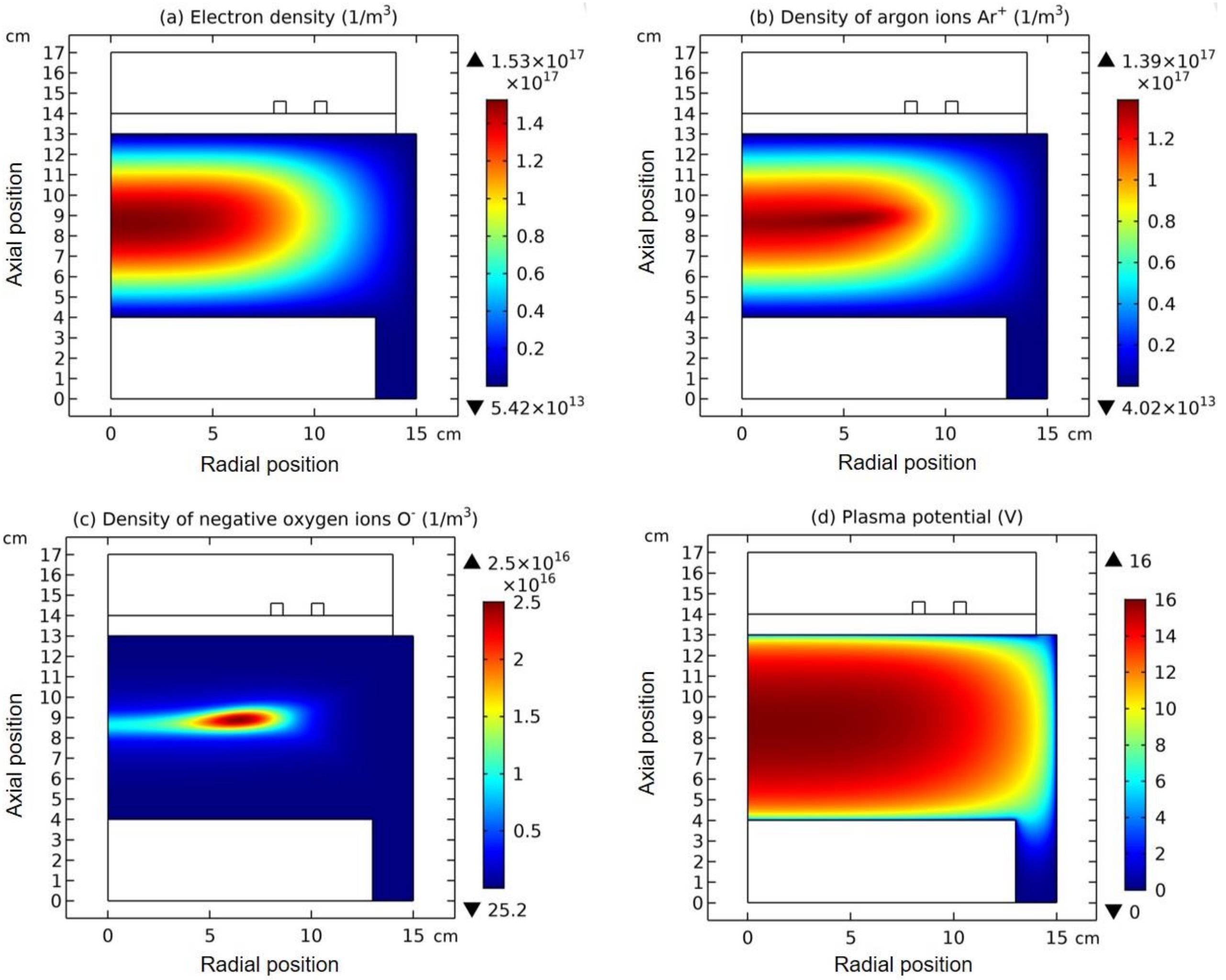

Figure 1 Densities of electron (a), argon ions (b), and negative oxygen ions (c), and the plasma potential (d) of inductively coupled $Ar/O_2$ plasma, given by the fluid model simulation at the discharge conditions of 300W, 30mTorr and 10% $O_2$ content.

At steady state, astonishing $O^-$ delta density is found in the fluid model simulation of inductively coupled $Ar/O_2$ plasma, at the discharge conditions of 300W, 30mTorr and 10% $O_2$ content, as shown in Fig. 1(c). Accordingly, the negative chemical source of $O^-$ species is found in Fig. 2(b), which is also very astonishing and rarely seen. To interpret the forming mechanism of $O^-$ delta, it is needed to investigate the temporal dynamics of this fluid model simulation.

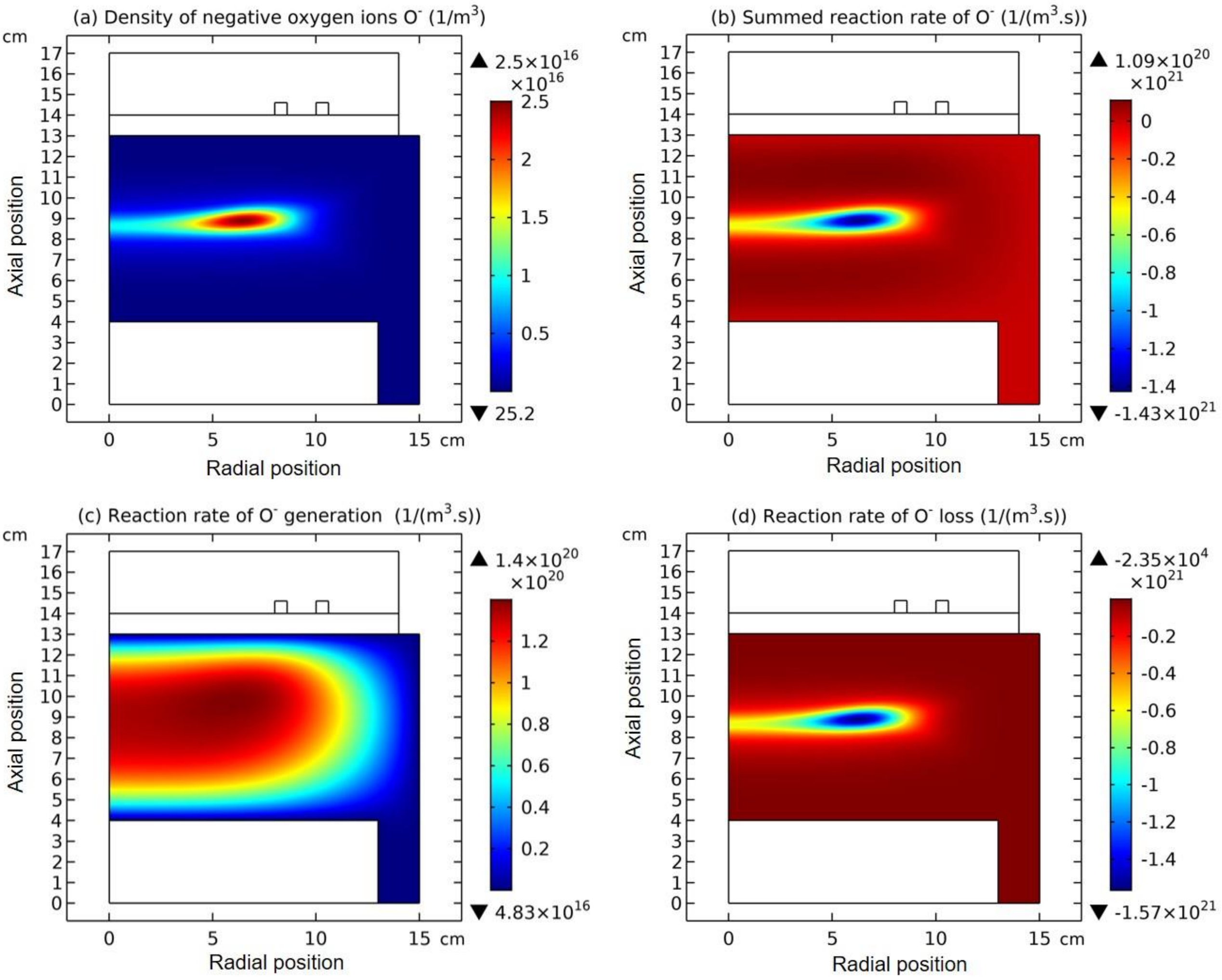

Figure 2 Density of negative oxygen ions $O^-$ (a), and the summed reaction rate (b), pure generation rate (c) and pure depletion rate (d) of $O^-$ of inductively coupled $Ar/O_2$ plasma, respectively, given by the fluid model simulation at the discharge conditions of 300W, 30mTorr and 10% $O_2$ content.

## (b) Temporal dynamics

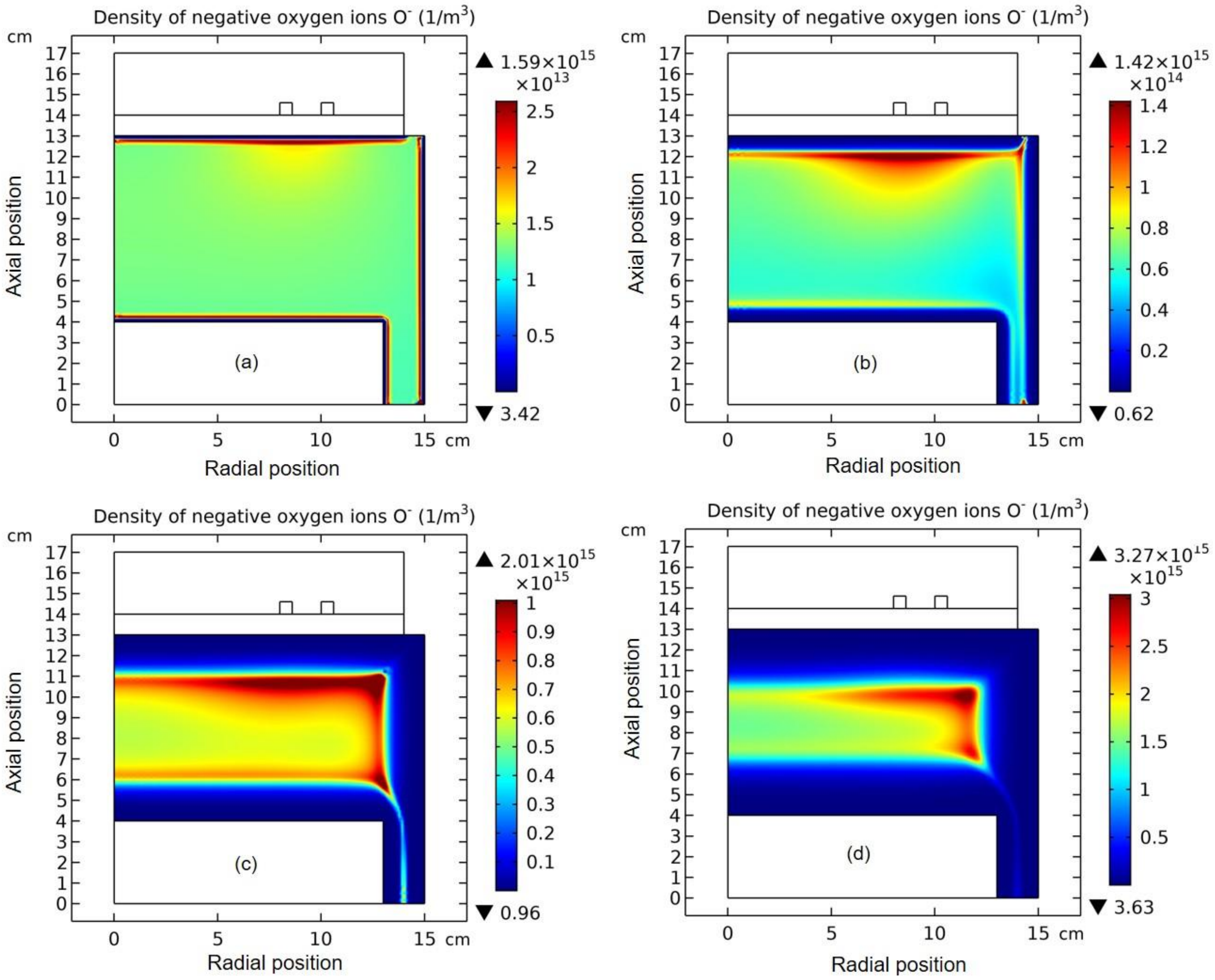

Figure 3 Evolution of $O^-$ density with time at the stage of drift accumulating $O^-$ species. The selected time points are sequentially (a) $1.0\times10^{-7}s$, (b) $1.0\times10^{-6}s$, (c) $5.109\times10^{-6}s$, and (d) $1.0\times10^{-5}s$. Data are given by the fluid model simulation of inductively coupled $Ar/O_2$ plasma at the discharge conditions of 300W, 30mTorr and 10% $O_2$ content.

The whole process for the $O^-$ delta formation in the weakly electronegative plasma can be divided into several parts. At first, the $O^-$ species are drifted and accumulated by the ambi-polar diffusion potential given by the electron and ions at small creation rate of $O^-$, as shown in Fig. 3. Then, as the $O^-$ species are pushed into the potential bottom, its negative chemical source ascribed to the substantial recombination reactions of $O^-$ with ions, $Ar^+$, $O^+$ and $O_2^+$, is formed, and accordingly the prototype $O^-$ delta is generated, as shown in Figs. 4 and 5. Besides for the negative source, the free diffusion of $O^-$ species is still needed for forming delta, which is presented in Fig. 6, where the $O^-$ species assemble in the potential bottom, upon comparison. In Fig. 7, the theory of self-coagulation arisen from the quasi-Helmholtz equation that consists of free diffusion and negative chemistry source is illustrated, which explains well the forming mechanism of $O^-$ delta. After the prototype, the $O^-$ delta still experiences the walk and refinement stages in Figs. 8 and 9, which finally produces the steady state $O^-$ delta structure.

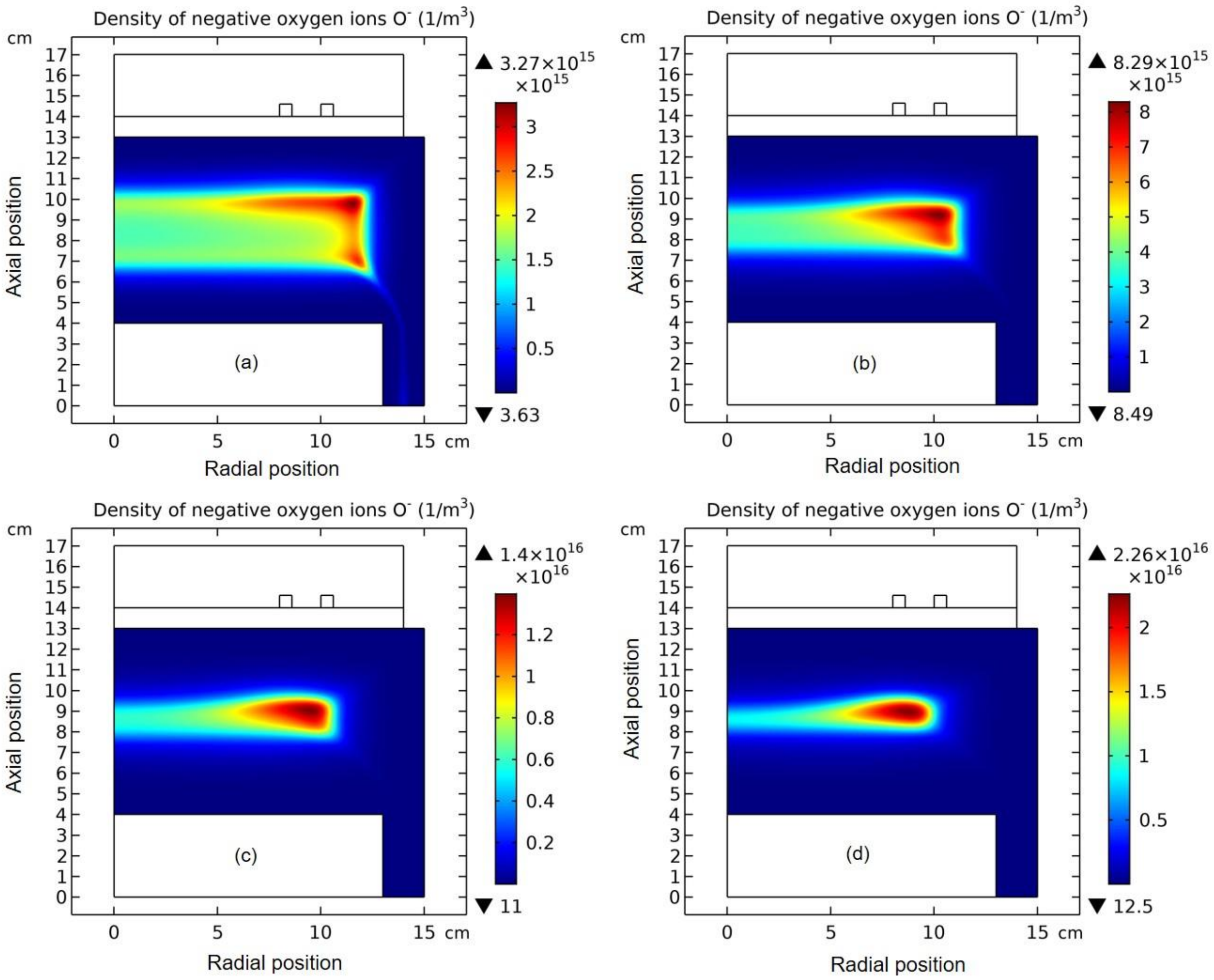

Figure 4 Evolution of $O^-$ density with time at the stage of prototype delta $O^-$ profile forming. The selected time points are sequentially (a) $1.0\times10^{-5}s$, (b) $1.778\times10^{-5}s$, (c) $2.371\times10^{-5}s$, and (d) $3.162\times10^{-5}s$. Data are given by the fluid model simulation of inductively coupled $Ar/O_2$ plasma at the discharge conditions of 300W, 30mTorr and 10% $O_2$ content.

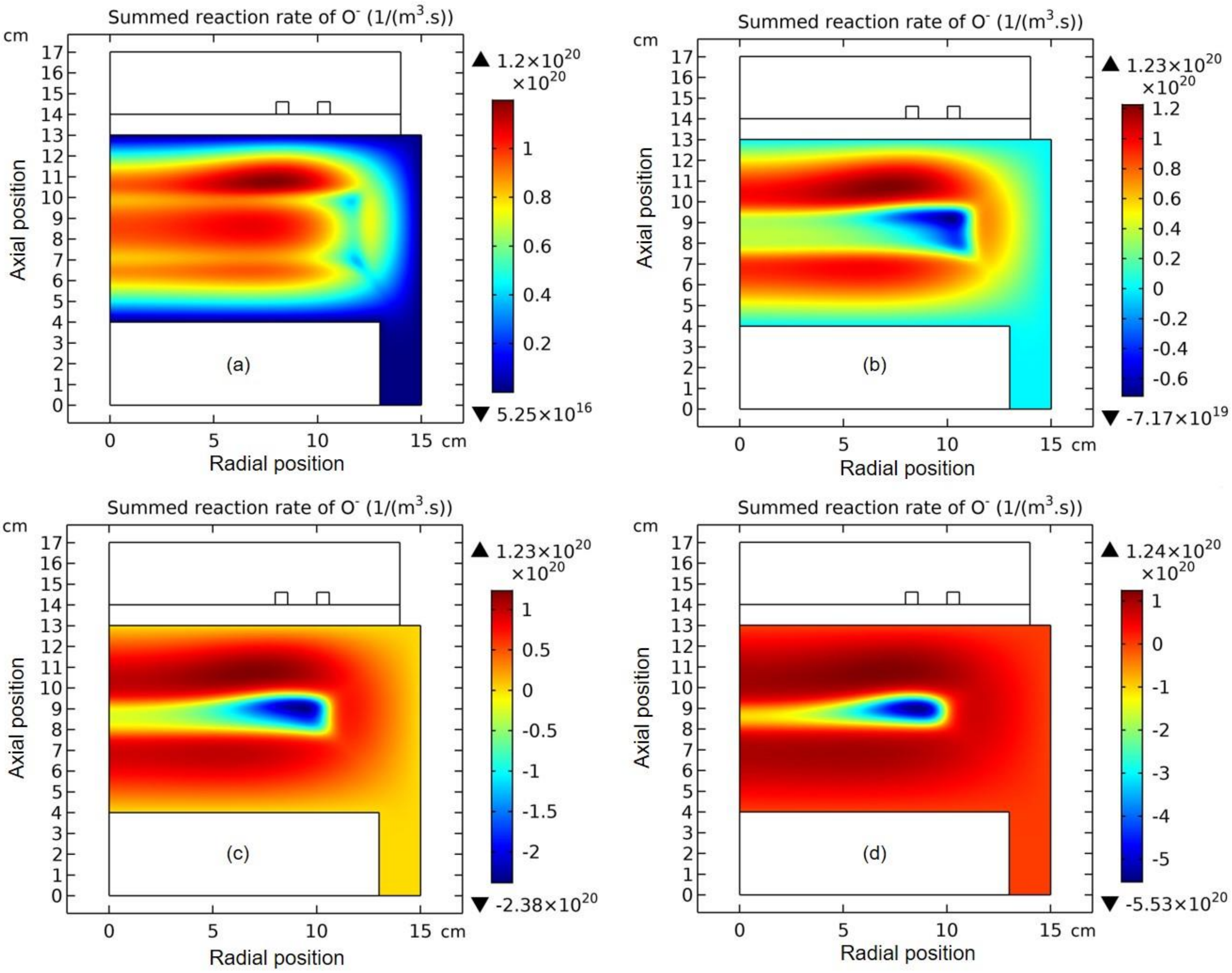

Figure 5 Evolution of $O^-$ chemical source with time at the stage of prototype delta $O^-$ profile forming. The selected time points are sequentially (a) $1.0\times10^{-5}s$, (b) $1.778\times10^{-5}s$, (c) $2.371\times10^{-5}s$, and (d) $3.162\times10^{-5}s$. Data are given by the fluid model simulation of inductively coupled $Ar/O_2$ plasma at the discharge conditions of 300W, 30mTorr and 10% $O_2$ content. This figure is used to illustrate that when the negative source is formed, the $O^-$ species starts to self-coagulate.

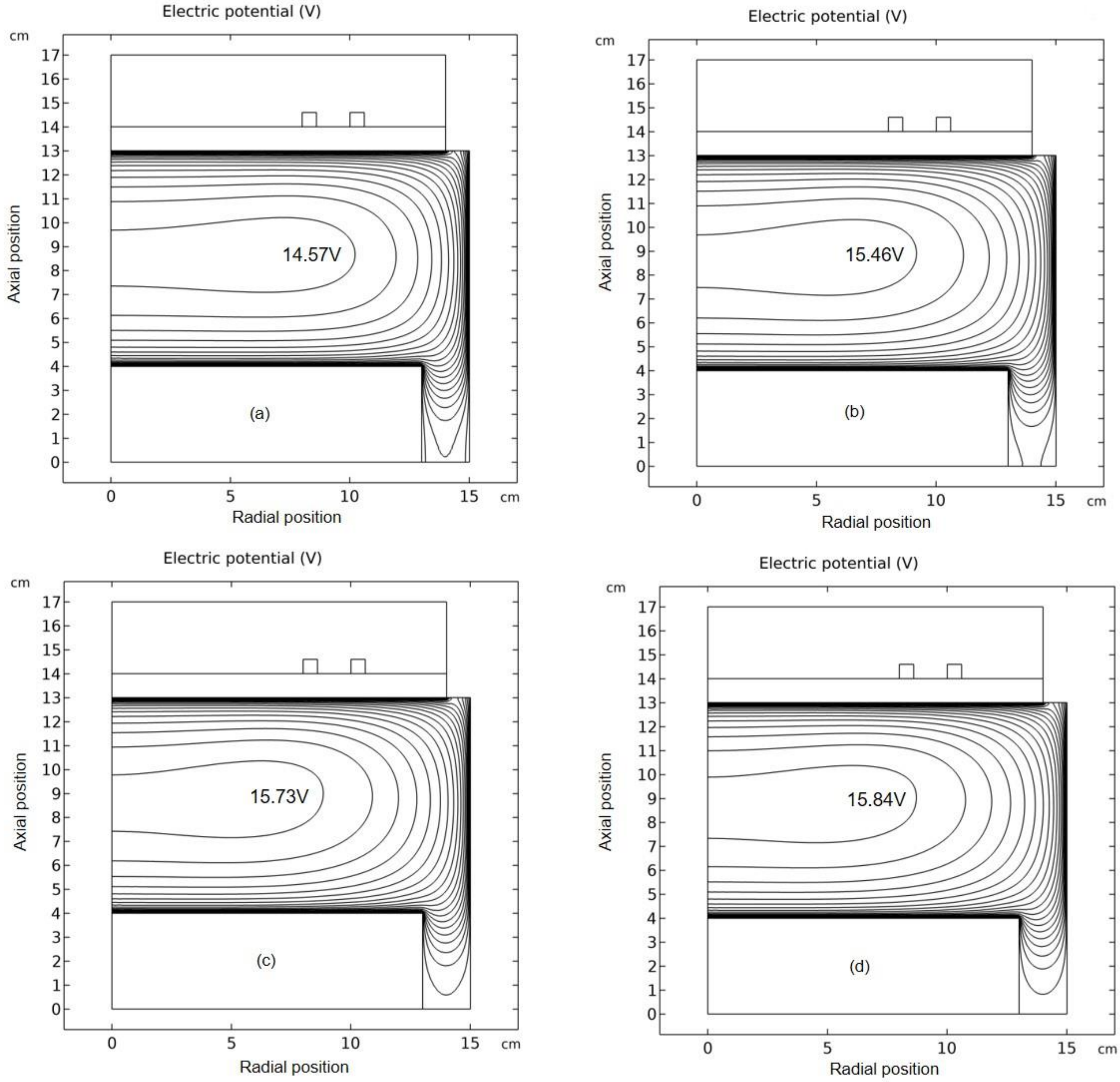

Figure 6 Evolution of plasma potential contour with time at the stage of prototype delta $O^-$ profile forming. The selected time points are sequentially (a) $1.0\times10^{-5}s$, (b) $1.778\times10^{-5}s$, (c) $2.371\times10^{-5}s$, and (d) $3.162\times10^{-5}s$. Data are given by the fluid model simulation of inductively coupled $Ar/O_2$ plasma at the discharge conditions of 300W, 30mTorr and 10% $O_2$ content. This figure indicates that the self-coagulation happens in the potential bottom, thus unveiling the second necessary condition of self-coagulation, free diffusion.

Combination of negative source and free diffusion

$$-D_{O-}\nabla^2 n_{O-} = -n_{O-} n_{Ar+} k_{rec.} = -n_{O-} \nu_{rec.}.$$

Quasi-Helmholtz equation deduction

$$\nabla^2 n_{O-} - n_{O-}\frac{\nu_{rec.}}{D_{O-}} = \nabla^2 n_{O-} - n_{O-} k^2 = 0,$$
$$\nabla^2 n - n k^2 = 0.$$

Variable separation method

$$\frac{1}{\rho}\frac{\partial}{\partial \rho}\left(\rho\frac{\partial n}{\partial \rho}\right) + \frac{\partial^2 n}{\partial z^2} - k^2 n = 0,$$
$$n(\rho, z) = R(\rho) Z(z),$$
$$Z'' + \nu^2 Z = 0,$$
$$\frac{d^2 R}{d\rho^2} + \frac{1}{\rho}\frac{dR}{d\rho} - \left(k^2 + \nu^2\right) R = 0.$$

Solution at the definitions of axially homogeneousness and radial nature (finite value of axis)

$${\nu_m}^2 = m^2\pi^2 / l^2,$$
$$Z_m = \sin(m\pi z / l),$$
$$Z = \sum_{m=0}^{\infty} c_m Z_m = \sum_{m=0}^{\infty} c_m \sin(m\pi z / l).$$
$$\frac{d^2 R}{d\rho^2} + \frac{1}{\rho}\frac{dR}{d\rho} - \left(k^2 + {\nu_m}^2\right) R = 0,$$
$$R = d_m I_0(\sqrt{k^2 + {\nu_m}^2}\rho) = d_m I_0(\sqrt{k^2 + m^2\pi^2 / l^2}\rho).$$
$$n(\rho, z) = R(\rho) Z(z) = \sum_{m=0}^{\infty} c_m \sin(m\pi z / l)\cdot d_m I_0(\sqrt{k^2 + {\nu_m}^2}\rho)$$
$$= \sum_{m=0}^{\infty} a_m \sin(m\pi z / l)\cdot I_0(\sqrt{k^2 + m^2\pi^2 / l^2}\rho).$$

Limit solution at the radial homogeneousness of border

$$n(\rho, z) = R(\rho) Z(z) = \sum_{m=0}^{\infty} a_m \sin(m\pi z / l)\cdot I_0(\sqrt{k^2 + m^2\pi^2 / l^2}\rho)$$
$$= \lim_{m\to\infty}\left[a_m \sin(m\pi z / l)\cdot\infty\right] = \lim_{m\to\infty}\left[a_m \sin(m\pi z / l)\cdot\lim_{z\to 0}\frac{1}{z}\right]$$
$$= \lim_{z\to 0}\left[\lim_{m\to\infty} a_m \sin(m\pi z / l)\cdot\frac{1}{z}\right] = \lim_{z\to 0}\left[\lim_{m\to\infty} a_m \cdot\frac{\sin(m\pi z / l)}{z\pi / l}\cdot\frac{\pi}{l}\right]$$
$$= \lim_{\zeta\to 0}\left[\lim_{m\to\infty} a_m' \cdot\frac{1}{\pi}\frac{\sin(m\zeta)}{\zeta}\right] = \lim_{\zeta\to 0}\left[a_\infty' \lim_{m\to\infty}\frac{1}{\pi}\frac{\sin(m\zeta)}{\zeta}\right]$$
$$= a_\infty' \lim_{\zeta\to 0}\delta(\zeta).$$

Figure 7 Self-coagulation analytic theory deduced from fluid model simulation

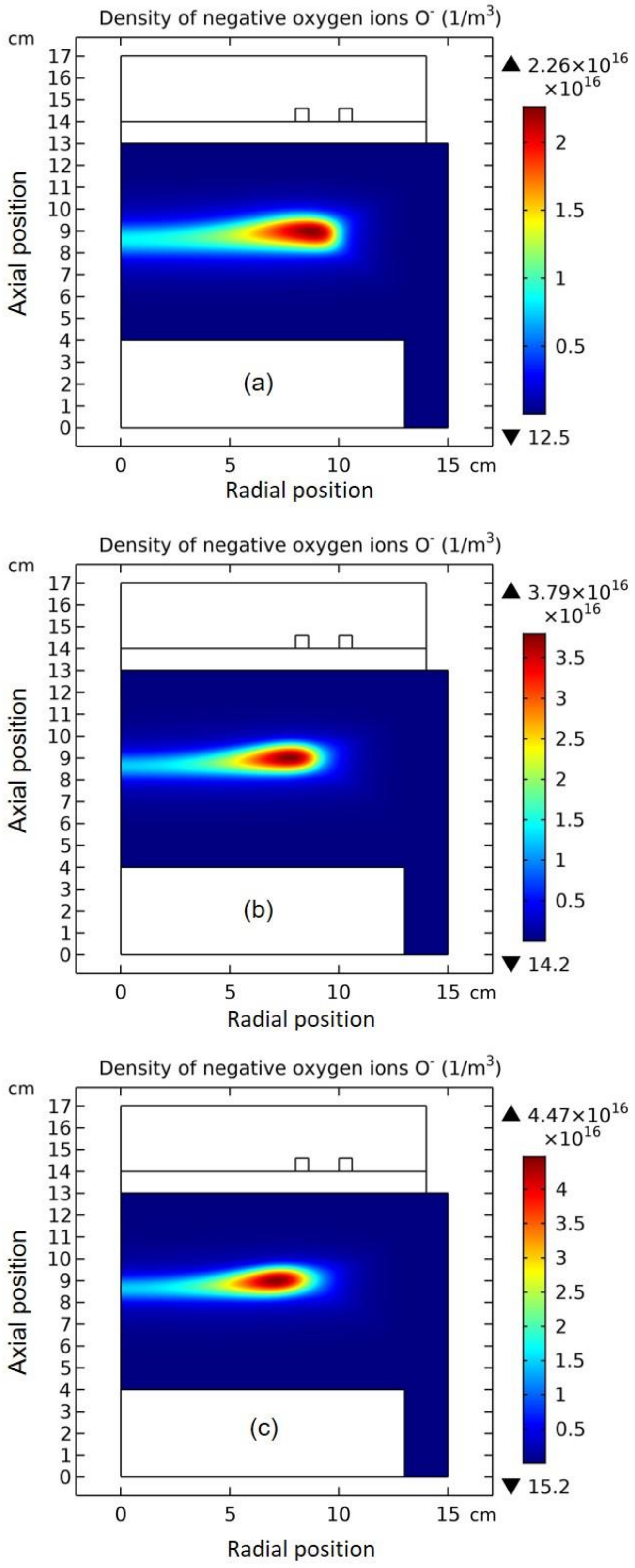

Figure 8 $O^-$ density evolution with time in the delta walk stage. The selected time points are sequentially (a) $3.162\times10^{-5}s$, (b) $5.109\times10^{-5}s$, and (c) $7.499\times10^{-5}s$. The data are given by the fluid model simulation of inductively coupled Ar/$O_2$ plasma at the discharge conditions of 300W, 30mTorr and 10% $O_2$ content.

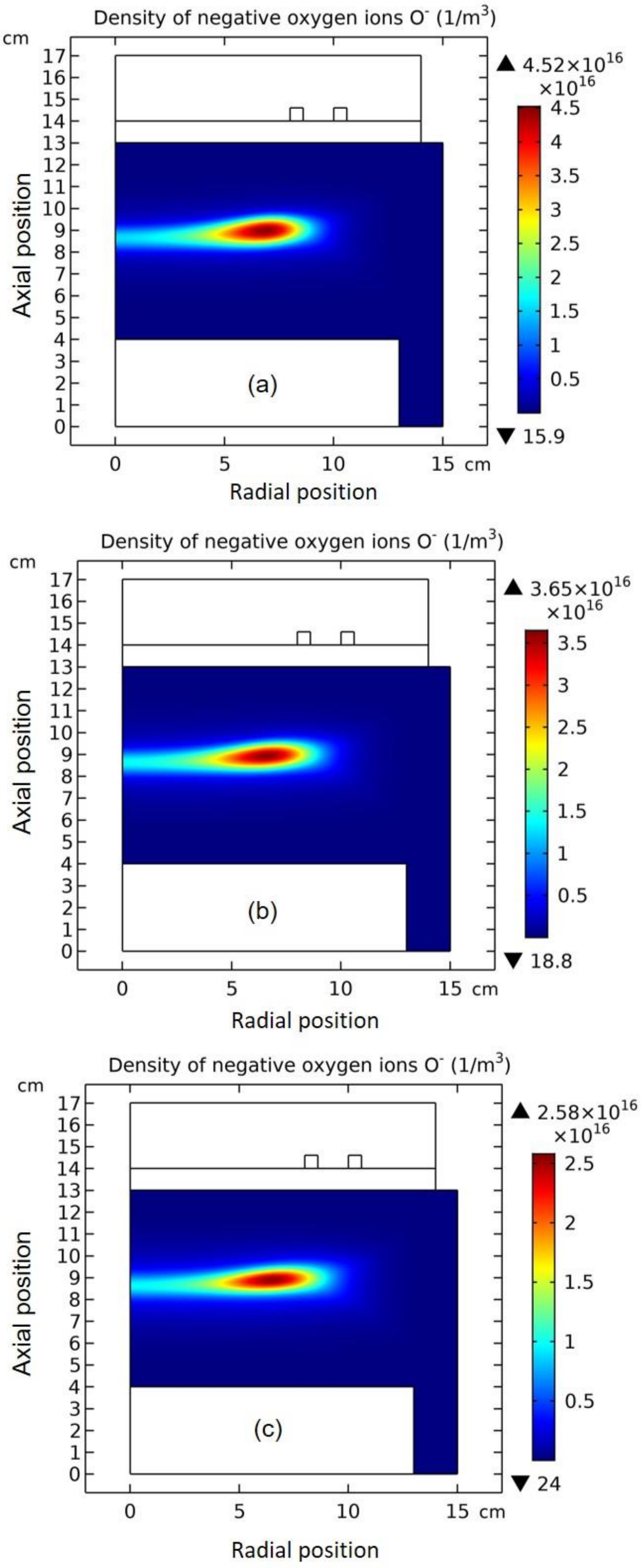

Figure 9 $O^-$ density evolution with time in the delta refining stage. The selected time points are sequentially (a) $1.0\times10^{-4}s$, (b) $5.109\times10^{-4}s$, and (c) $3.162\times10^{-3}s$. The data are given by the fluid model simulation of inductively coupled $Ar/O_2$ plasma at the discharge conditions of 300W, 30mTorr and 10% $O_2$ content.

### (III.1.B) Space plasma characteristic

In this section, the space plasma characteristic that is produced in the laboratory is exhibited. Besides, the condition that self-coagulation is not occurred is discussed. More details can be found in our previous publication. The connection between the laboratory and space plasmas are built; see more example in Section III.2(B).

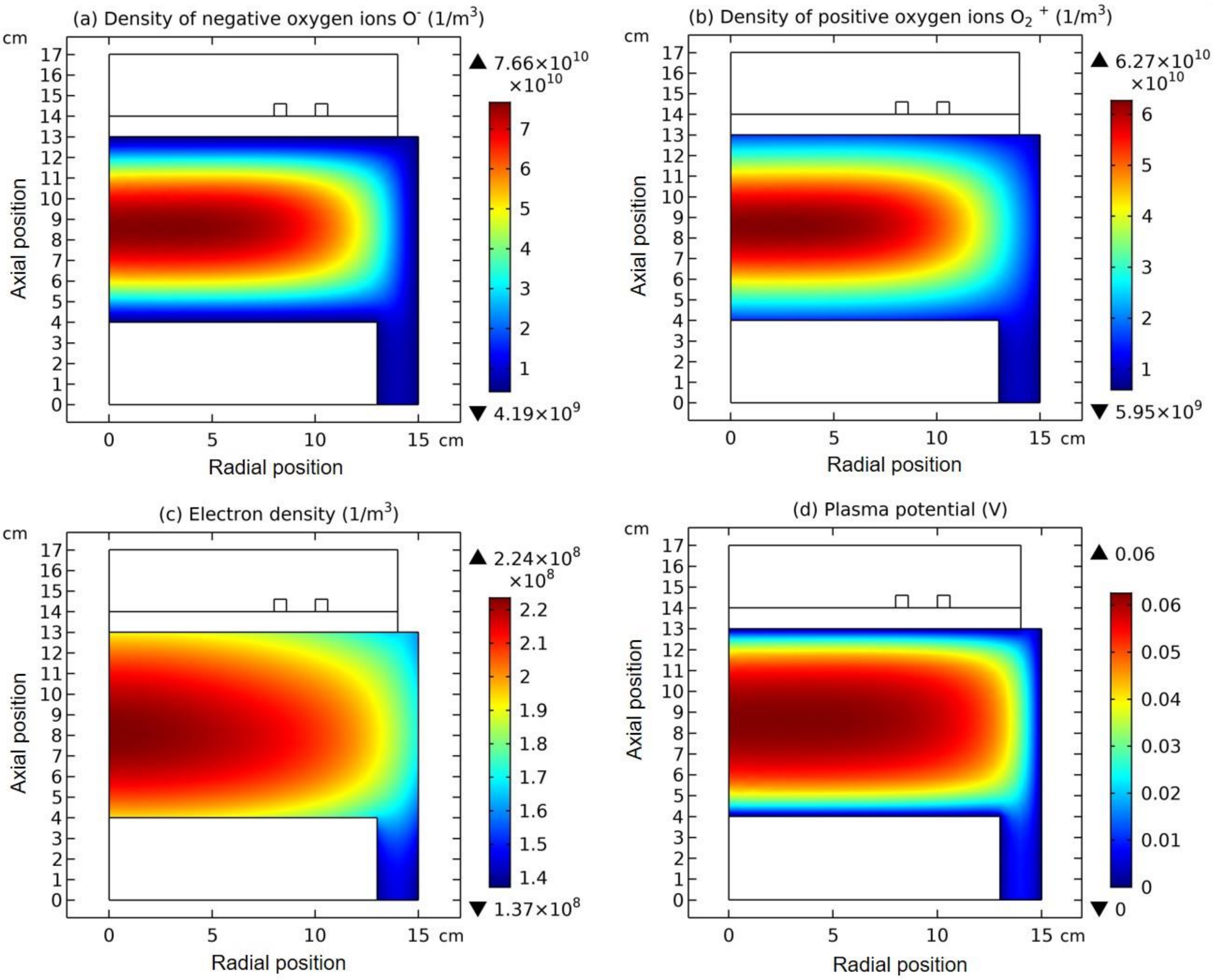

Figure 10 Densities of negative oxygen atomic ions $O^-$ (a), negative oxygen molecular ions $O_2^-$ (b), and electron (c), and the plasma potential (d) of inductively coupled Ar/$O_2$ plasma, given by the fluid model simulation at the discharge conditions of 300W, 10mTorr and **90%** $O_2$ content.

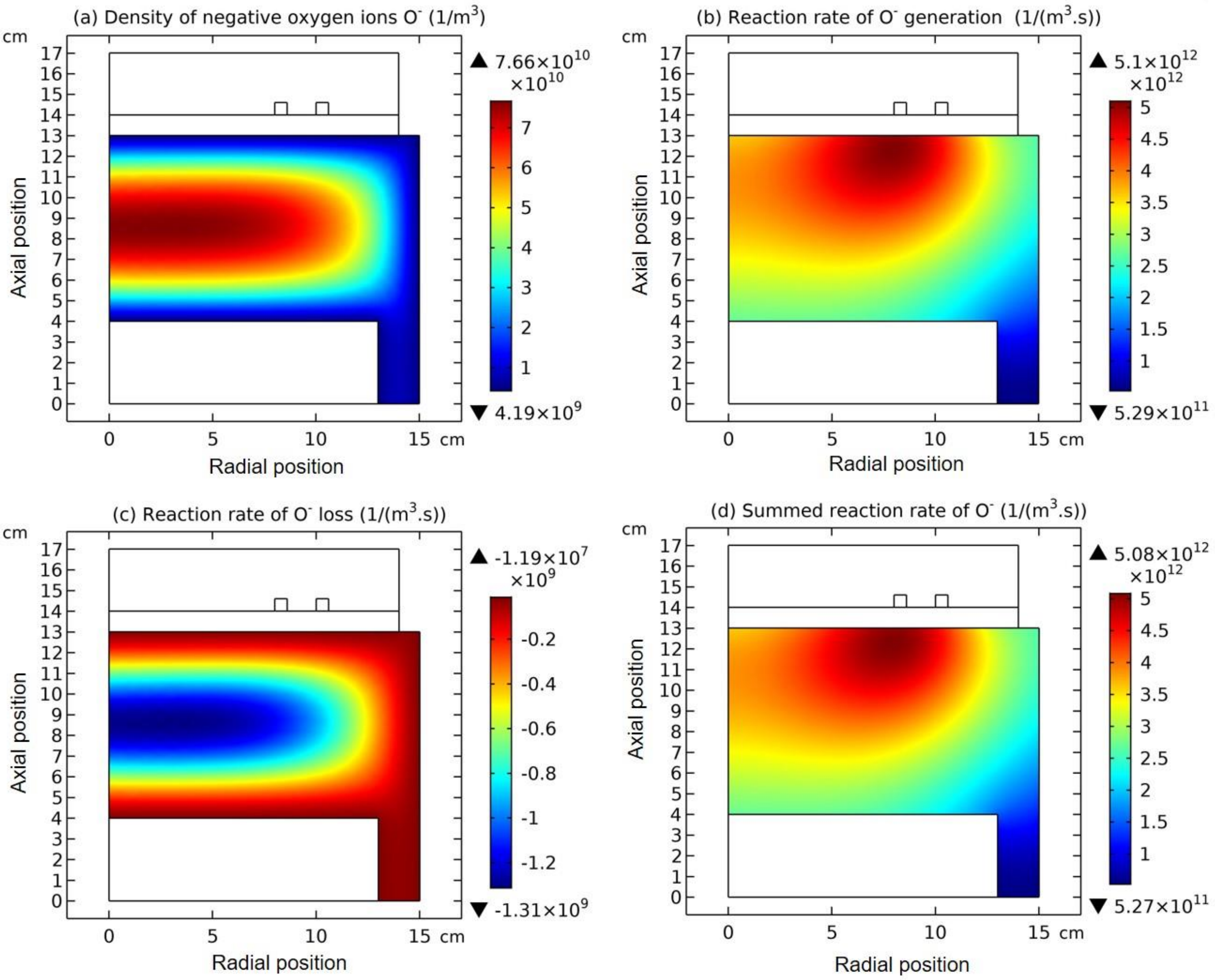

Figure 11 Density of negative oxygen atomic ions $O^-$ (a), and the generation rate (b), depletion rate (c) and the summed reaction rate (d) of $O^-$, in the inductively coupled $Ar/O_2$ plasma simulated by the fluid model at the discharge conditions of 300W, 10mTorr and **90%** $O_2$ content.

The analytic theory above implies one of the disappearing conditions of self-coagulation, *i.e.*, the negative chemistry source is unable to form. This is indeed verified by another $Ar/O_2$ inductive plasma fluid simulation at the discharge conditions of 300W, 10mTorr and 90% $O_2$ content. The $O^-$ species density profile exhibits parabolic feature at the positive chemistry source in Figs. 10 and 11. Of more significance is that this gaseous discharge that should originally produce the low-temperature laboratory plasma now exhibits space plasma characteristic, by means of the very low plasma density, $10^4$ $cm^{-3}$, analogous to the aurora. This is logic since only at such low ions densities, the recombination loss rate can be negligible, regarding to its quadratic function of plasma density.

## (III.2) Ar/$SF_6$ plasma

Variable definition declaration: in this section, the term, *ions*, is used to define the general charged heavy species. It thereby includes both the positive and negative ions. For better distinguishing the positive and negative ions, the terms, *cations and anions*, are used. Instead, the word, *ions*, means the sum of cations and anions. A new convention.

### (III.2.A) Space-stratification phenomenon

Many early works pointed out that electronegative plasma bulk consists of the electronegative core and electropositive halo, at high enough electronegativities. The fluid model simulation of electronegative plasma presents more details for the formation of stratification. It arises from the discharge SEQUENCE. At initial, the discharge mainly produces electron and cation and the ambi-polar diffusion potential is established. At this time, the anion generation rate and its density are low, and the Boltzmann relation cannot be reached and the anion drifts inward and hence accumulates. When ionization creates enough electrons, the attachment rate grows and anion source is strong enough for establishing its Boltzmann relation. As this balance is built, the space is stratified naturally into electronegative (core) and electropositive (edge) parts. Besides for the stratification, the electronegative discharge displays the spatial characteristics of parabola and self-coagulation in the core, given by both the fluid model and analytic theory.

#### (a) Stratification, anion Boltzmann relation and parabola theory

As seen in Figs. 12 and 13, the space-stratification structure is well predicted by the fluid model in an Ar/$SF_6$ inductive plasma at the discharge conditions of 300W, 10mTorr and 10% $SF_6$ content. In addition, the parabola and self-coagulation appear along the ions densities profiles, and the anion Boltzmann relation is given in Fig. 14, all in the electronegative core. At the Boltzmann relations of both electron and anion, the parabola theory of Lichtenberg *et al* predicts well the parabola profile illustrated in Fig. 15, at the assumption of ambi-polar diffusion of triple-species system (electron, cation and anion) of strong electronegativity. The anions are all expelled in the edge, a conventional electropositive plasma halo, connected to sheaths. It implies nonlinear dynamics at the interface of the core and halo.

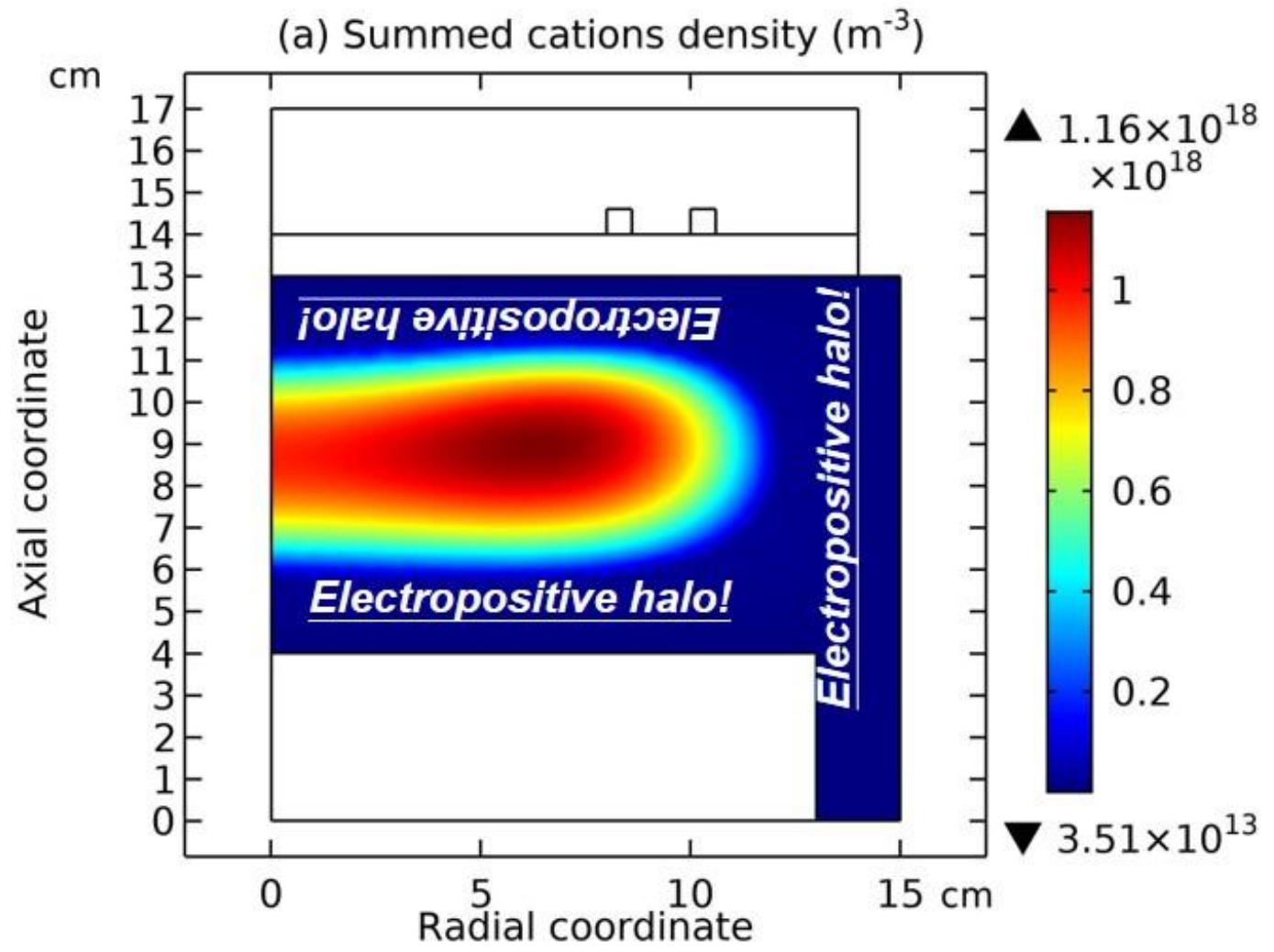

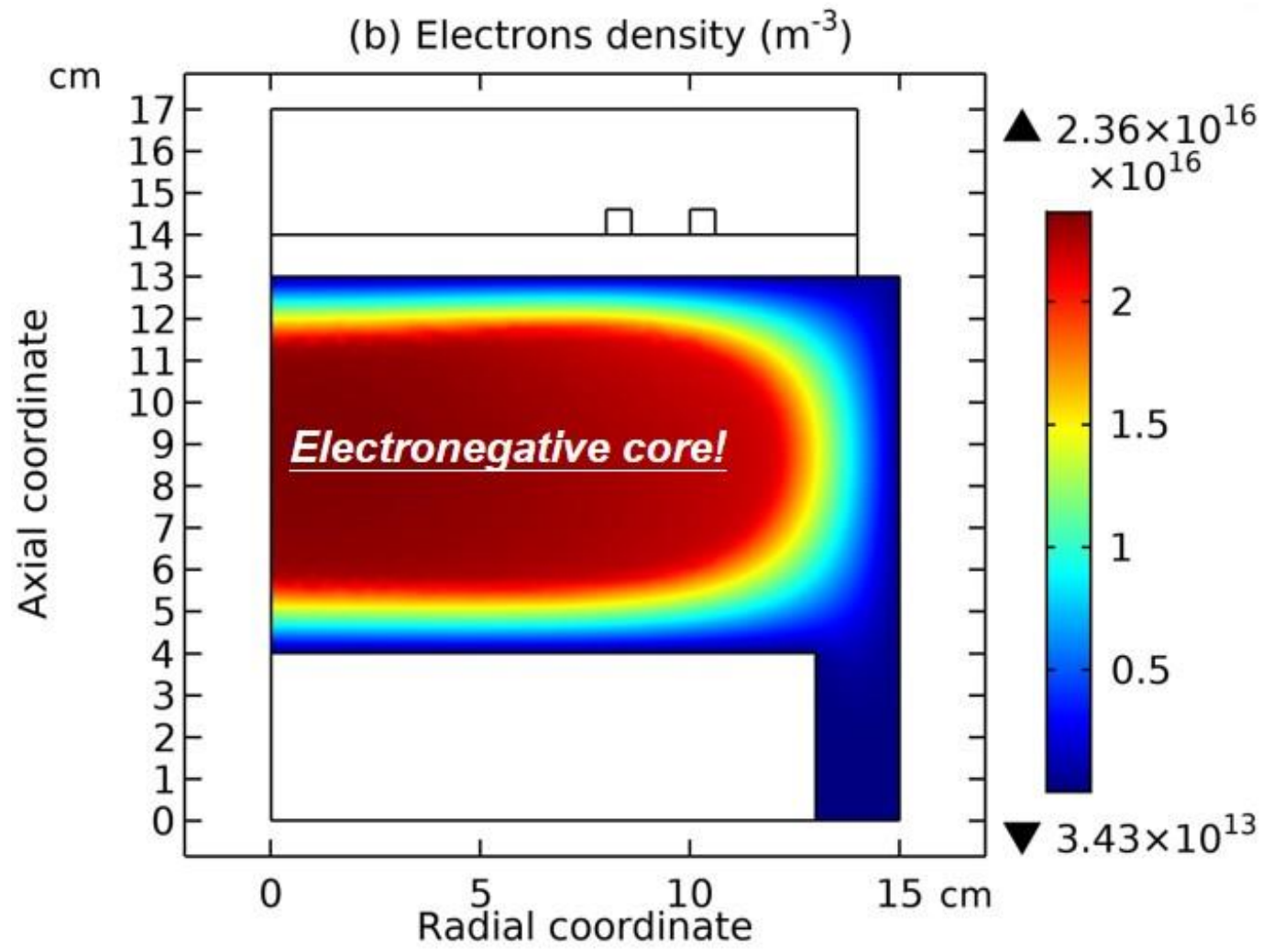

Figure 12 Densities of summed cations (a) and electron (b) of $Ar/SF_6$ inductively coupled plasma given by the fluid model simulation at the discharge conditions of 300W, 10mTorr and 10% $SF_6$ content.

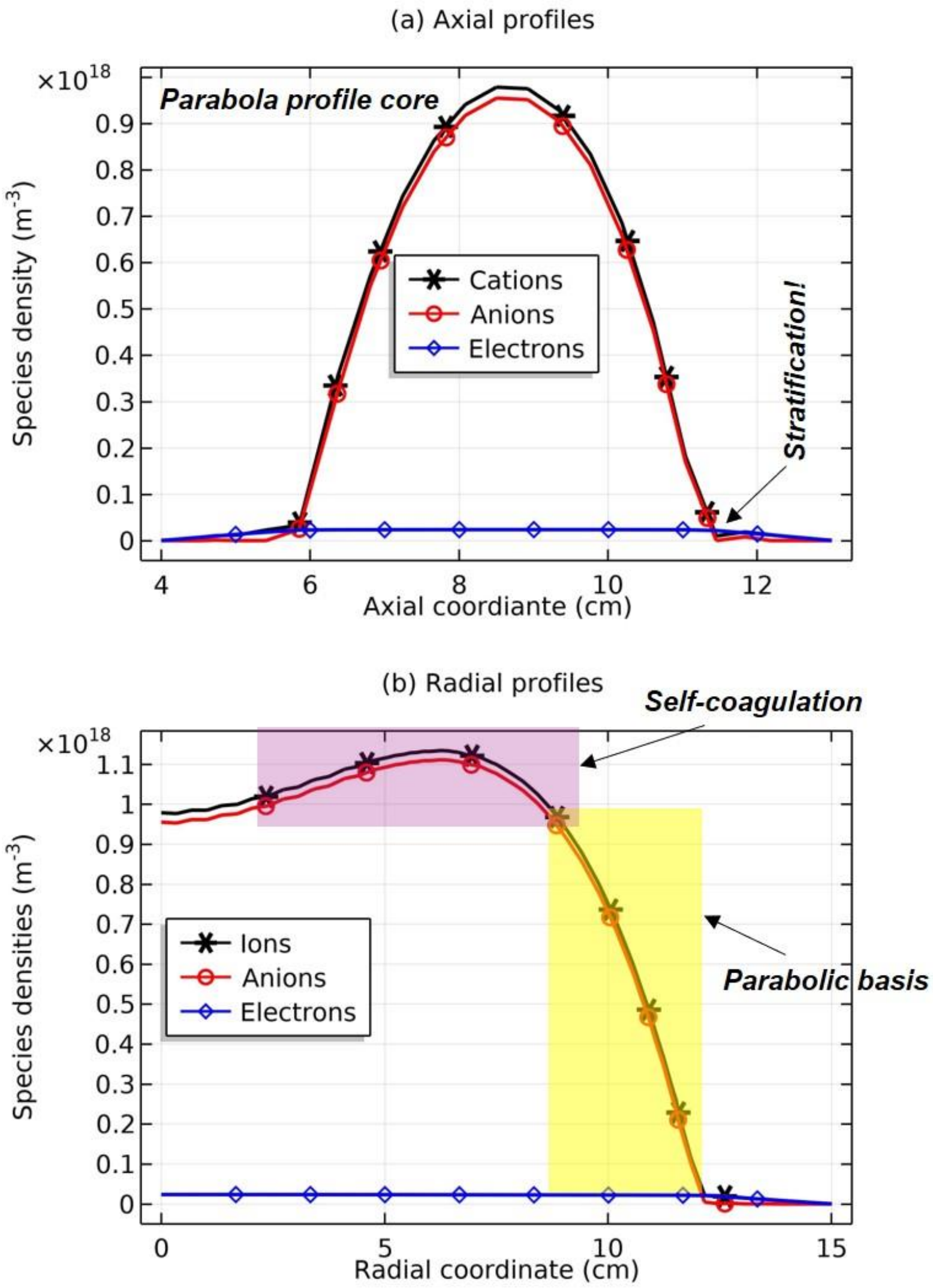

Figure 13 Axial (a) and radial (b) profiles of the summed cations and anions densities, and the electron density, in the Ar/$SF_6$ inductively coupled plasma simulated by the fluid model at the discharge conditions of 300W, 10mTorr and 10% $SF_6$ content.

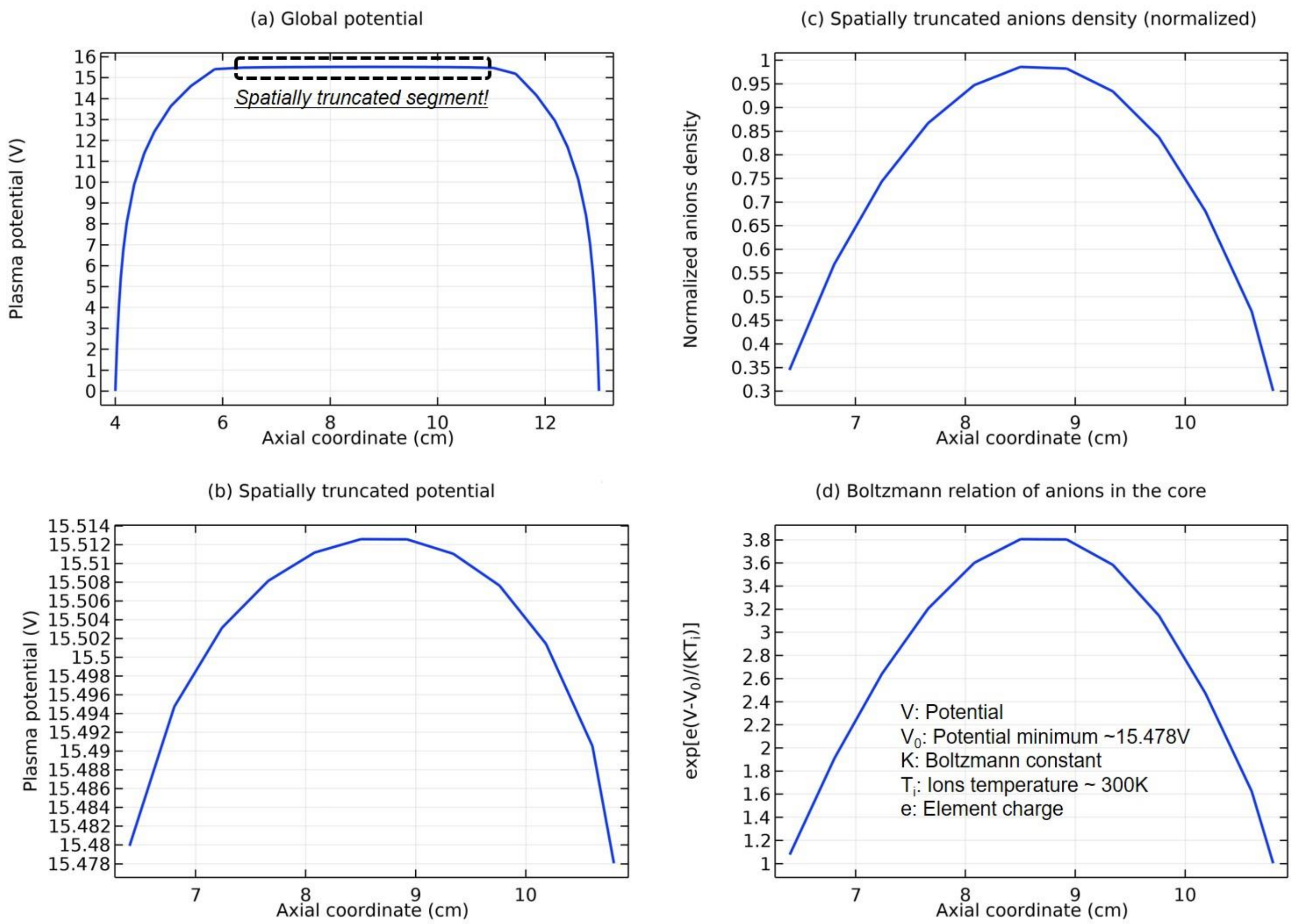

Figure 14 Boltzmann relation of anions in the electronegative core when the basic discharge structure is stratified parabolic profile, at the same discharge conditions as in Fig. 12. In (a), the global axial plasma potential is exhibited. During the central flatten part of global potential curve (essentially the electronegative core part), a segment is truncated, and in this truncated segment, the detail of potential (tiny change) (b), normalized anions density (c) and the Boltzmann equilibrium of anions with potential (d) are sequentially shown. The discrepancy in the magnitudes of (c,d) is ascribed to the robust truncation operation, and it is believed not to influence the reasonability of anions Boltzmann equilibrium illustrated. In the present fluid model, the temperatures of ions and anions are both assumed to room temperature.

Cation flux at drift and diffusion approximation

$$\Gamma_+ = \Gamma_- + \Gamma_e,$$

$$n_+ = n_- + n_e, \quad \alpha = n_-/n_e,$$

$$\Gamma_+ = -D_+\nabla n_+ + n_+\mu_+E,$$

$$\Gamma_- = -D_-\nabla n_- - n_-\mu_-E,$$

$$\Gamma_e = -D_e\nabla n_e - n_e\mu_eE.$$

$$\Gamma_+ = -\frac{(\mu_e+\mu_-\alpha)D_+ + \mu_+(1+\alpha)D_e(\nabla n_e/\nabla n_+) + \mu_+(1+\alpha)D_-(\nabla n_-/\nabla n_+)}{\mu_e+\mu_-\alpha+\mu_+(1+\alpha)}\nabla n_+.$$

Ambi-polar diffusion coefficient at Boltzmann relation of anion and electron

$$\Gamma_+ = -D_{a+}\nabla n_+,$$

$$\gamma = T_e/T_i,$$

$$\frac{\nabla n_-}{n_-} = \gamma\frac{\nabla n_e}{n_e}.$$

$$\nabla n_+ = \nabla n_- + \nabla n_e,$$

$$\frac{\nabla n_e}{\nabla n_+} = \frac{1}{1+\gamma\alpha}, \quad \frac{\nabla n_-}{\nabla n_+} = \frac{\gamma\alpha}{1+\gamma\alpha}.$$

$$\frac{D_-}{D_+} = \frac{\mu_-}{\mu_+}, \quad \frac{D_e}{D_+} = \gamma\frac{\mu_e}{\mu_+},$$

$$D_{a+} = D_+\frac{(1+\gamma+2\gamma\alpha)\left(1+\alpha\frac{\mu_-}{\mu_e}\right)}{(1+\gamma\alpha)\left(1+\frac{\mu_+}{\mu_e}(1+\alpha)+\frac{\mu_-}{\mu_e}\right)},$$

Four approximations on cation transport equation

$$\mu_-/\mu_e,\ \mu_+/\mu_e \ll 1, \quad D_{a+} \cong D_+\frac{1+\gamma+2\gamma\alpha}{1+\gamma\alpha}.$$

$$\alpha \gg 1, \quad D_{a+} \cong 2D_+$$

$$\frac{n_e}{n_{e0}} = \left(\frac{n_-}{n_{-0}}\right)^{1/\gamma}, \quad n_e = n_{e0}$$

$$-\frac{d}{dx}\left(D_{a+}(\alpha)\frac{dn_+}{dx}\right) = K_{iz}n_0n_e - K_{rec}n_+n_-,$$

Parabola solution at four approximations above

$$-2D_+\frac{d^2n_+}{dx^2} = K_{iz}n_0n_{e0},$$

$$\frac{n_+}{n_{e0}} = \alpha_0\left(1-\frac{x^2}{l^2}\right)+1,$$

$$\alpha_0 = n_{-0}/n_{e0},$$

where $l$ is the nominal position where $\alpha = 0$.

Figure 15 Parabola theory of electronegative core region, given by Lichtenberg *et al*.

### (b) Double layer and discontinuity theory

The interface of core and edge separates two transport schemes, *i.e.*, the triple-species ambi-polar diffusion at the interface left and normal electron-cation (electron-ion, defined precedingly in electropositive plasma) ambi-polar diffusion at the right. In the analytic theories, the ambi-polar diffusion coefficients of two- and triple- species systems are quite different, about $\gamma D_+$ versus $2D_+$. To ensure the operation of two different transports, a double layer structure is formed at the interface, modelled as dipole in Fig. 16. Moreover, the dipole at the limit that the length tends to zero is indeed a breaking point inserted in the density and potential profiles, illustrated in Fig. 17 and Eq. (3.1).

$$E_o = \frac{q}{\pi\varepsilon_0 l},$$

$$\lim_{l\to 0} E_0 = \lim_{l\to 0}\frac{q}{\pi\varepsilon_0 l} = \infty,$$

$$E_A^2 = E_+^2 + E_-^2 - 2E_+E_-\cos\theta,$$

$$E_A^2 = \left(\frac{q}{4\pi\varepsilon_0}\right)^2\left(\frac{1}{r^2}+\frac{1}{r'^2}-\frac{2}{rr'}\cdot\frac{r^2+r'^2-l^2}{2rr'}\right),$$

$$\lim_{l\to 0} E_A^2 = \lim_{l\to 0}\left(\frac{q}{4\pi\varepsilon_0}\right)^2\left(\frac{1}{r^2}+\frac{1}{r'^2}-\frac{2}{rr'}\cdot\frac{r^2+r'^2-l^2}{2rr'}\right)\sim 0, \text{ at the condition of } l\sim 0, r\sim r'. \quad \text{Eq. (3.1)}$$

In Fig. 18, the axial profile of net charge density is plotted where two asymmetric dipoles form laterally. The asymmetry that negative pole is heavier verifies that fact the double layer arises from the accumulation of negative charge carrier, anions. In Figs. 19 and 20, the species densities, plasma potential and charge density in the processes of forming basic stratification and final convergence are given. The density and potential in Fig. 19 when stratification is being formed vibrate once with time. The vibration ascribed to the build and collapse period of anion Boltzmann relation does not transmit through the plasma since it lacks continue medium and is truncated by the discontinuity of dipole field at the limit. This is the physics essence of stratification and double layer, as we understood.

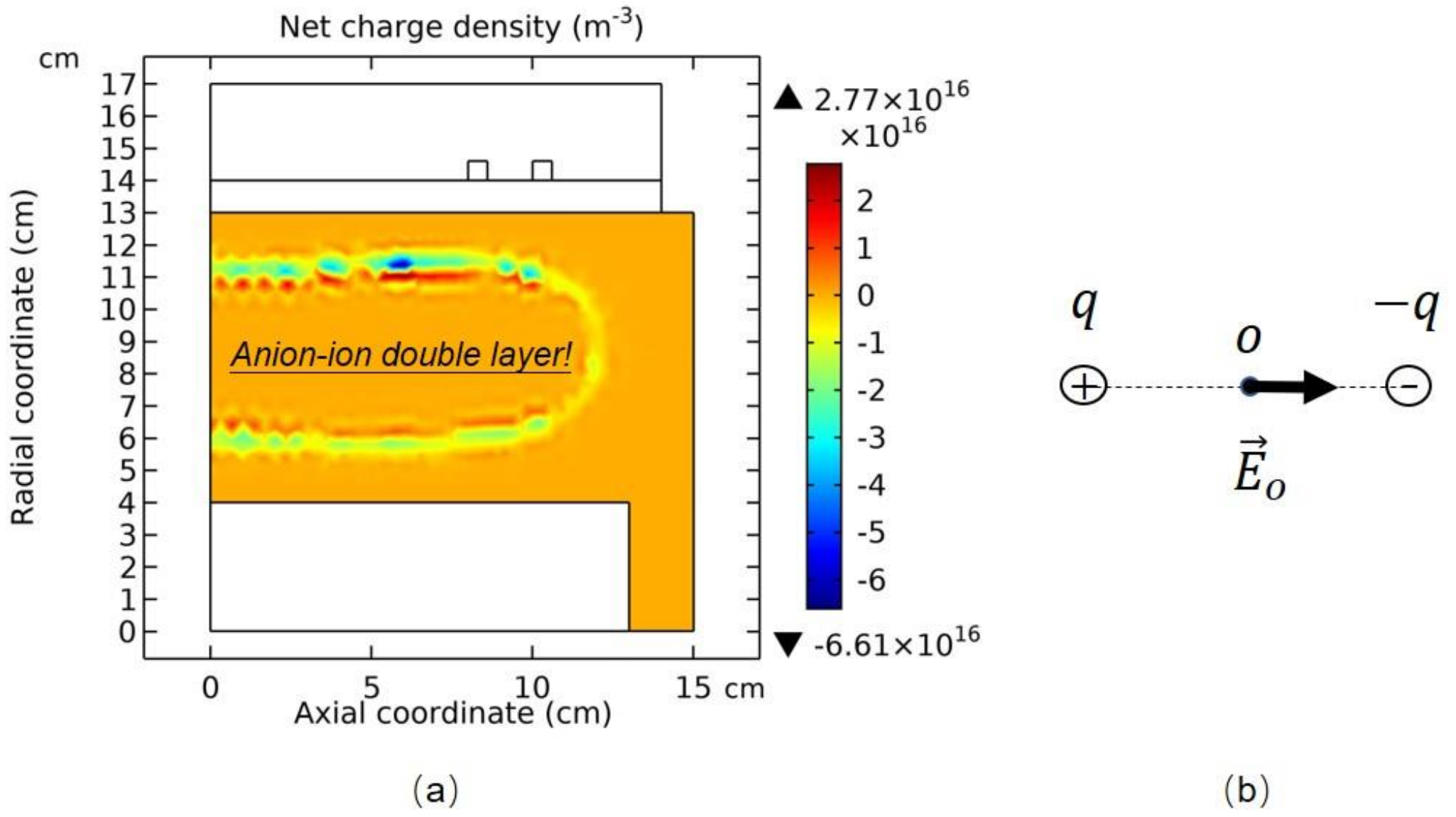

Figure 16 In (a), the net charge density of Ar/$SF_6$ inductive plasma is given, which exhibits double layer structure, *i.e.*, combination of two differently charged layers. In (b), this double layer is modelled as dipole of electromagnetics. The net charge density is given by the fluid model simulation at the discharge conditions of 300W, 10mTorr and 10% $SF_6$ content. The figure data is sampled at the time point that the stratification structure is basically formed in the simulation.

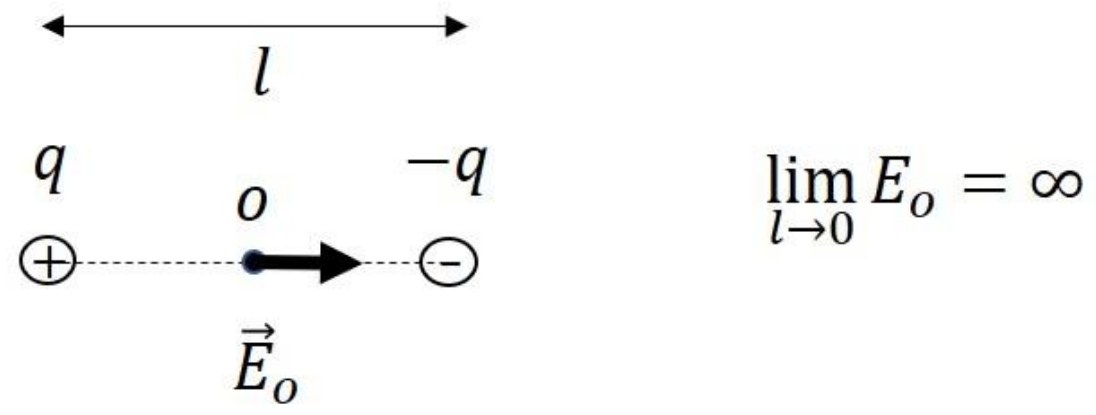

(a) Electric field intensity of dipole moment at the center along the dipole direction

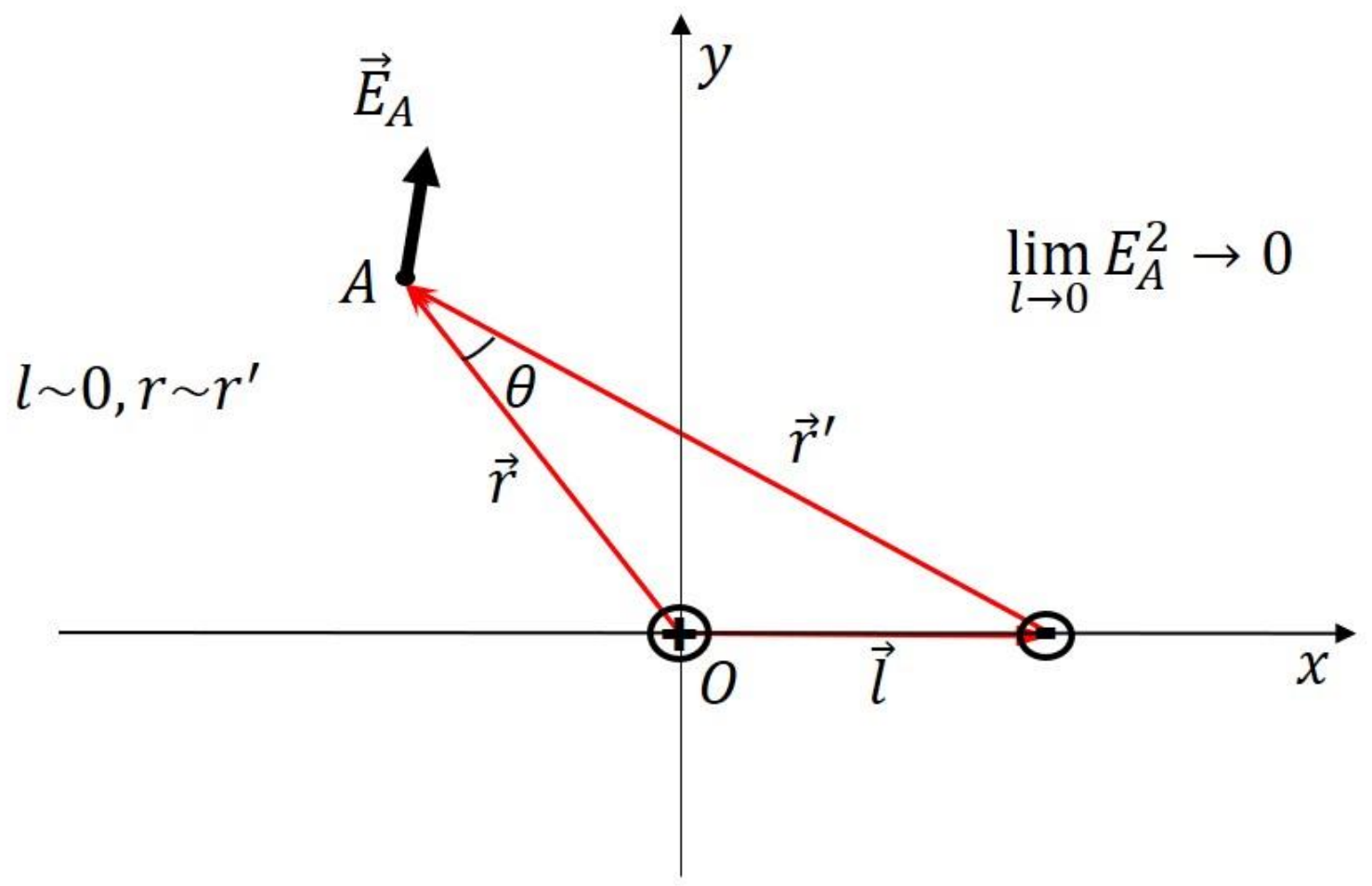

(b) Electric field intensity of dipole moment at arbitrary point of the space

Figure 17 Schematic of electric field intensities of dipole moment at two cases, *i.e.*, at (a) the dipole center and (b) one arbitrary location of the space. At the limit that the dipole length tends to zero, i.e., $l \to 0$, the field tends to infinite at the center in (a) and zero at arbitrary location in (b). The math processes of the two limits can be found in the text. The introduced dipole moment model is used to represent the essence of the *double layer* hidden in the stratification discharge structure simulated at low pressure, 10mTorr. As the dipole at this limit indicated, the double layer is actually a *breaking point* (*delta function*) among the smooth/continuous distribution/profile, in the view of math functions.

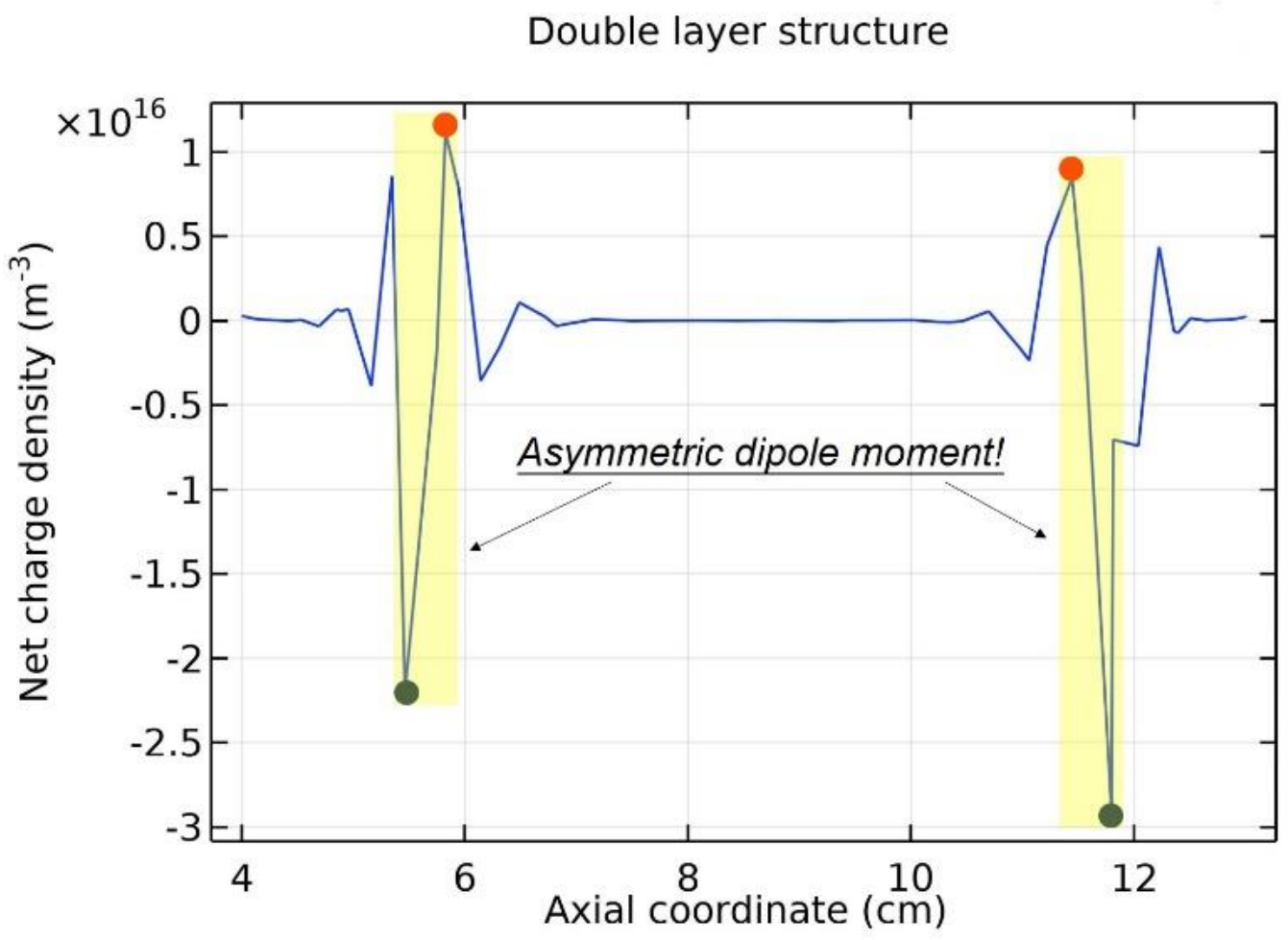

Figure 18 Axial profile of net charge density of 10mTorr $Ar/SF_6$ discharge, at the steady state. The other discharge conditions are the same as in Fig. 16. The double layer is clearly seen from the net charge density profile, where two asymmetric dipole moments are abstracted. The figure data is sampled at steady state.

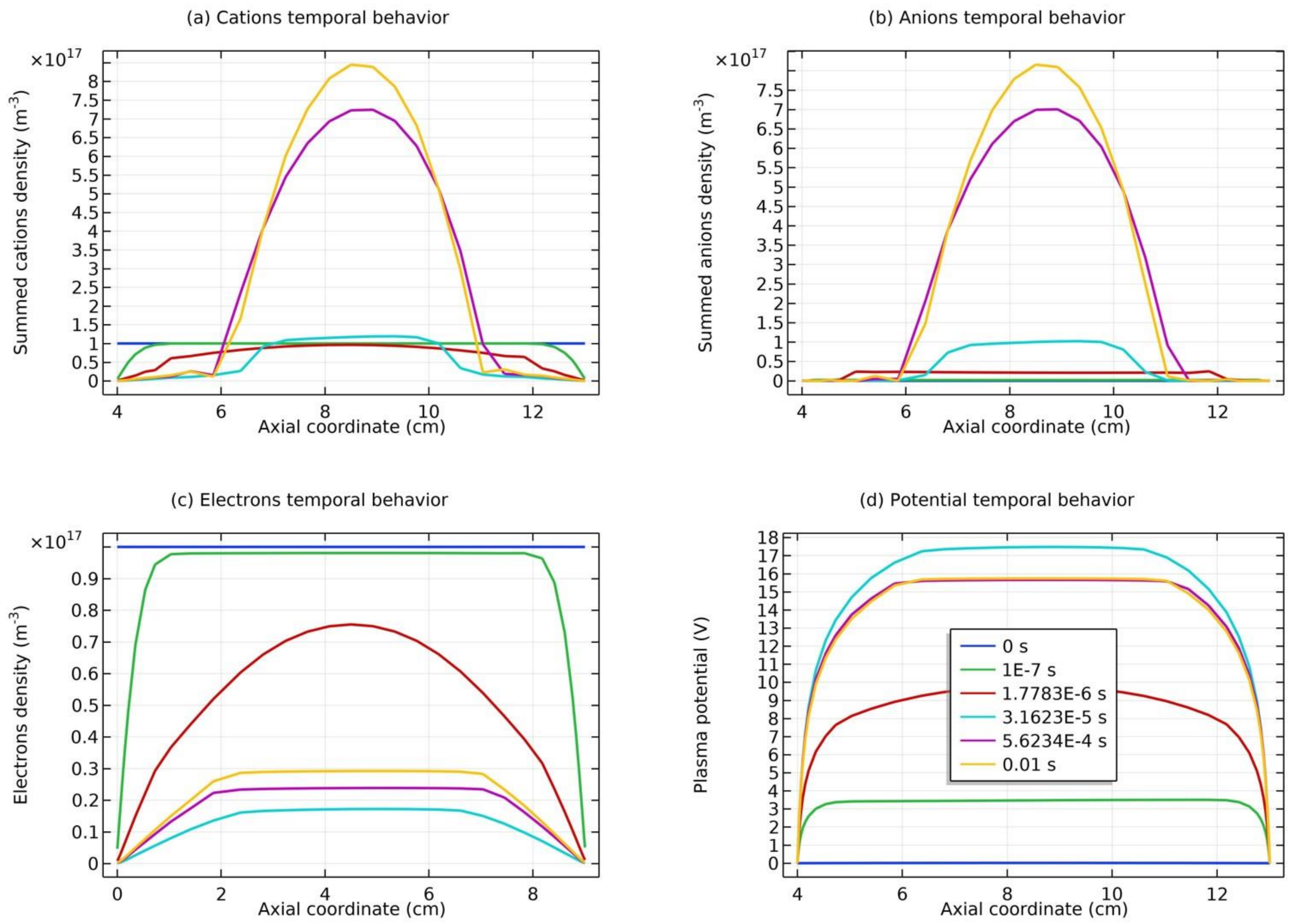

Figure 19 The temporal behaviors of (a) cations density, (b) anions density, (c) electrons density and (d) potential, when the basic stratification structure is formed. The three species densities share the same legend with the potential in (d). The discharge conditions are the same as in Fig. 16.

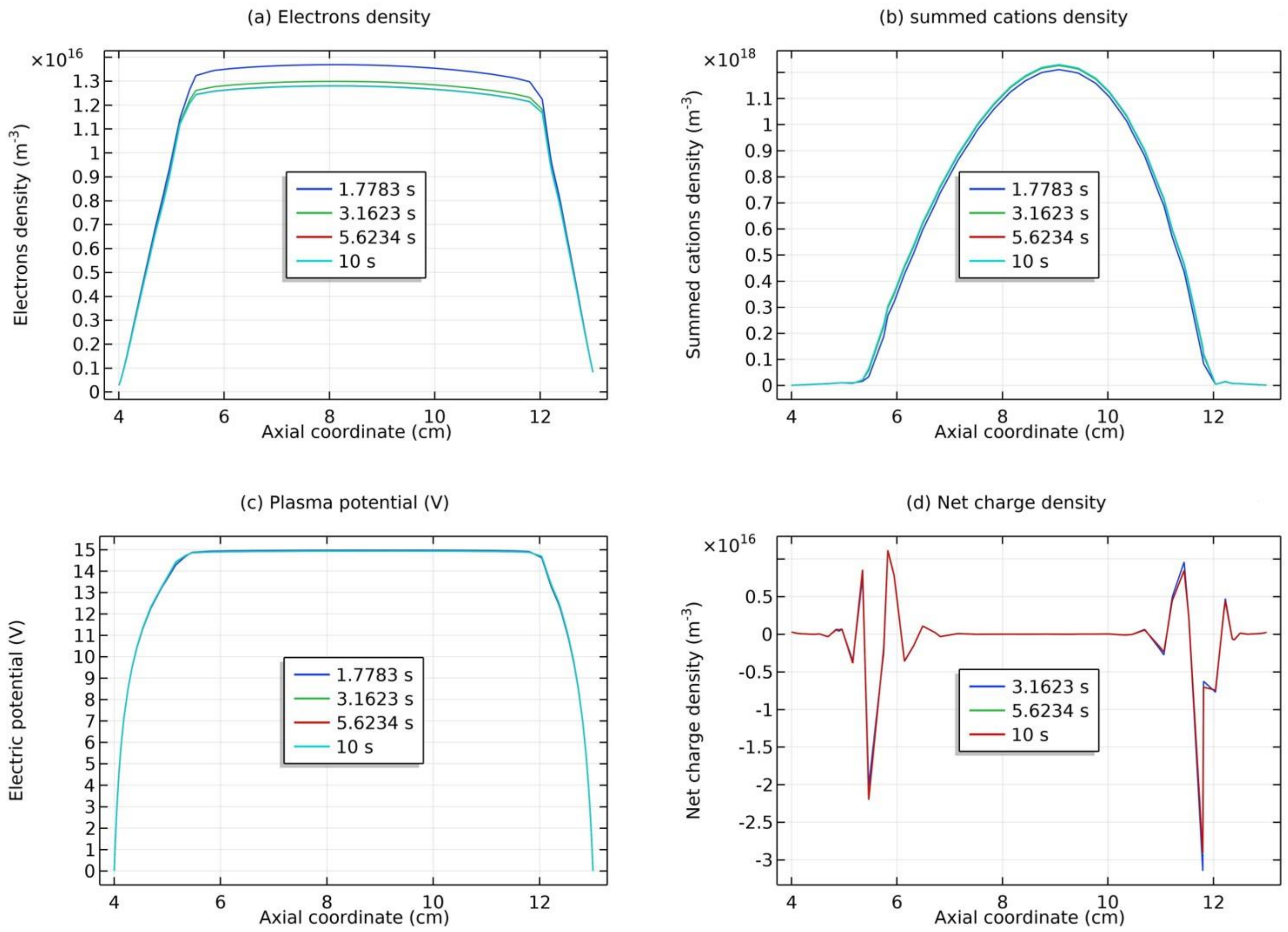

Figure 20 Convergence of mixed Ar/$SF_6$ discharge with the simulating time. The temporal behaviors of axial electrons density (a), summed cations density (b), plasma potential (c) and net charge density (d) are sequentially shown, when the discharge structure has been established and the discharge is approaching to steady state. The discharge conditions are the same as in Fig. 16.

### (c) Spontaneous self-coagulation of anions

In Fig. 21, the summed anions density in the core region exhibits self-coagulation and its net source is negative at the coagulation location. As indicated in the self-coagulation theory of Fig. 7, the free diffusion condition is still needed. It is automatically satisfied for the potential of core is flatten, as shown in Fig. 3(a), which thereby cannot drift the anions. The parabola theory above and the analytics of Fig.22 indicate that the axial and radial ions densities profiles should be both parabolic, WAS the influence of self-coagulation excluded. The combination of parabola and coagulation leads to complex discharge structure of electronegative plasmas. Without the squeeze of ambi-polar potential, we would like to call the self-coagulation spontaneous purely due to the chemistry source. Accordingly, the ambi-polar self-coagulation that assemble the cations for satisfying the neutrality is non-advective; see next in Section. III.2(B).

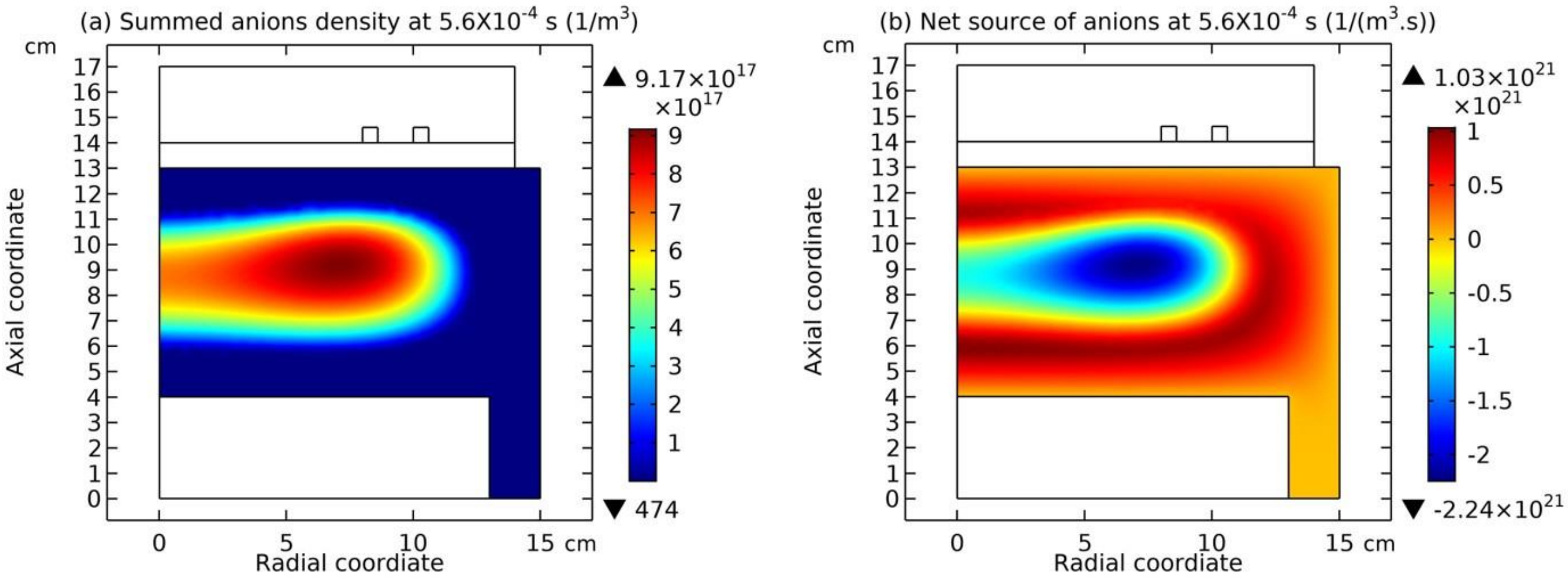

Figure 21 Summed anions density (a) and their net source (b) at the intermediate discharge process of Ar/$SF_6$ inductive plasma, given by the fluid model at the discharge conditions of 300W, 10mTorr and 10% $SF_6$ content.

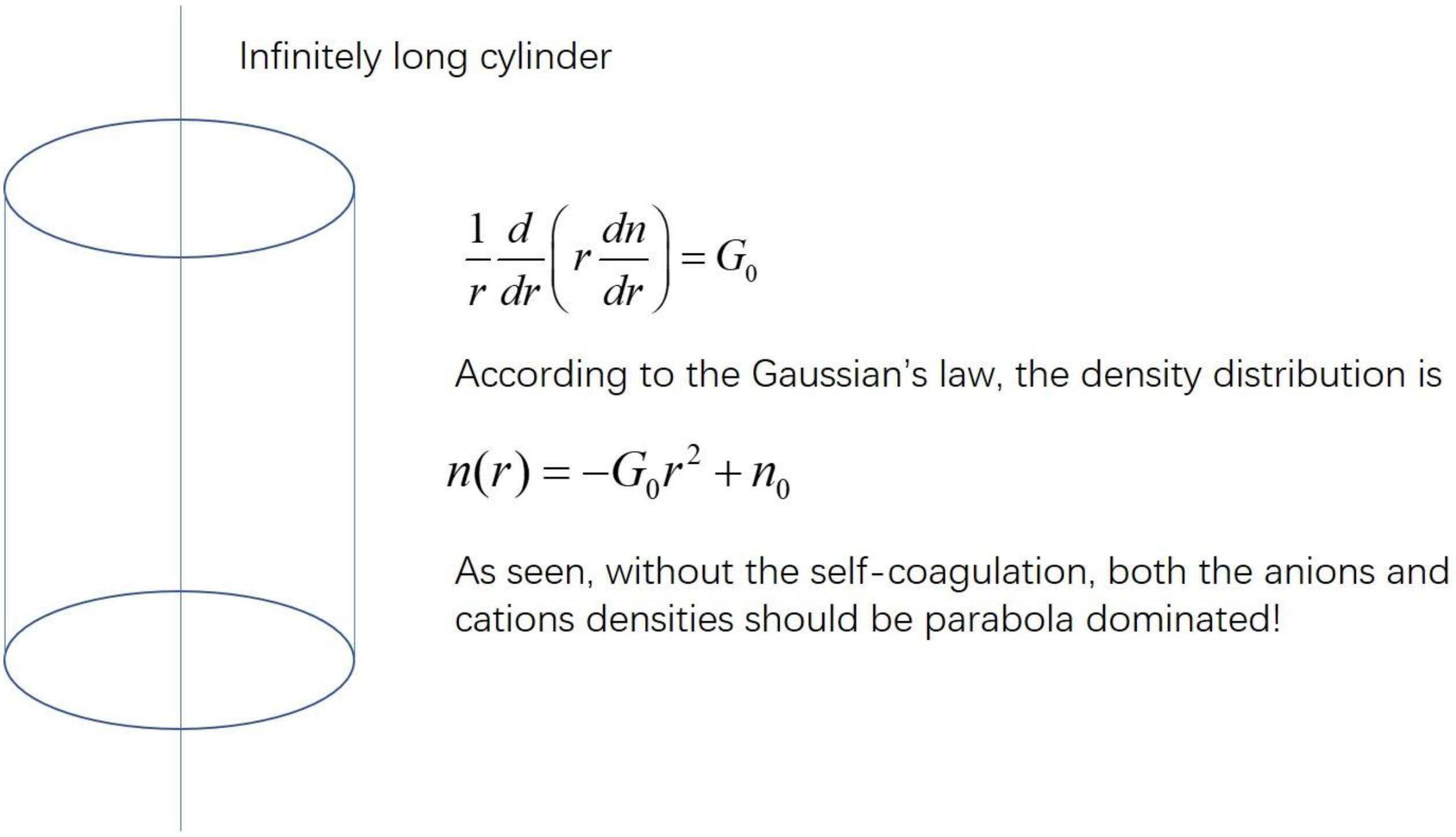

Figure 22 Without the self-coagulation scheme, both the axial and radial anions and cations densities profiles should be parabola. Here in this figure, the radial parabola feature is exhibited. The axial parabola feature can be found in the parabola theory of Fig. 15.

### (III.2.B) Self-coagulation-to-coil scheme of ions

As the analytic works predicted, the area of electropositive edge region shrinks with pressure, and finally the space-stratification disappears. This discharge feature is captured by the fluid simulation as well. At the previous stratification theory we proposed, it is natural for it disappearing at high pressure, because high pressure provides high electronegativity and hence high anion source, at which the anion Boltzmann relation is more easily built. At the high pressure, 90mTorr, after the stratification is disappeared, the self-coagulation-to-coil scheme of ions densities is discovered in the Ar/$SF_6$ inductive plasma fluid simulation. Its mechanism is hence illustrated in the section.

#### (a) Pre-condition of self-coagulation

As shown in Figs. 23-25, at the beginning of the discharge, the initially-set uniform electron density swiftly coagulates to the coil at strong attachment loss. Correspondingly, the ambi-polar diffusion potential of electron and cation is established, which is suppressed toward the coil as well at the electron Boltzmann relation illustrated in Fig. 26. At high pressure, source term dominates over transport term in the continuity equation. So, the electron coagulation arisen from chemistry source just happens at high pressure, not at low pressure in the previous section. Besides, this is not a self-coagulation behavior since the free diffusion is not achieved. But it offers pre-condition (see next).

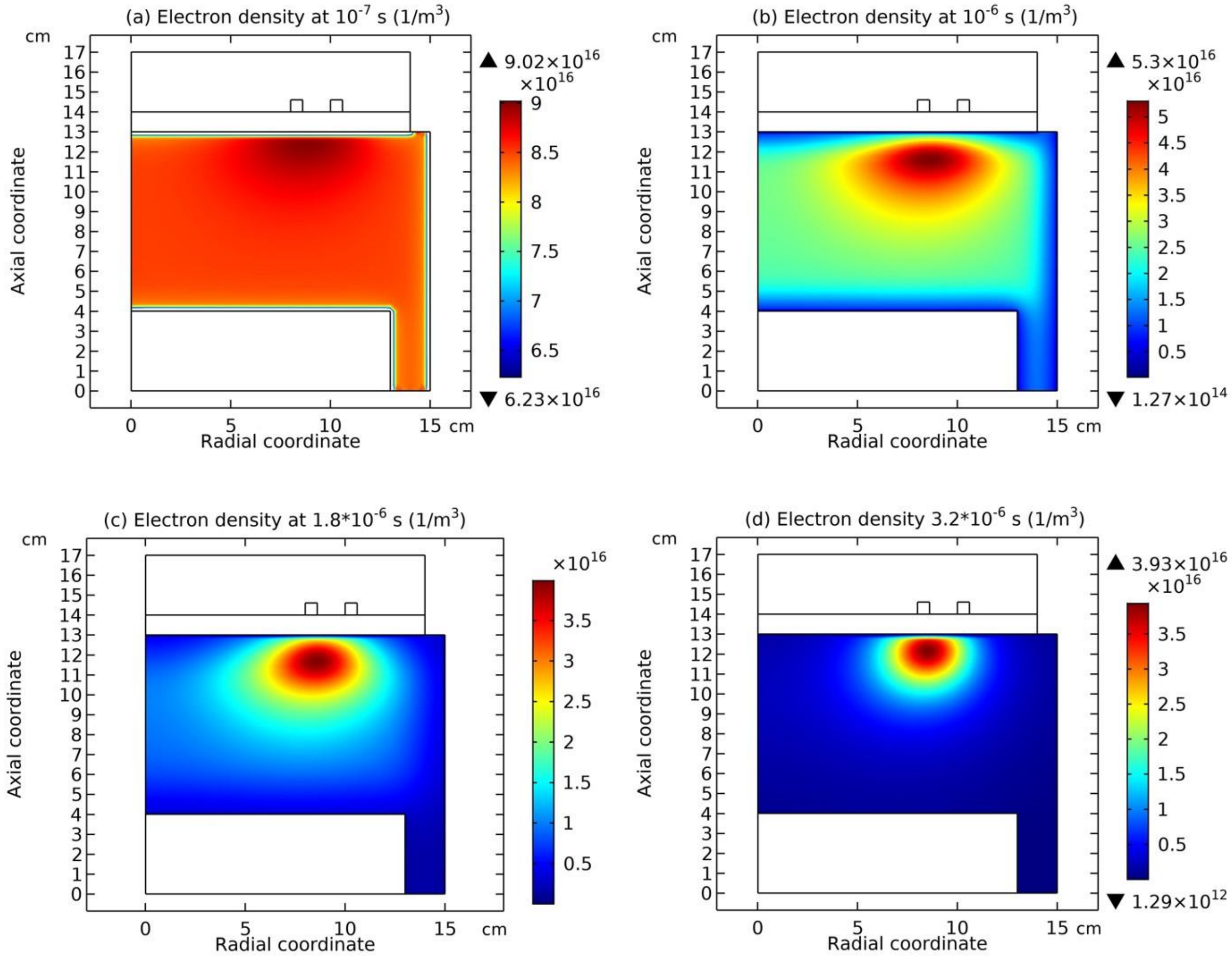

Figure 23 Electron density of Ar/$SF_6$ inductive plasma with the time in an early stage of discharge, plotted at the (a) $10^{-7}$ s, (b) $10^{-6}$ s, (c) $1.8 \times 10^{-6}$ s, and (d) $3.2 \times 10^{-6}$ s time points, respectively. The figure data is given by the fluid model simulation at the discharge conditions of 300W, 90mTorr and 10% $SF_6$ content.

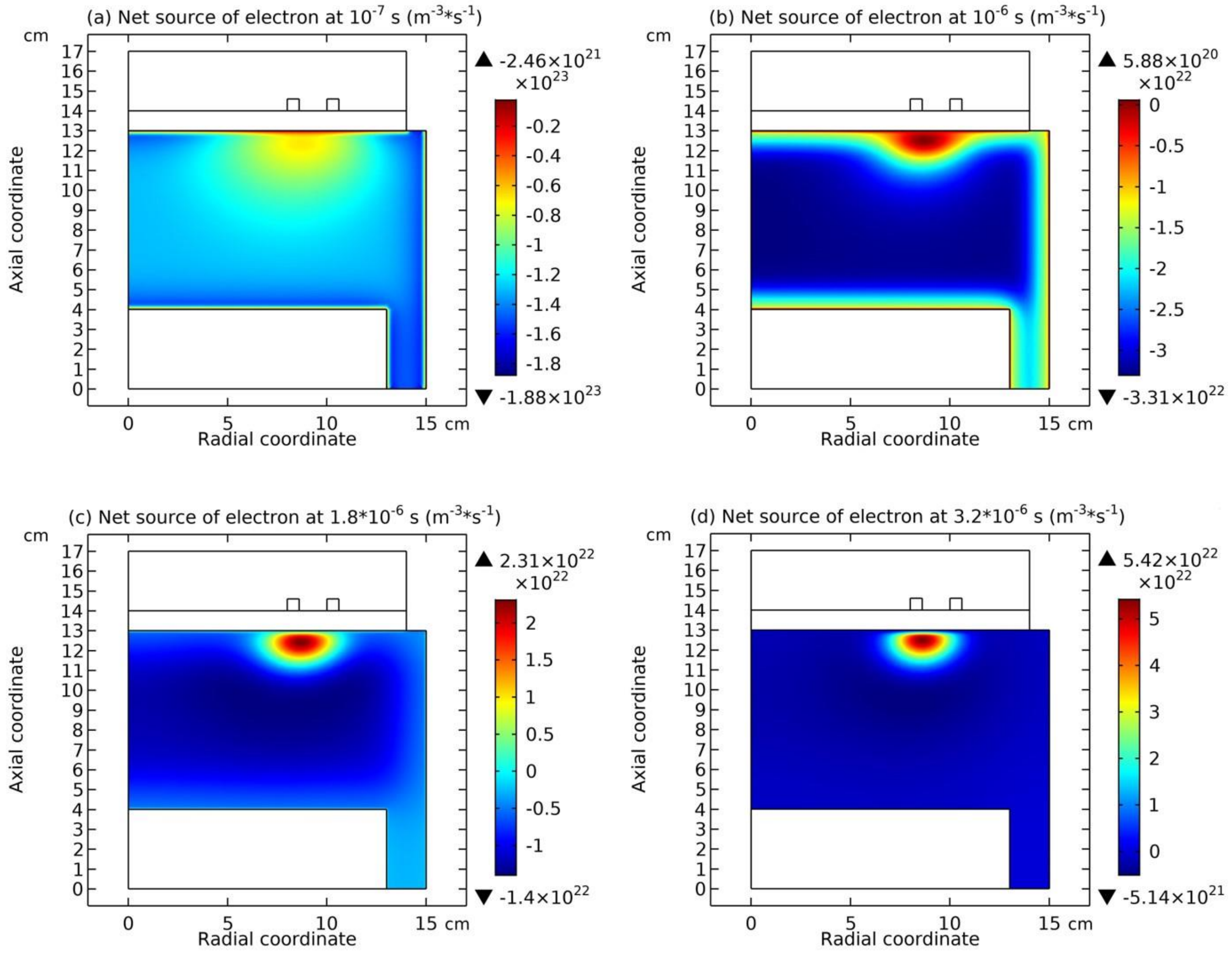

Figure 24 Electron source of Ar/$SF_6$ inductive plasma with the time in an early stage of discharge, plotted at the (a) $10^{-7}$ s, (b) $10^{-6}$ s, (c) $1.8 \times 10^{-6}$ s, and (d) $3.2 \times 10^{-6}$ s time points, respectively. The figure data is given by the fluid model simulation at the discharge conditions of 300W, 90mTorr and 10% $SF_6$ content.

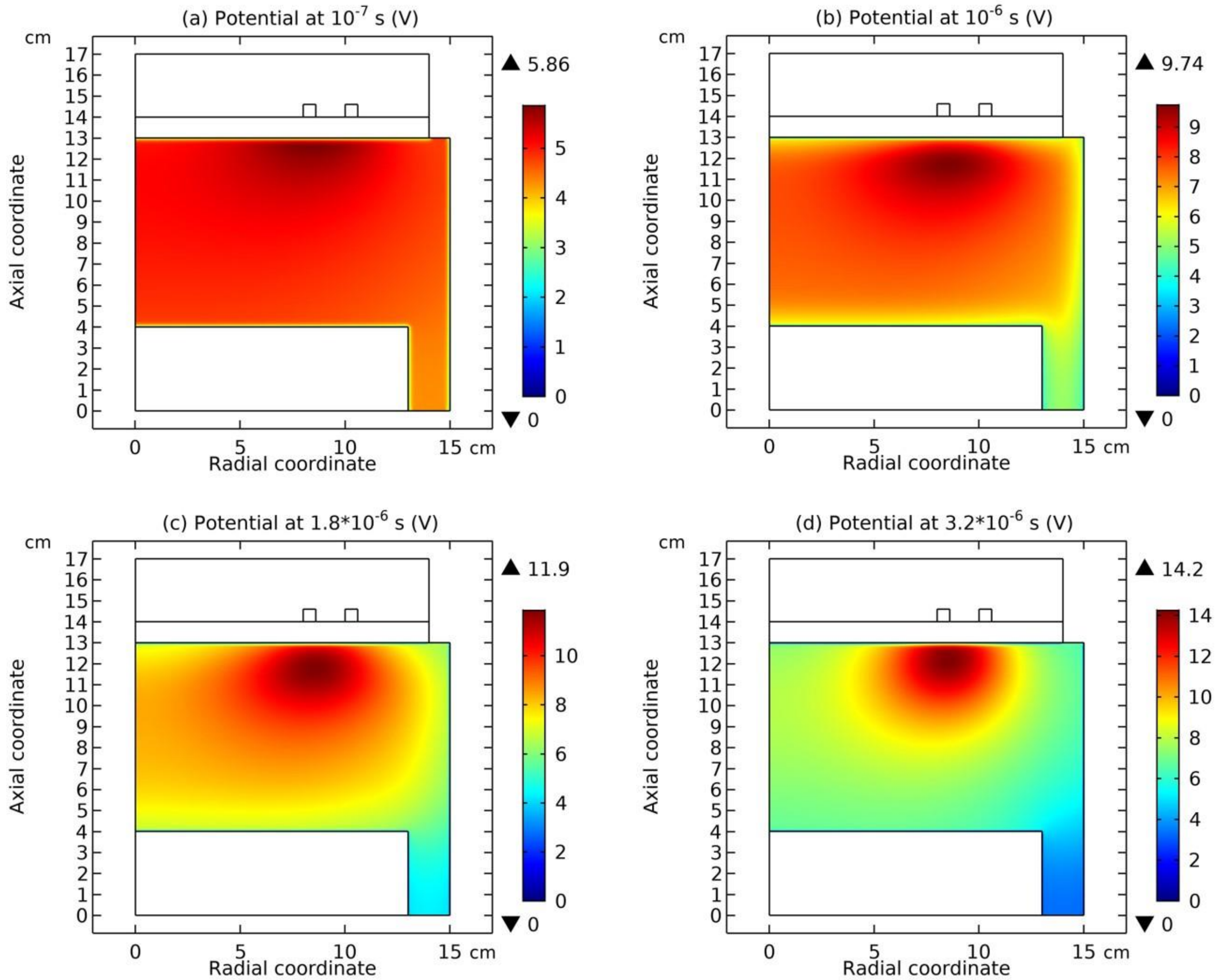

Figure 25 Plasma potential of $Ar/SF_6$ inductive plasma with the time in an early stage of discharge, plotted at the (a) $10^{-7}$ s, (b) $10^{-6}$ s, (c) $1.8 \times 10^{-6}$ s, and (d) $3.2 \times 10^{-6}$ s time points, respectively. The figure data is given by the fluid model simulation at the discharge conditions of 300W, 90mTorr and 10% $SF_6$ content.

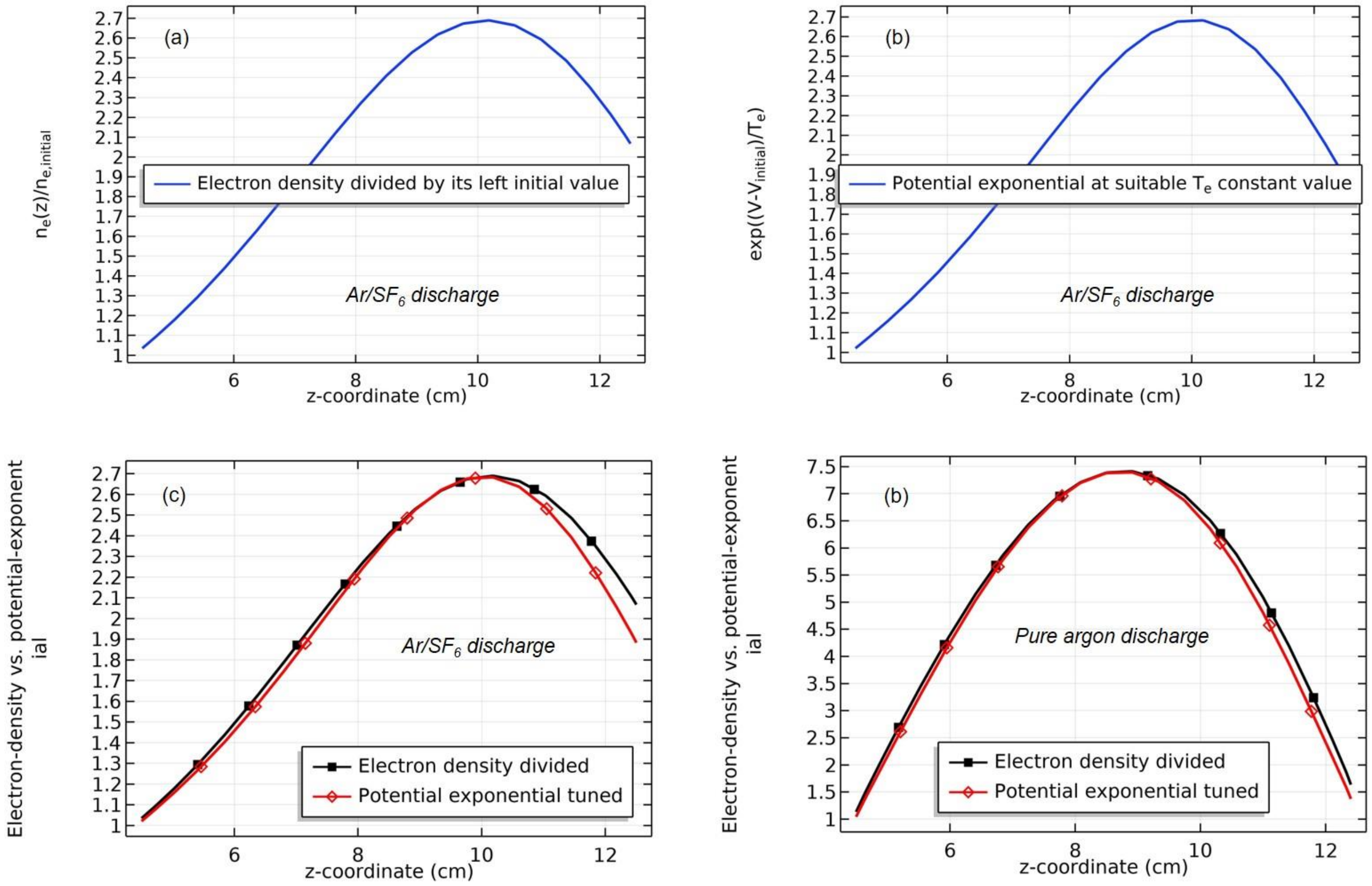

Figure 26 (a) Axial electron density divided by its left initial value, (b) potential exponential given at suitable constant electron temperature value, (c) comparison of the two quantities, in the 90mTorr and 300W mixture discharge (10% $SF_6$ content). And (d) comparison of the two quantities in the 90mTorr and 300W pure argon discharge. Radial position of the sampled electron density and potential is $r = 0$ cm. Statistics of electron in Ar/$SF_6$ discharge basically satisfies the Boltzmann relation, as shown in panel (c). The peripherical departure is caused by the positive source of sheath that ascends slightly the density curve, as in the case of pure argon plasma shown in panel (d). The simulated data of Ar/$SF_6$ plasma is sampled at the time of $3.2 \times 10^{-6}$ s, when the ambi-polar potential is being established. The simulated data of pure Ar plasma is given by the fluid model at the steady state.

(b) Self-coagulation stage of anions

At the strong ambi-polar potential surrounding the coil, it is believed that the anion Boltzmann relation is not reached, temporally. So, the anions are accumulated by drift. And at high enough density, the negative chemical source is formed and accordingly, the self-coagulation of anions at the potential bottom is happened, in Figs. 27 and 28. This is very similar to the $Ar/O_2$ plasma case in Section III. 2(A). In Fig. 29, before and after the self-coagulation, the anions density borders under the coil are compared. Border is soften at the self-coagulation, implying a chemistry process, as compared to the physics process, *i.e.*, ambi-polar diffusion. The self-coagulation is happened very fast, like an instantaneous behavior, which can only be predicted by unsteady-state solution.

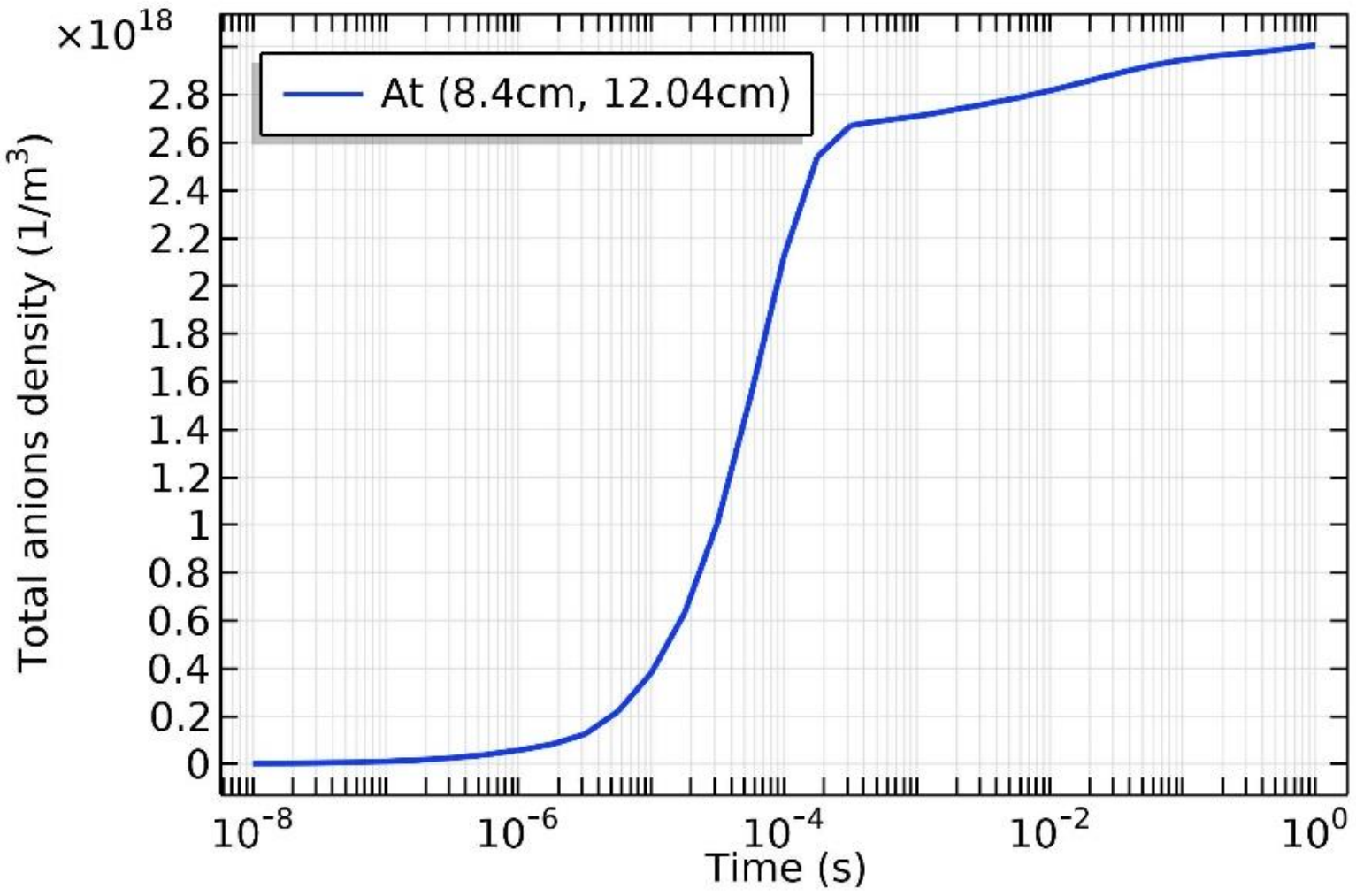

Figure 27 Total anions density versus simulating time, sampled at the location of (8.4cm, 12.04cm). The sampled point is in the heating region of coil electric field. Anions density therein increases strikingly within certain time segment, implying abrupt discharge mode transition, *i.e.*, from ambi-polar diffusion drift to self-coagulation of anions. The figure data are given by the fluid model simulation of $Ar/SF_6$ inductive plasma at the discharge conditions of 300W, 90mTorr and 10% $SF_6$ content.

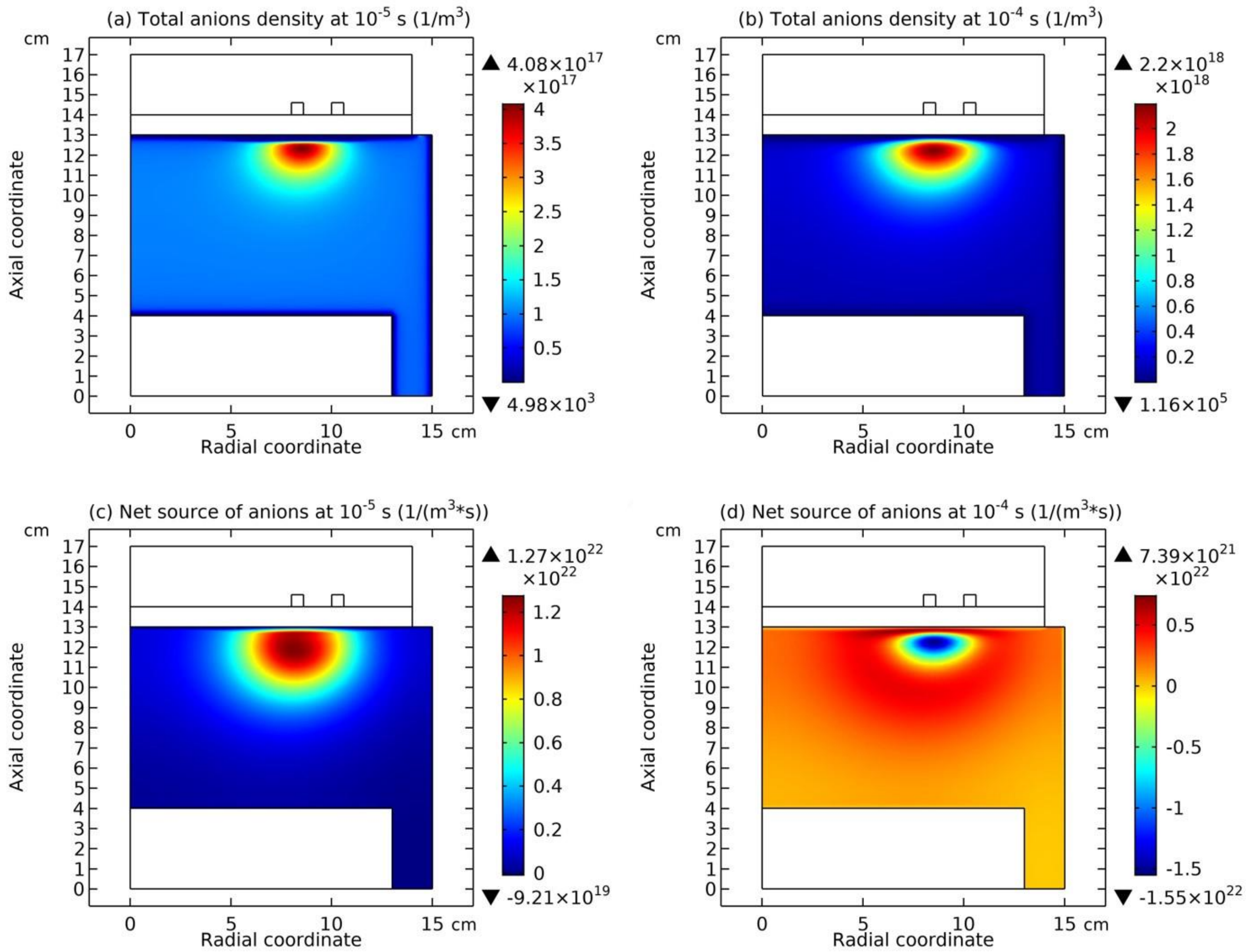

Figure 28 Total anions density at (a) $10^{-5}$ s and (b) $10^{-4}$ s, in the time of abrupt mode transition. And the net source of anions (c,d) at the above two times. The figure data are given by the fluid model simulation of Ar/$SF_6$ inductive plasma at the discharge conditions of 300W, 90mTorr and 10% $SF_6$ content.

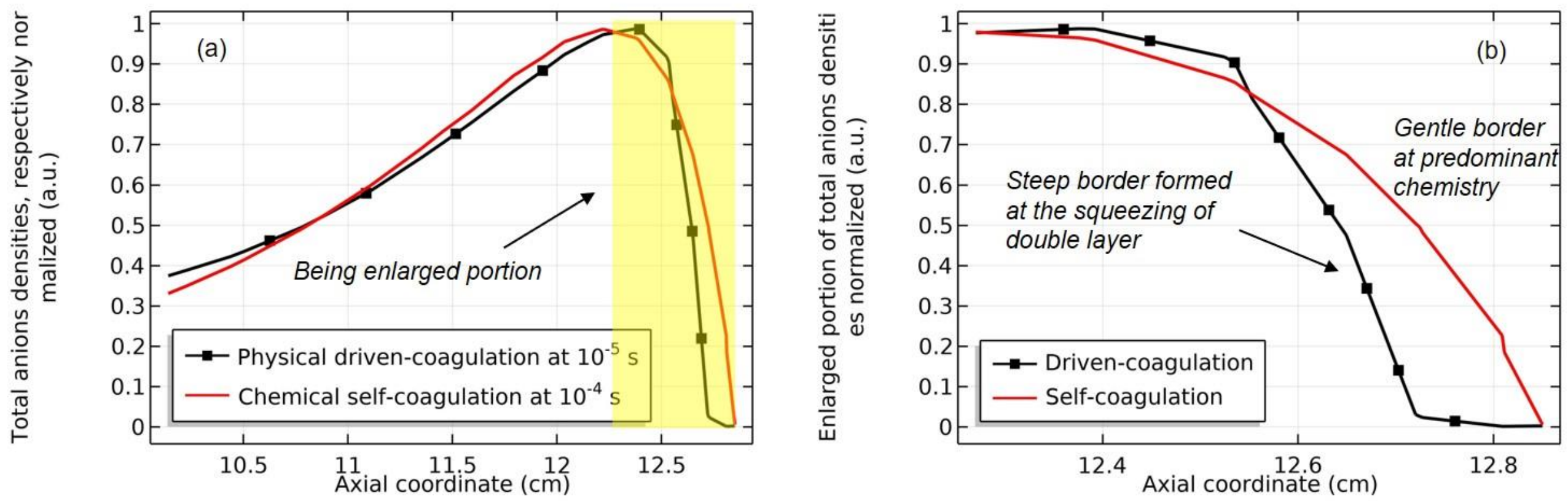

Figure 29 (a) Total anions densities normalized at their own maximum at two times, *i.e.*, $10^{-5}$ s and $10^{-4}$ $s$. In (b), the enlarged portion of anions densities close to the dielectric window is shown. The figure data are given by the fluid model simulation of Ar/$SF_6$ inductive plasma at the discharge conditions of 300W, 90mTorr and 10% $SF_6$ content.

### (c) Collapse of ambi-polar potential and electron transport change

As seen in Figs. 30 and 31, at the self-coagulation, the potential collapses. This is caused by the fact that anion density significantly increases after self-coagulation. At high enough anion density, the original two-species system (electron and cation) ambi-polar diffusion turns into triple-species (electron, cation and anion) system ambi-polar diffusion. Here, perpendicular to the potential contour, two transport mechanisms coexist, ambi-polar self-coagulation in the bottom and ambi-polar diffusion of triple-species system beyond the bottom. Out of the bottom, the potential barrel is attenuated at high electronegativity and Boltzmann anion (which is now reached at substantial anion density gradient after self-coagulation). As shown in the parabola theory of Fig. 15, at these conditions, the potential is flattened, illustrating the potential collapse herein.

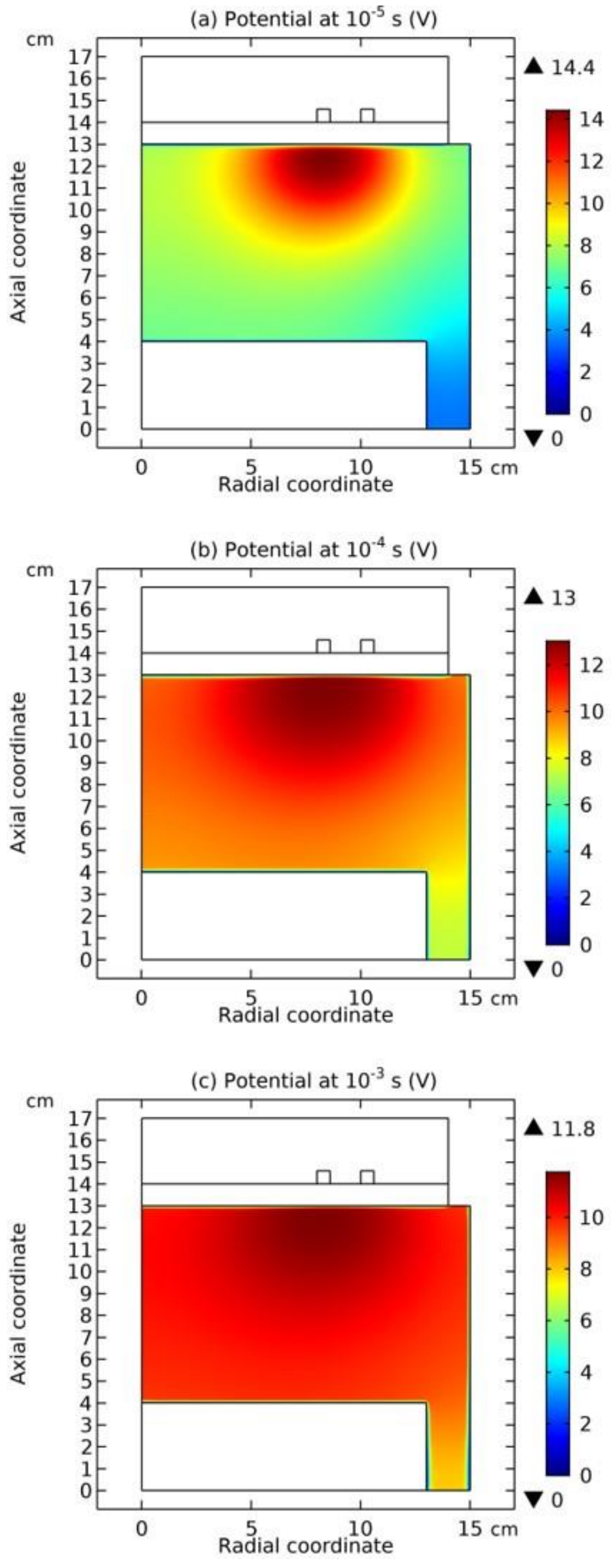

Figure 30 Plasma potential evolution in the anion mode transition process, with its profile exhibited at (a) $10^{-5}$ s, (b) $10^{-4}$ s, and (b) $10^{-3}$ s, respectively. Potential magnitude decreases with time and the gradient is attenuated meanwhile, implying that the ambi-polar diffusion potential collapse at the anion mode

transition. The figure data are given by the fluid model simulation of Ar/$SF_6$ inductive plasma at the discharge conditions of 300W, 90mTorr and 10% $SF_6$ content.

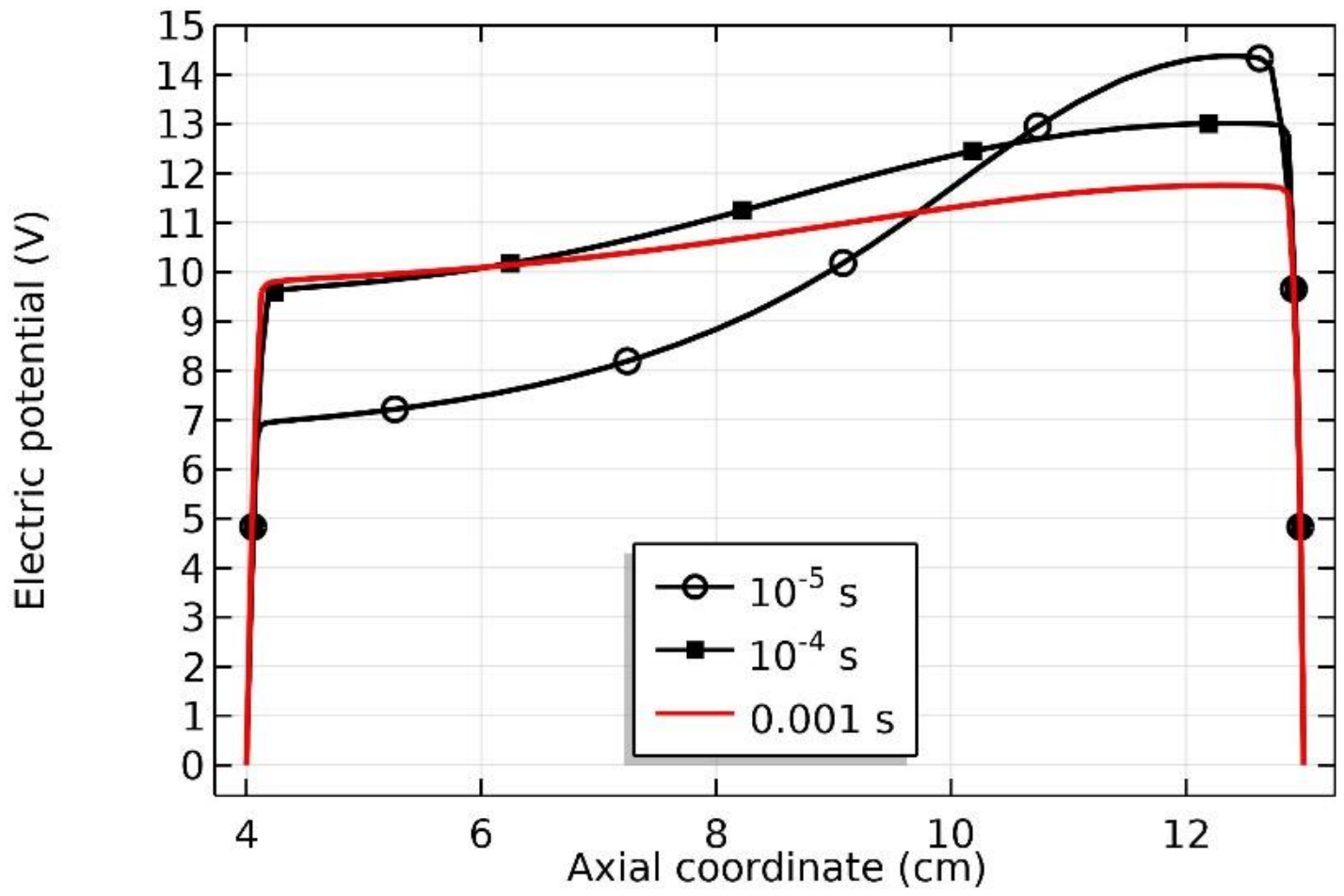

Figure 31 Axial potential profile at $10^{-5}$ s, $10^{-4}$ s, and $10^{-3}$ s, respectively, at the radial location of r = 8cm. The figure data are given by the fluid model simulation of Ar/$SF_6$ inductive plasma at the discharge conditions of 300W, 90mTorr and 10% $SF_6$ content.

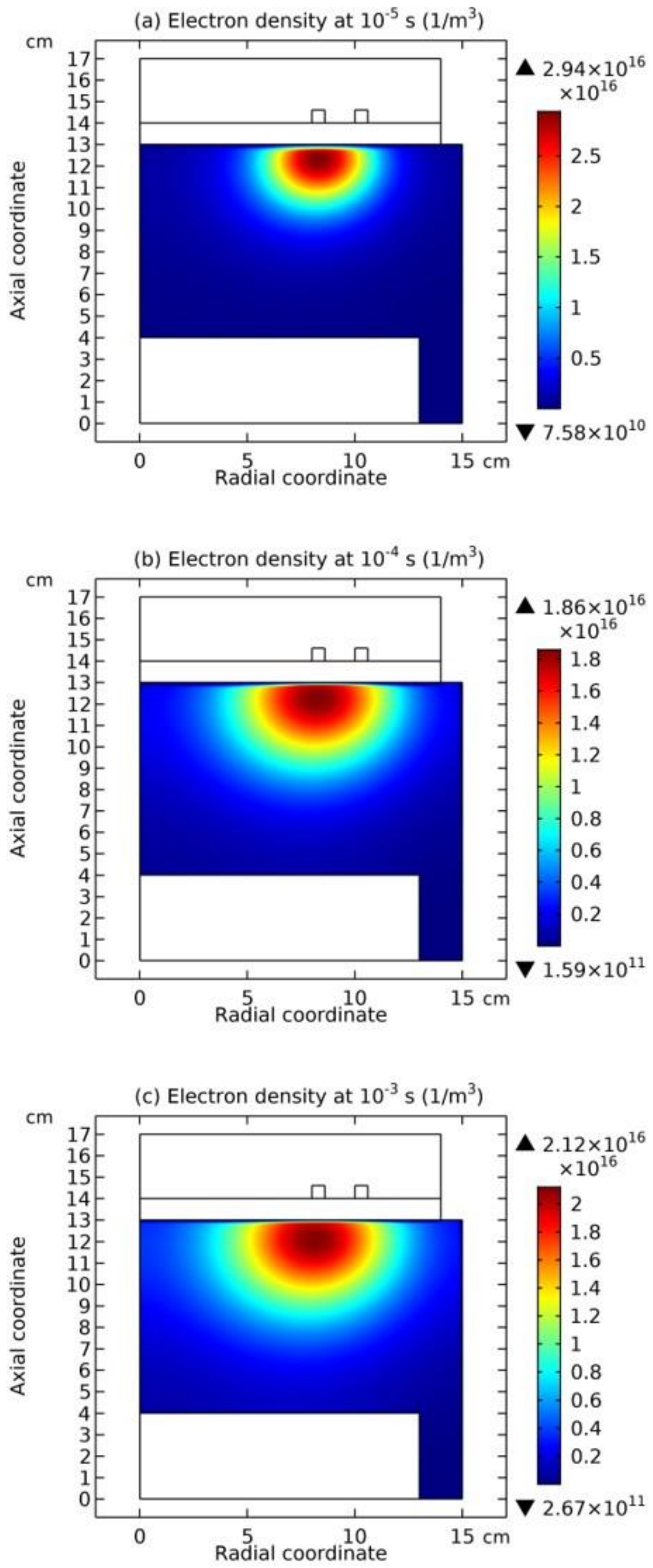

Figure 32 Electron density at $10^{-5}\ s$, $10^{-4}\ s$, and $10^{-3}\ s$, respectively, after self-coagulation of anion. The figure data are given by the fluid model simulation of Ar/$SF_6$ inductive plasma at the discharge conditions of 300W, 90mTorr and 10% $SF_6$ content.

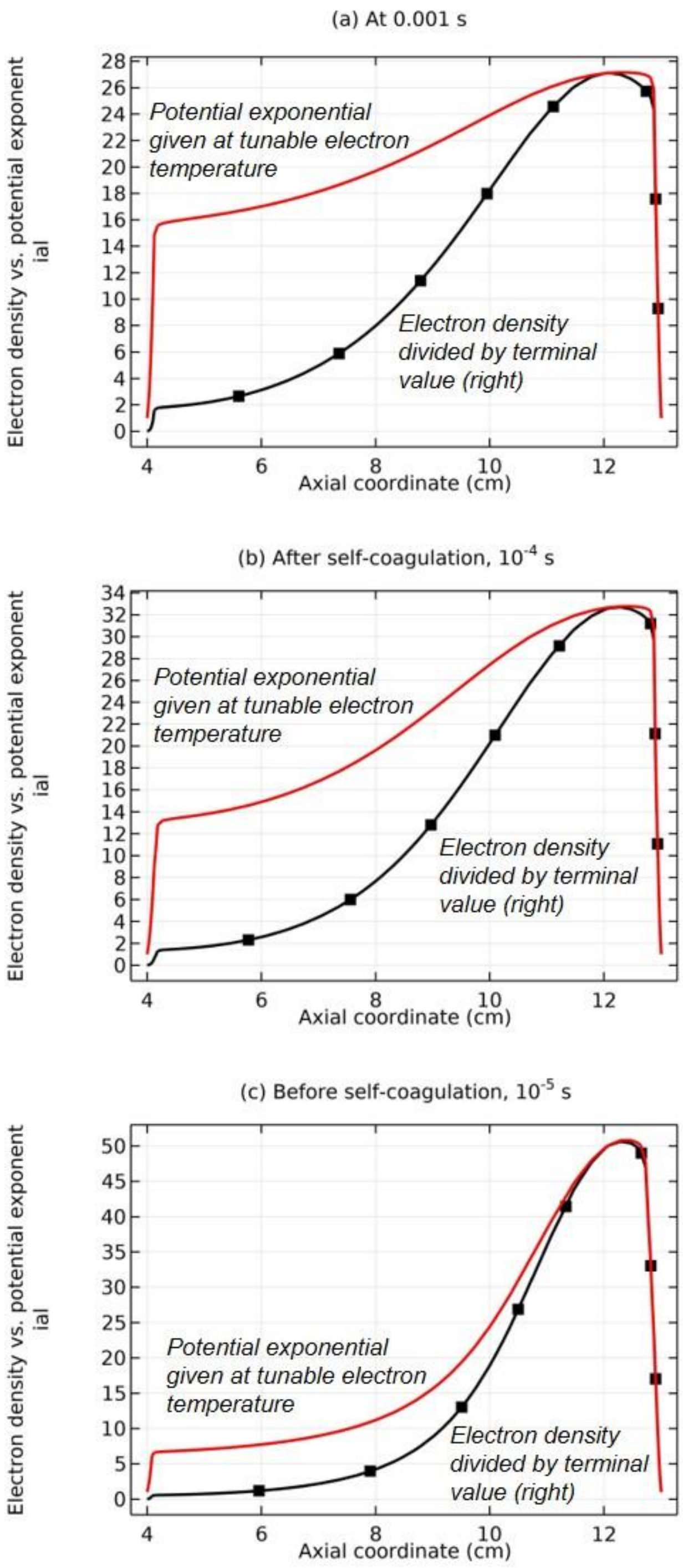

Figure 33 Electron density and Boltzmann balance at different times, after self-coagulation of anion. The figure data are given by the fluid model simulation of Ar/$SF_6$ inductive plasma at the discharge conditions of 300W, 90mTorr and 10% $SF_6$ content.

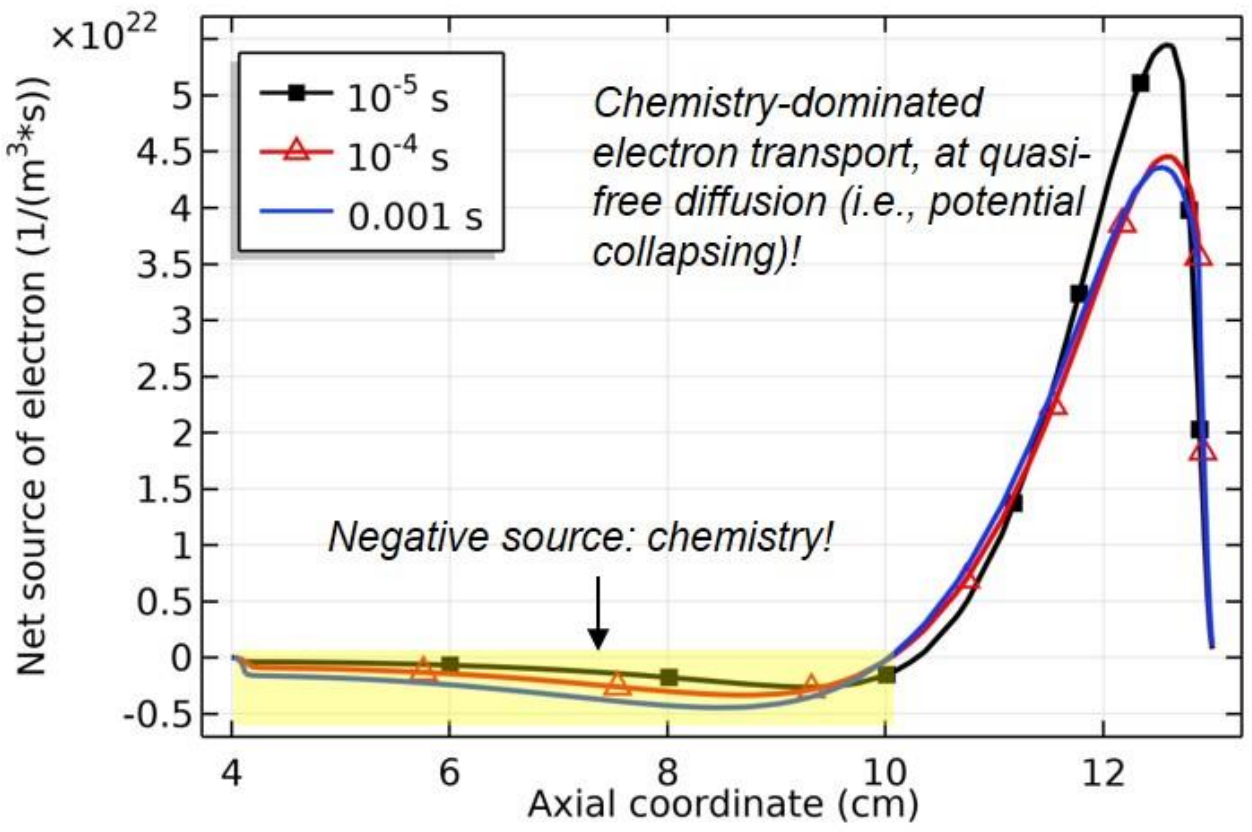

Figure 34 Axial profile of net source of electron at different times, after self-coagulation of anion. The figure data are given by the fluid model simulation of Ar/$SF_6$ inductive plasma at the discharge conditions of 300W, 90mTorr and 10% $SF_6$ content.

Along with the potential collapse, the electron density profile is expanded, as shown in Fig. 32. Meanwhile, the electron density deviates from the Boltzmann relation in Fig. 33. This deviation is very interesting, for it represents the self-coagulation of ELECTRON. The potential collapses and electrons tend to quasi-freely diffuse. Outside of the positive electron source under the coil, clearly negative chemical source is seen from the Fig. 34, as it is known a strong electronegative discharge case. The two conditions lead to the self-coagulation of electrons. Self-coagulation is one self-organization behavior, regardless to the polarity and mass of considered species. More analysis of the electron Boltzmann deviation is shown in Section III.2(C).

(d) Mass point behavior and theory of post-self-coagulation stage

Ar/$SF_6$ discharge creates adverse ions type. At the self-coagulation of anion and ambi-polar self-coagulation of ions (see further), the multiple ions exhibit mass point behavior. It is analyzed that self-coagulation does not relate to the species polarity. However, after the self-coagulation, the expelling forces DO exist between the individual self-coagulated ion mass points with the same polarities. As shown in Fig. 35, the originally coagulated $SF_3^+$ cation density is dispersed by the $Ar^+$ cation. The $SF_3^+$ coagulation is destroyed because the it is heavier and its density is comparable to the $Ar^+$ (illustrating in Fig. 36). The self-coagulation related theory in Ref. [15] indicates that the lighter the species is, the easier it is self-coagulated. Moreover, when the total charge amount that two elements carry fixed, the maximum expelling force between them is given at averagely distributing the total charge amount, shown in Fig. 37. Very simple issue. The expelling phenomenon of mass points is also occurred between anions. In Fig. 38, the $SF_5^-$ and $SF_6^-$ anions coagulations are disturbed by the light $F^-$ mass point. As compared, shown in Fig. 39, the density of $F^-$ is comparable to the sum of the above two heavy anions. The dispersion and disturbance of ions coagulation are gradual and temporal integrating effects. The expelling effect of anions is not that strong as the cations, because anions are confined by self-coagulation mechanism while cations are constrained by ambi-polar self-coagulation scheme, which is essentially electric interaction and weaker than the chemical restriction. As a result, the positive charge density (yellow and radiative) is given when the cation coagulation is destroyed, illustrated in Figs. 40 and 41. This is individual mass point behavior, and the plasma which is collective cannot respond. It is noted that it is not belonged to the sheaths which ARE the results of plasma collective interactions.

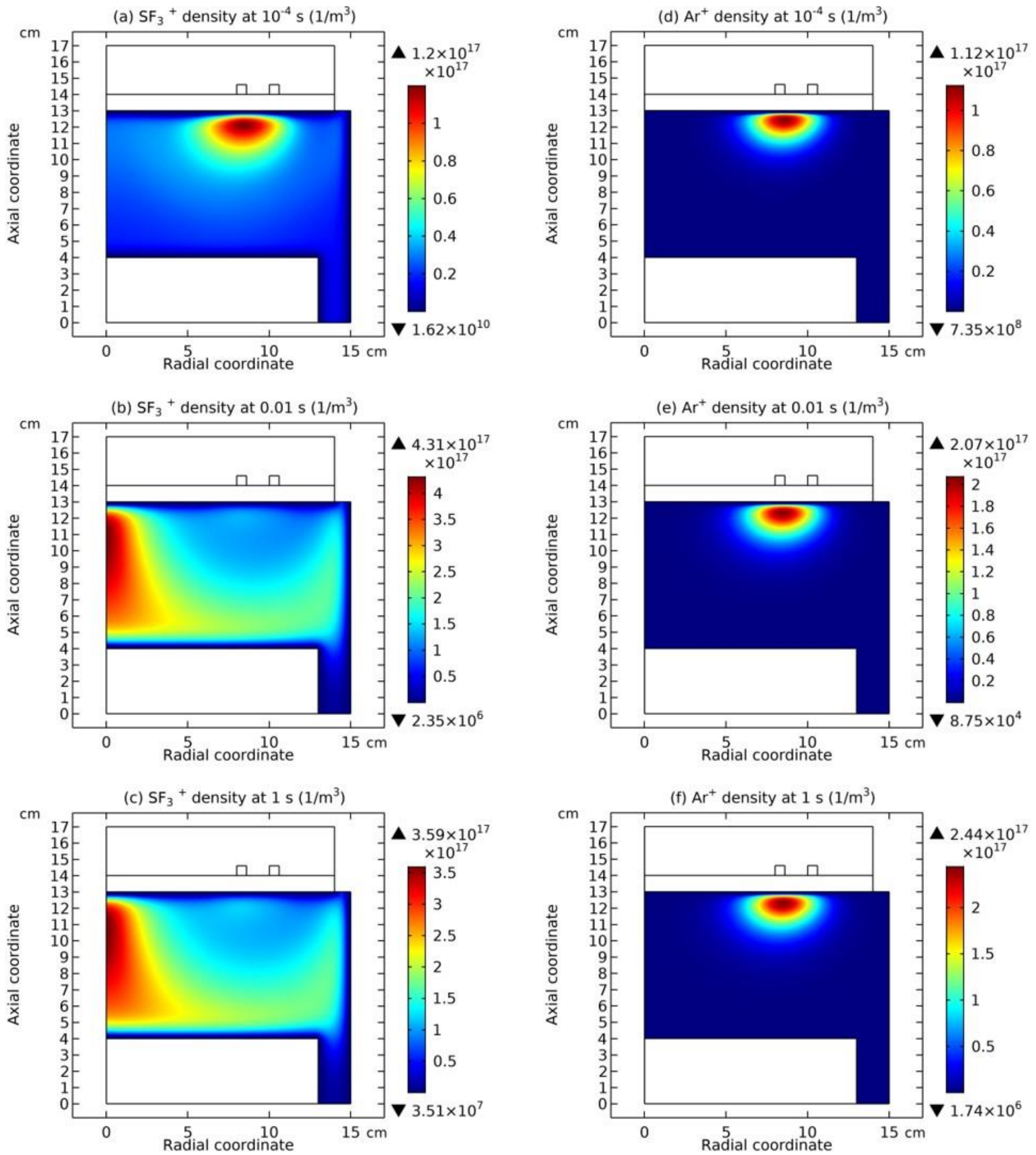

Figure 35 Evolutions of $SF_3^+$ and $Ar^+$ cations densities with the time after the self-coagulation process, respectively. In this plot, the coagulating appearance of heavy cation, $SF_3^+$, is blown up by the relatively light cation, $Ar^+$, illustrated as a *point mass phenomenon* inside the plasma. The figure data are given by the fluid model simulation of Ar/$SF_6$ inductive plasma at the discharge conditions of 300W, 90mTorr and 10% $SF_6$ content.

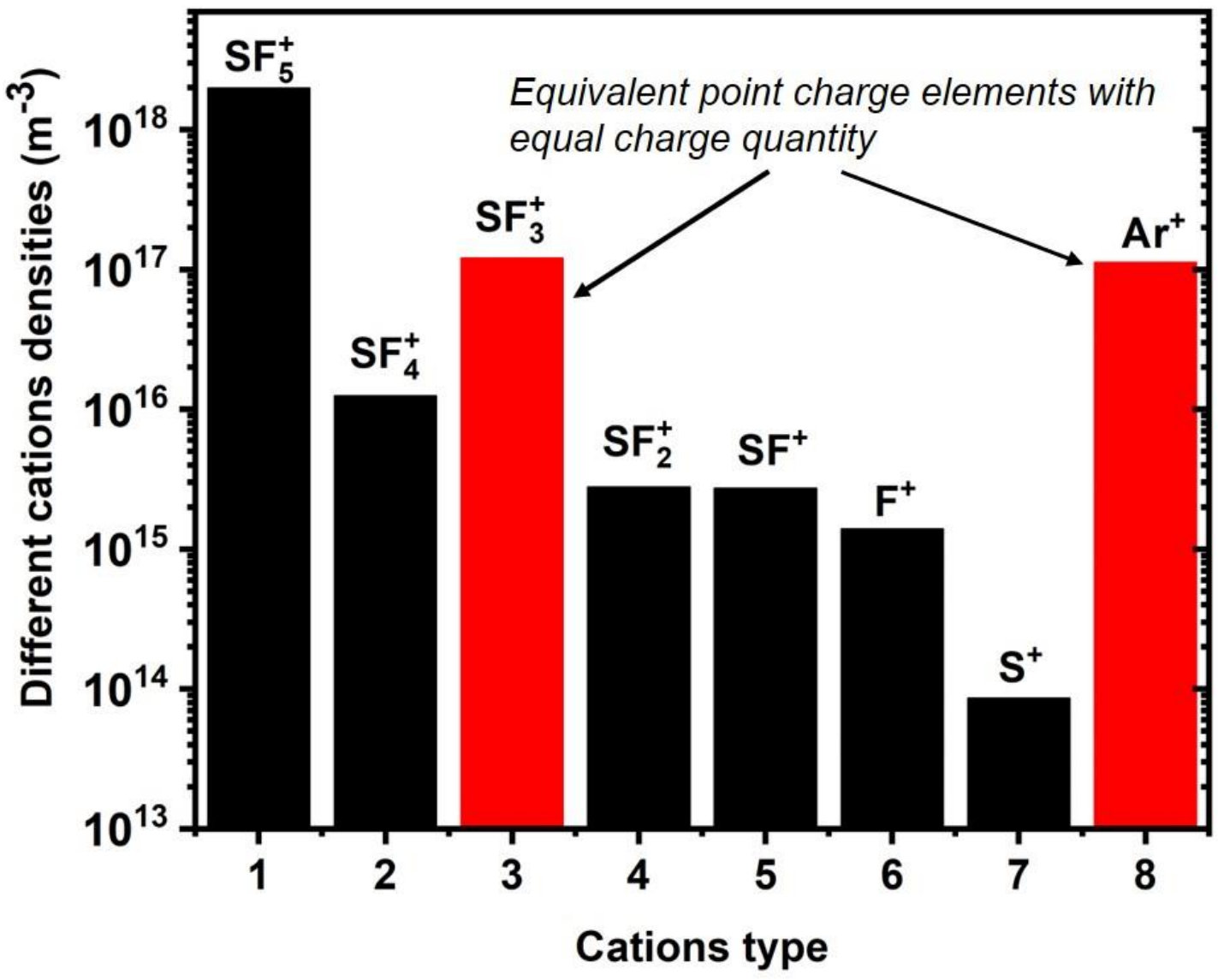

Figure 36 Plot of different peak cations densities at the time of $10^{-4}$ s, where the equivalent point charge elements with almost equal charge quantity is deduced. The figure data are given by the fluid model simulation of Ar/$SF_6$ inductive plasma at the discharge conditions of 300W, 90mTorr and 10% $SF_6$ content.

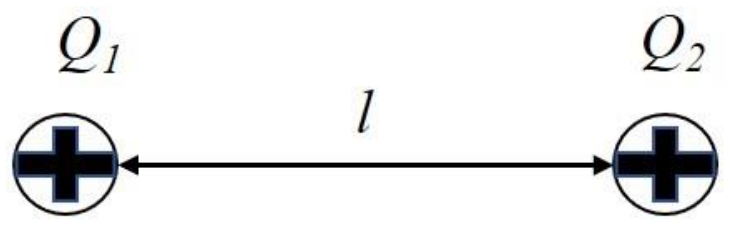

$$Q_1 + Q_2 = Q_T.$$

$$F = \frac{Q_1 \cdot Q_2}{4\pi\varepsilon_0 l^2} = \frac{Q_1 \cdot (Q_T - Q_1)}{4\pi\varepsilon_0 l^2} = \frac{-Q_1^2 + Q_T \cdot Q_1}{4\pi\varepsilon_0 l^2} = \frac{y(Q_1)}{4\pi\varepsilon_0 l^2}.$$

(a) Equivalent positive point charge models and the electric field force between them.

$y' = 0$ produces $x = Q_T / 2$, where the maximum is achieved.

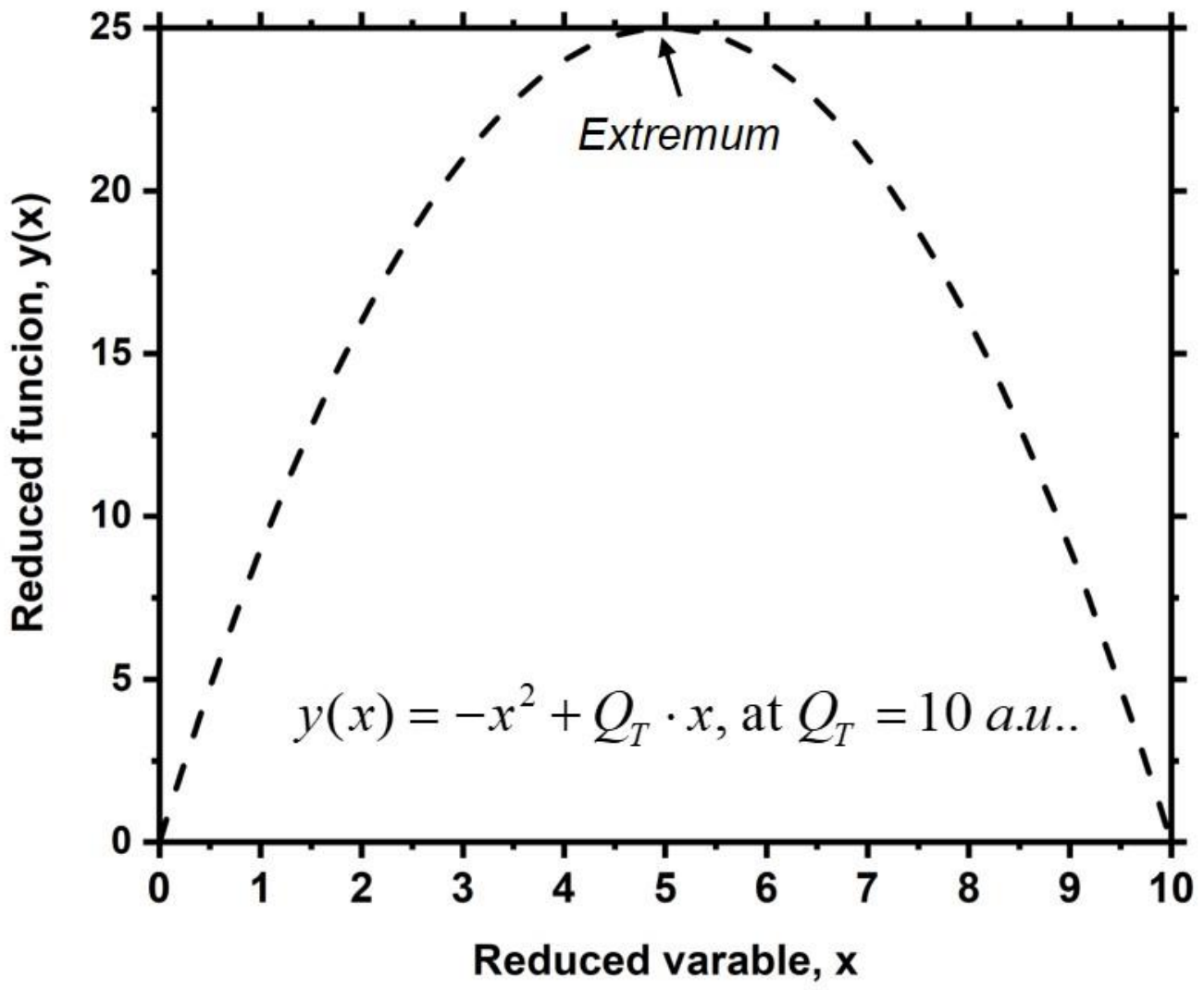

(b) Force versus the charge of first point charge model illustrated by a reduced function, and the extreme case, assuming specific total charge amount.

Figure 37 Transformed positive point charge models given by self-coagulation behavior, and the electric field force between them. The maximum force is given at averagely distributing the total charge amount. At such strong electric field force, the *heavy* $SF_3^+$ self-coagulation appearance is blown up by the equivalent *light* $Ar^+$ point charge model.

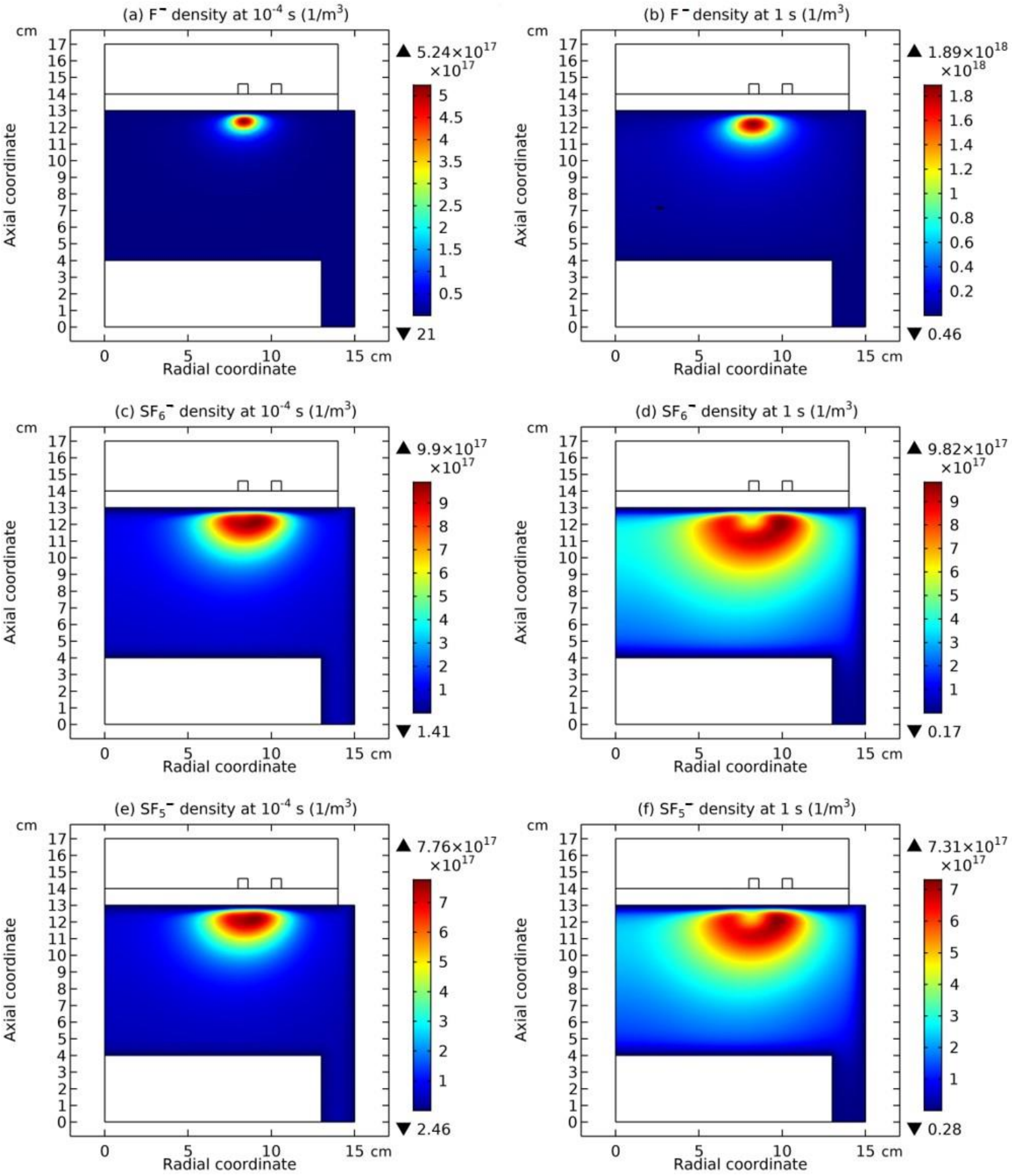

Figure 38 Evolutions of $F^{-}$, $SF_6^{-}$ and $SF_5^{-}$ anions densities with the time after the self-coagulation process, respectively. In this plot, the mass point phenomenon among anions is described. The figure data are given by the fluid model simulation of $Ar/SF_6$ inductive plasma at the discharge conditions of 300W, 90mTorr and 10% $SF_6$ content.

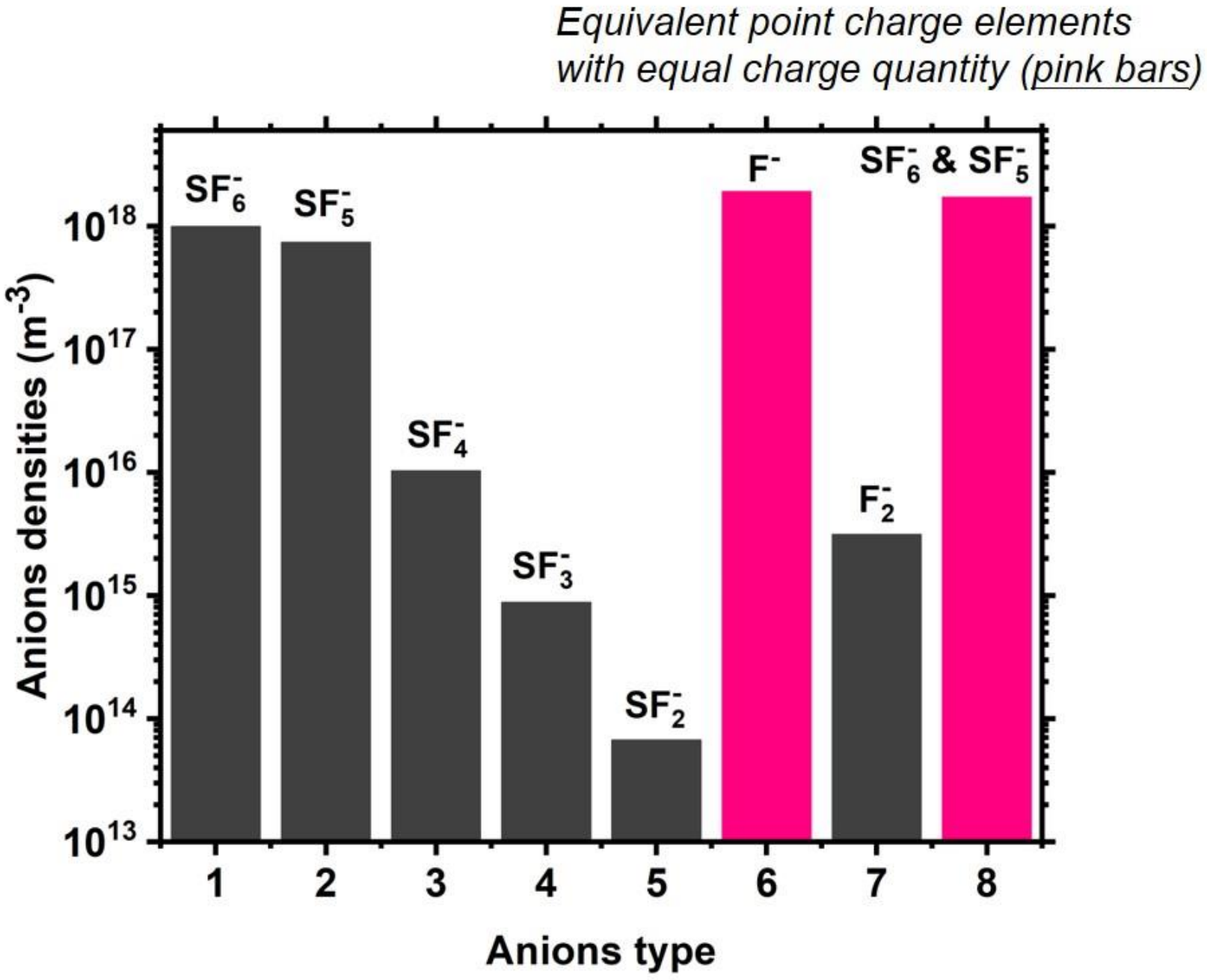

Figure 39 Plot of different peak anions densities at the time of 1s, where the equivalent point charge elements with almost equal charge quantity is deduced. The figure data are given by the fluid model simulation of Ar/$SF_6$ inductive plasma at the discharge conditions of 300W, 90mTorr and 10% $SF_6$ content.

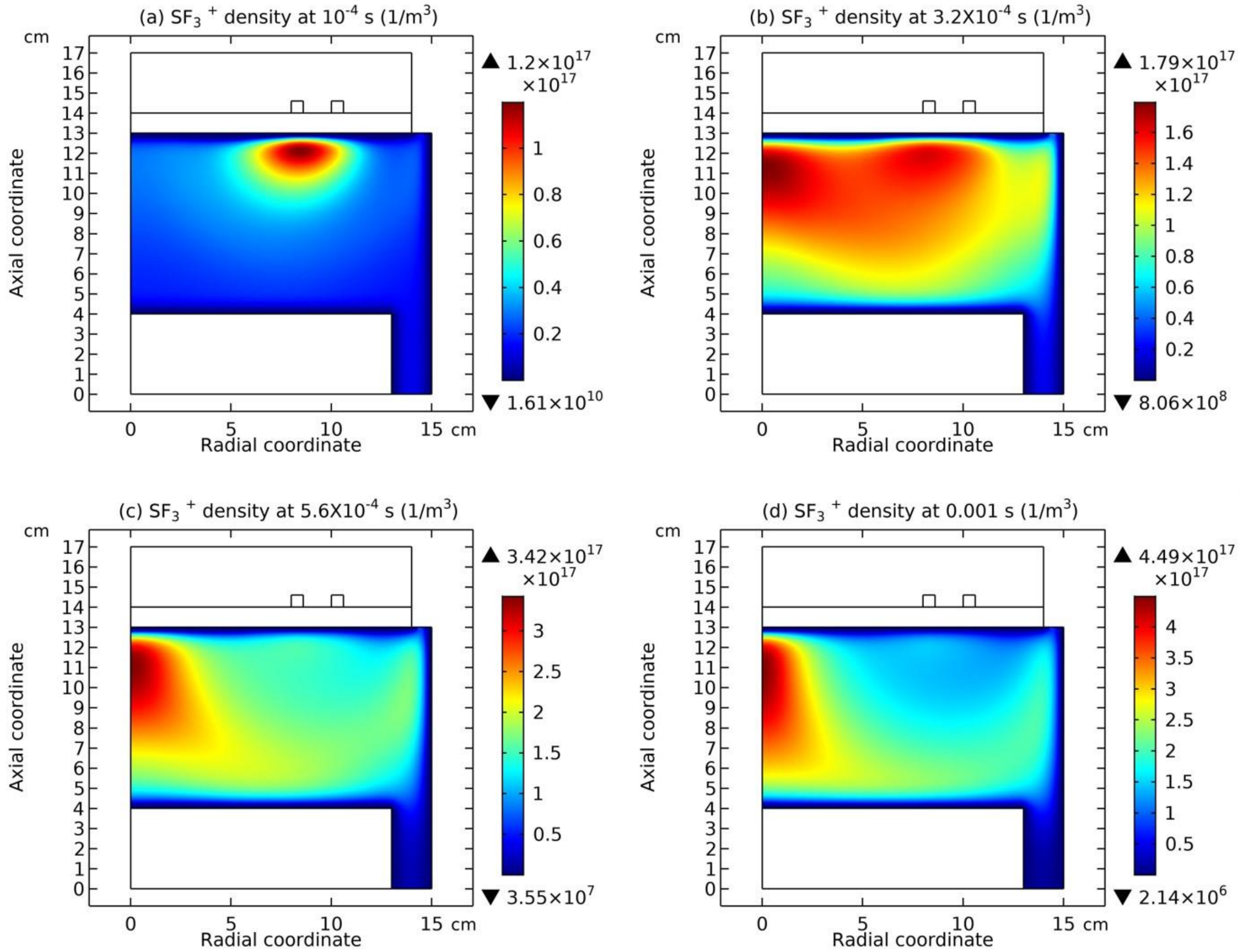

Figure 40 Process of $SF_3^+$ cation coagulation being destroyed. The figure data are given by the fluid model simulation of Ar/$SF_6$ inductive plasma at the discharge conditions of 300W, 90mTorr and 10% $SF_6$ content.

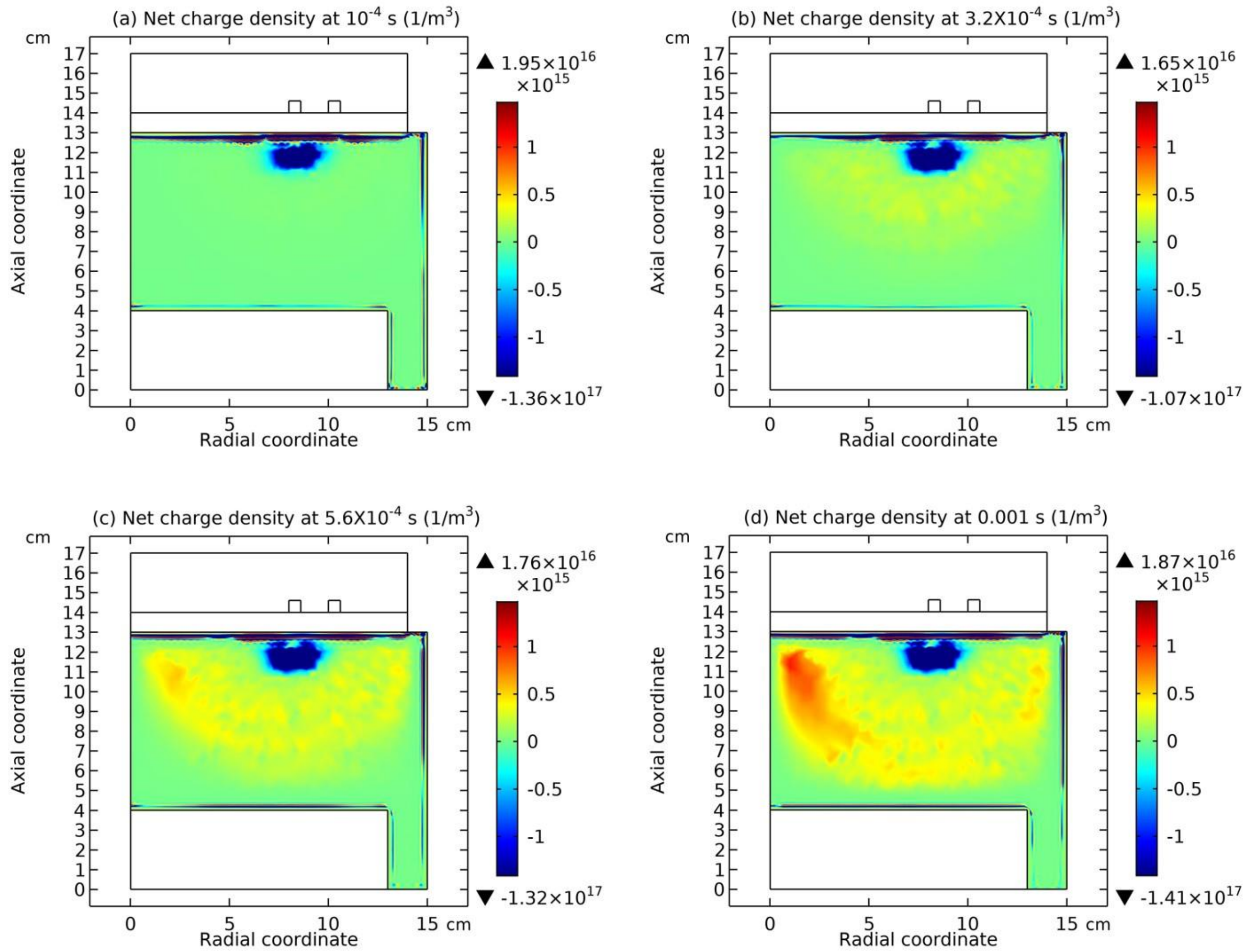

Figure 41 Process of positive and radiative charge density appearance when the cation coagulation is destroyed. The figure data are given by the fluid model simulation of $Ar/SF_6$ inductive plasma at the discharge conditions of 300W, 90mTorr and 10% $SF_6$ content. The blue and negative net charge density correlates to the ambi-polar self-coagulation concept, which will be explained in Sec. (f).

## (e) Re-self-coagulation of minor cations and its astronomy significance

In this section, it is shown that the minor cations re-self-coagulate after the mass point expelling effect, for instance the $F^+$ cation in Fig. 42. The position of re-self-coagulation is selected at the discharge axis where it is easier for species assembling (geometric effect), hence forming negative source as shown in Fig. 43. In addition, the drift of ambi-polar potential is not reached in this region, providing free diffusion condition. It is noted that the original $F^+$ coagulation under the coil is not influenced, where the ambi-polar self-coagulation holds. One more minor cation, $S^+$, re-self-coagulates in the Figs. 44 and 45. The difference is that it experiences two attempts for finishing the re-self-coagulation. The first attempt is failed for the forming negative source location is far away from the positive source at the coil. The negative source arises from the positive source normal transport. WAS their communication channel cut, the self-coagulation halts. The finally formed negative source of $S^+$ is presented in Fig. 46. One more distinction is that the $S^+$ re-self-coagulation is not that strong as the $F^+$ minor cation, again caused by the mass difference. Other minor cations, *e.g.*, $SF^+$, $SF_2^+$, fails to re-self-coagulate also because of their relatively large masses. In Fig. 47, it is shown that the minor cations re-self-coagulation has important astronomic significance, correlating the formation of celestial bodies. In the interstellar space, the plasma is quite sparse and its density is quite low. The re-self-coagulation behavior here is suitable to the space plasma, while the collective interactions of DENSE (relatively) low-temperature laboratory plasma is not applicable. Namely, the ambi-polar self-coagulation cannot happen in the sparse space plasma at low densities. Therefore, to counteract the general self-coagulation force inward, the outward centrifugal force is needed. That's why the celestial bodies all rotate. This is, we hope, a reasonable hypothesis from laboratory plasma studies, waiting for relevant validations of astronomers.

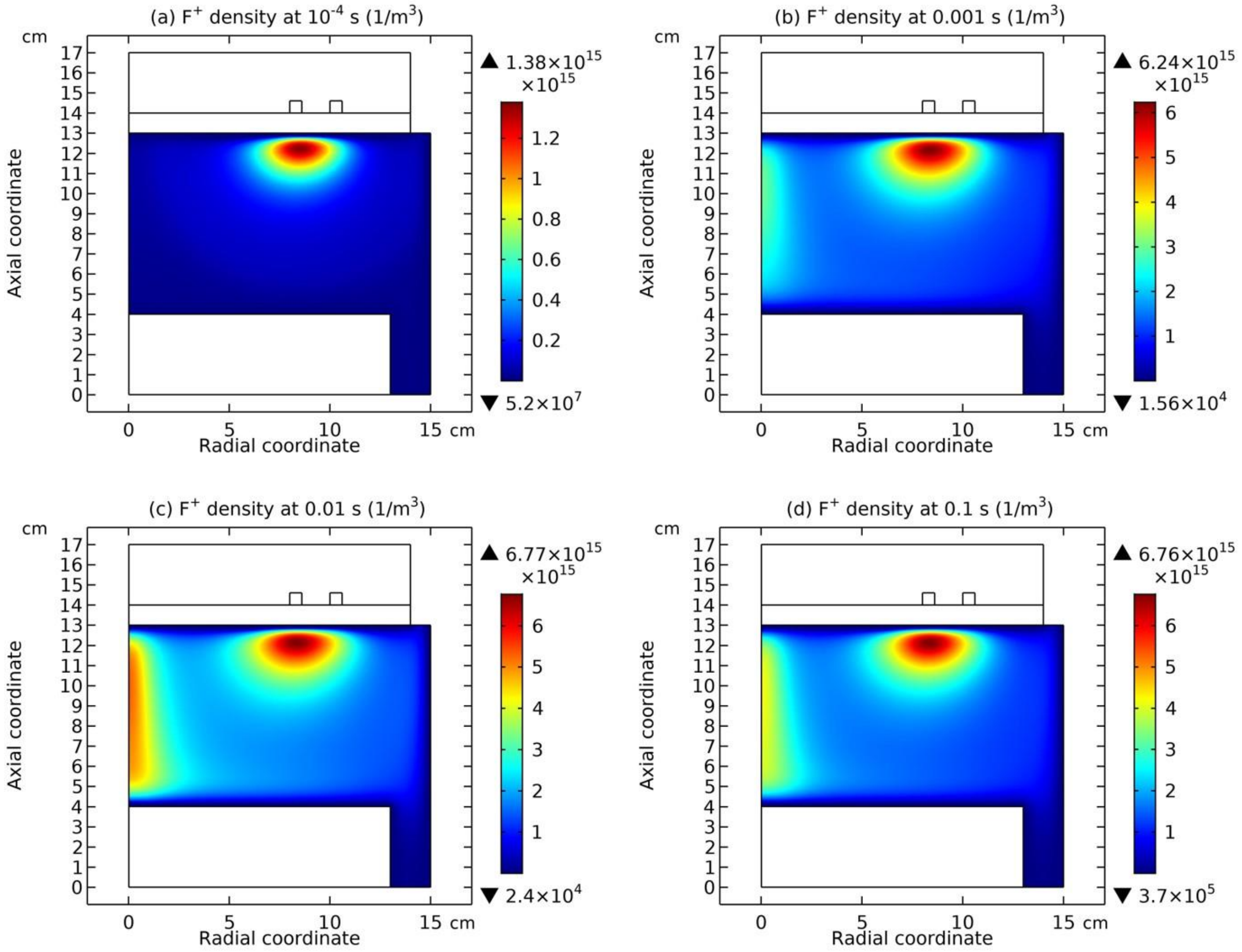

Figure 42 Process of minor cation $F^+$ re-self-coagulation after the mass-point expelling effect. The figure data are given by the fluid model simulation of $Ar/SF_6$ inductive plasma at the discharge conditions of 300W, 90mTorr and 10% $SF_6$ content.

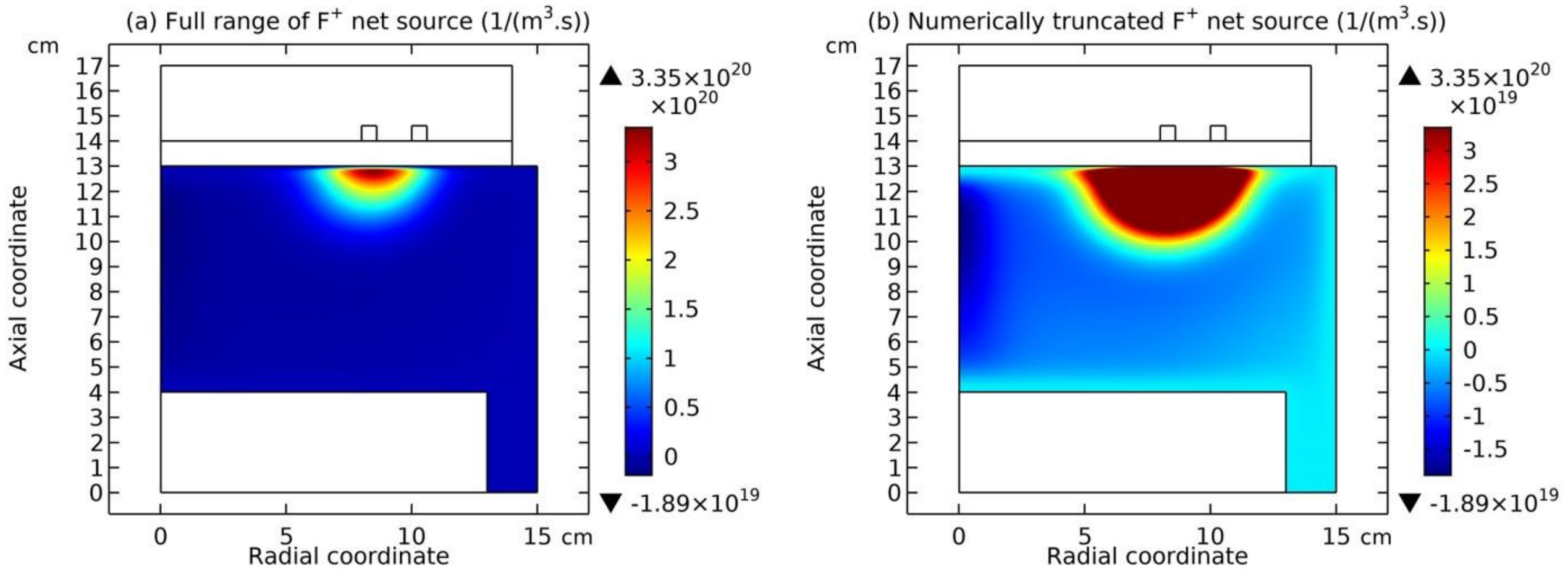

Figure 43 Net source of minor cation $F^+$ with (a) full range and (b) numerical truncation. The figure data are given by the fluid model simulation of $Ar/SF_6$ inductive plasma at the discharge conditions of 300W, 90mTorr and 10% $SF_6$ content.

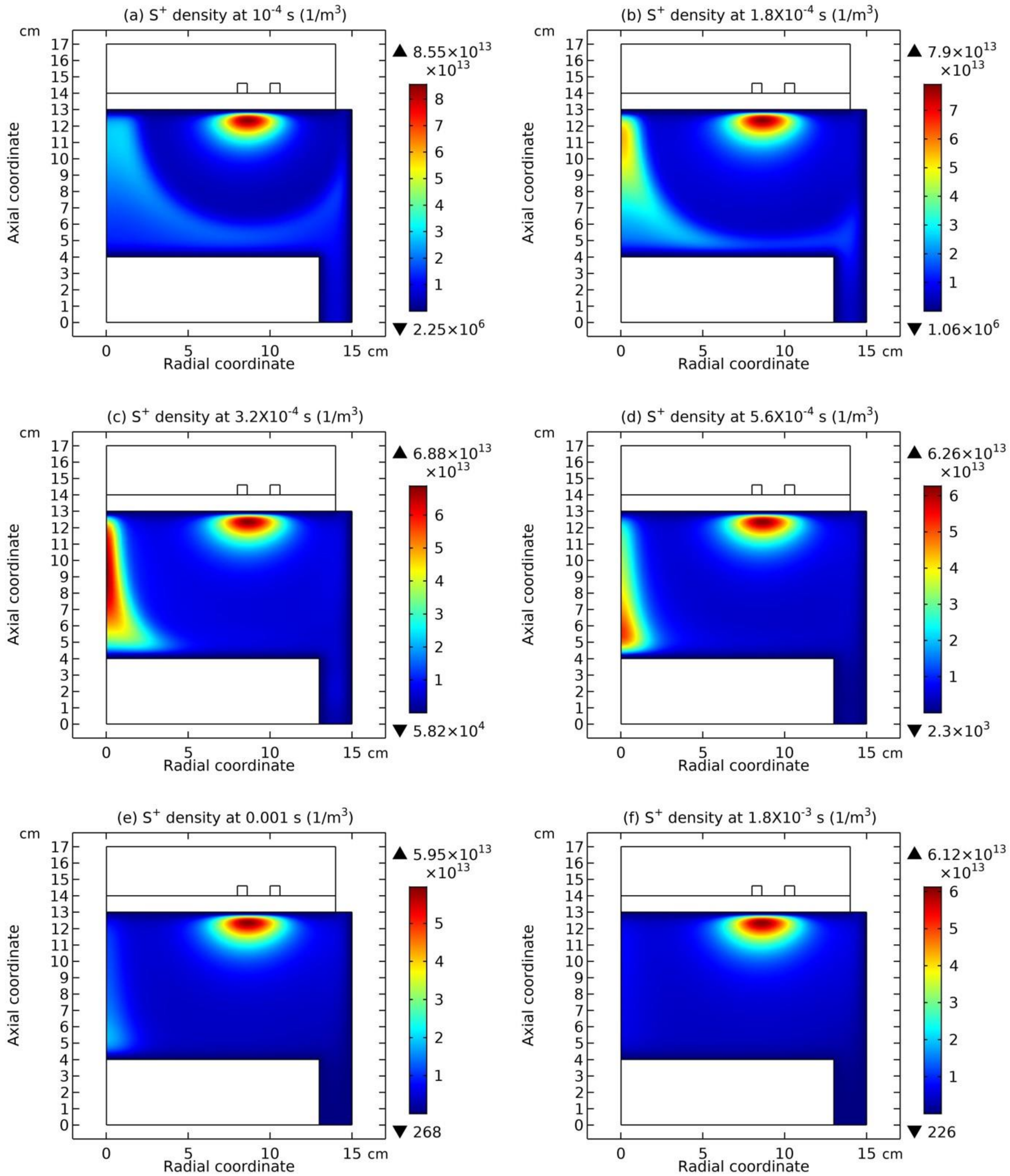

Figure 44 The first attempt for the minor cation $S^+$ re-self-coagulation. It fails at the end of this time. The figure data are given by the fluid model simulation of Ar/$SF_6$ inductive plasma at the discharge conditions of 300W, 90mTorr and 10% $SF_6$ content.

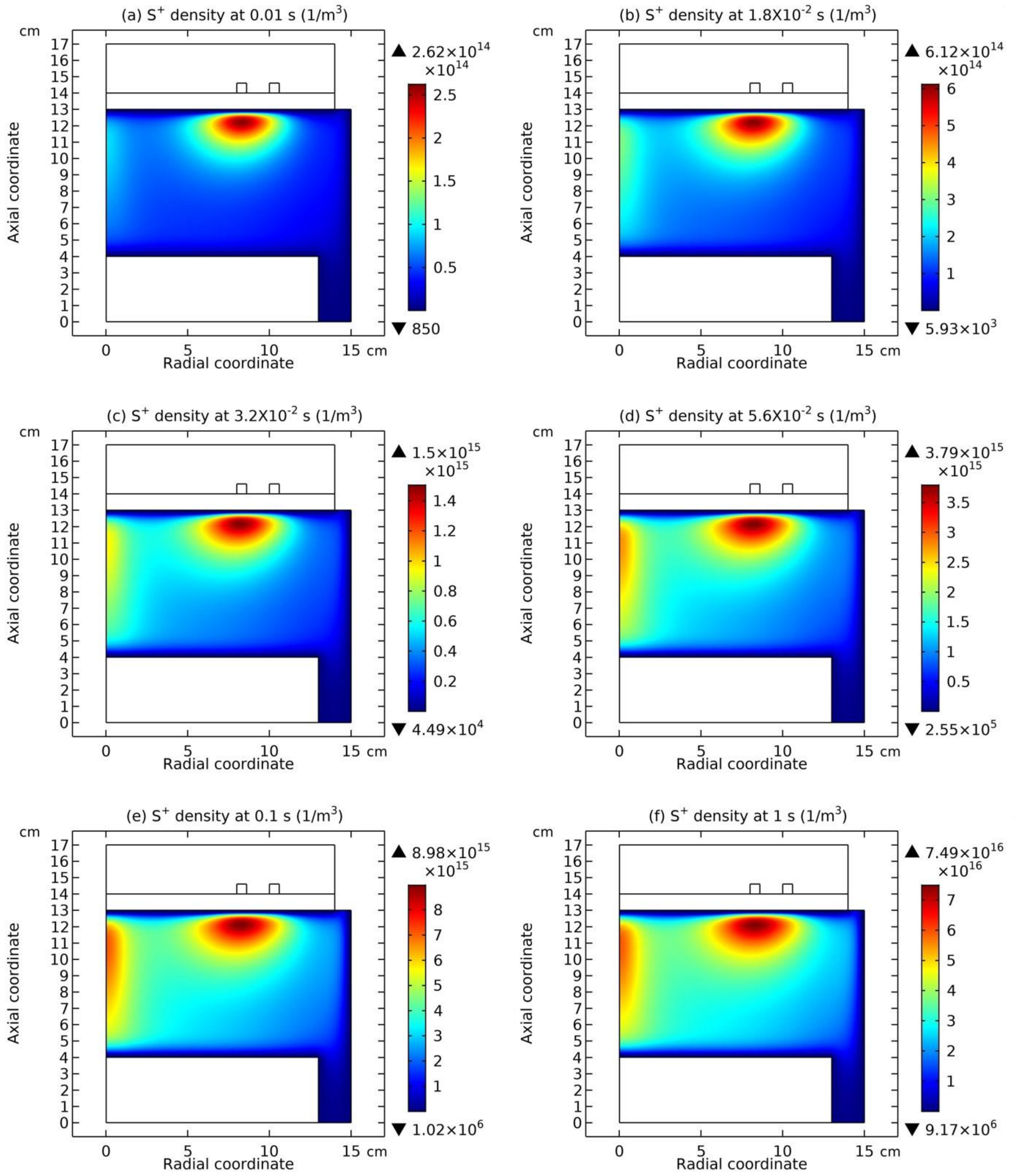

Figure 45 The second attempt of minor cation $F^+$ re-self-coagulation. It succeeds this time. The figure data are given by the fluid model simulation of Ar/$SF_6$ inductive plasma at the discharge conditions of 300W, 90mTorr and 10% $SF_6$ content.

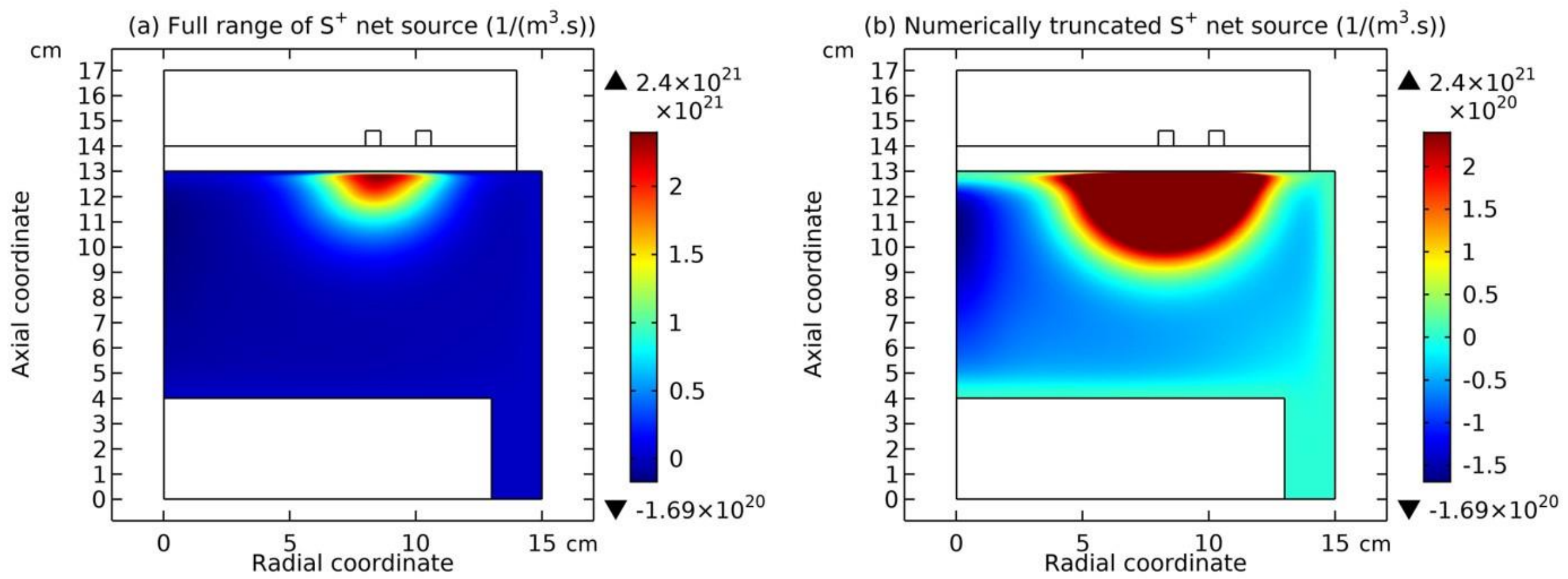

Figure 46 Net source of minor cation $S^+$ with (a) full range and (b) numerical truncation. The figure data are given by the fluid model simulation of $Ar/SF_6$ inductive plasma at the discharge conditions of 300W, 90mTorr and 10% $SF_6$ content.

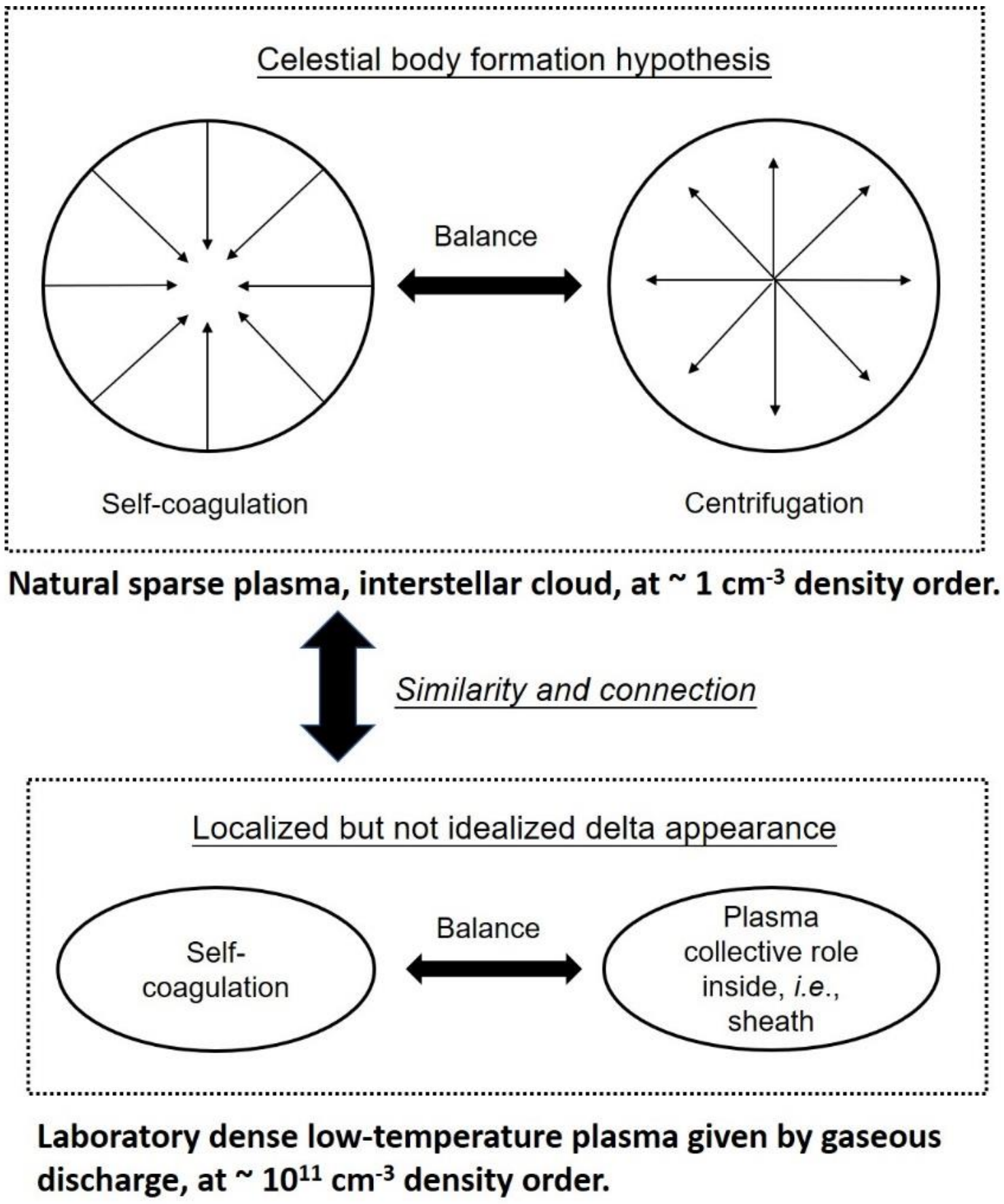

Figure 47 Hypothesis of the celestial body formation after cosmic explosion: A balance of core-oriented self-coagulation against rotational centrifugal force, and its connection with the presently investigated laboratory electronegative plasma case, where the self-coagulation of plasma is balanced by its own collective interaction results, *i.e.*, negative sheaths (see next).

### (f) Advective ambi-polar self-coagulation

The electron and summed cations densities distributions in Fig. 48 are different. In Fig. 49, obvious non-neutrality of plasm inside under the coil is observed (more obvious in Fig. 50). We believe it is located at the interface of two transport mechanisms, ambi-polar diffusion (triple-species system) and ambi-polar self-coagulation. The ambi-polar potential of triple-species system at high electronegativity is feeble and its influence on the self-coagulation is weak. For finishing the ambi-polar self-coagulation alone, the negative sheaths are needed. This is logic, since the anions are drifted before the self-coagulation, so their velocity can exceed over the Bohm threshold. This satisfies the sheath forming criterion. This type of ambi-polar self-coagulation is therefore defined as advective type. As shown in Fig. 51, the forming mechanism of delta anion in the inductive $Ar/O_2$ plasma belongs to the advective ambi-polar self-coagulation, as it holds the blue negative sheaths. Nevertheless, the 10mTorr $Ar/SF_6$ discharge is not, because their anions are not drifted, but are spontaneously self-coagulated at flat potential.

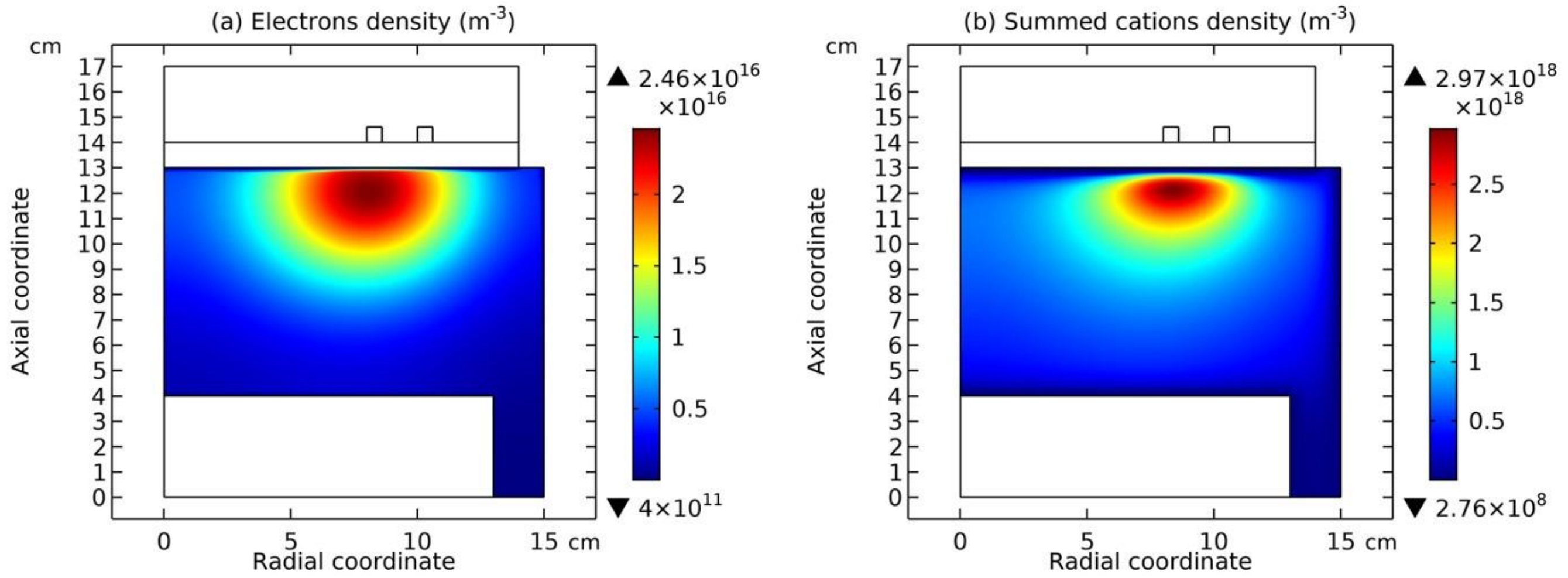

Figure 48 Electrons (a) and summed cations (b) densities of $Ar/SF_6$ inductive plasma, at the discharge conditions of 90mTorr, 300W and 10% $SF_6$ content, given by fluid model simulation.

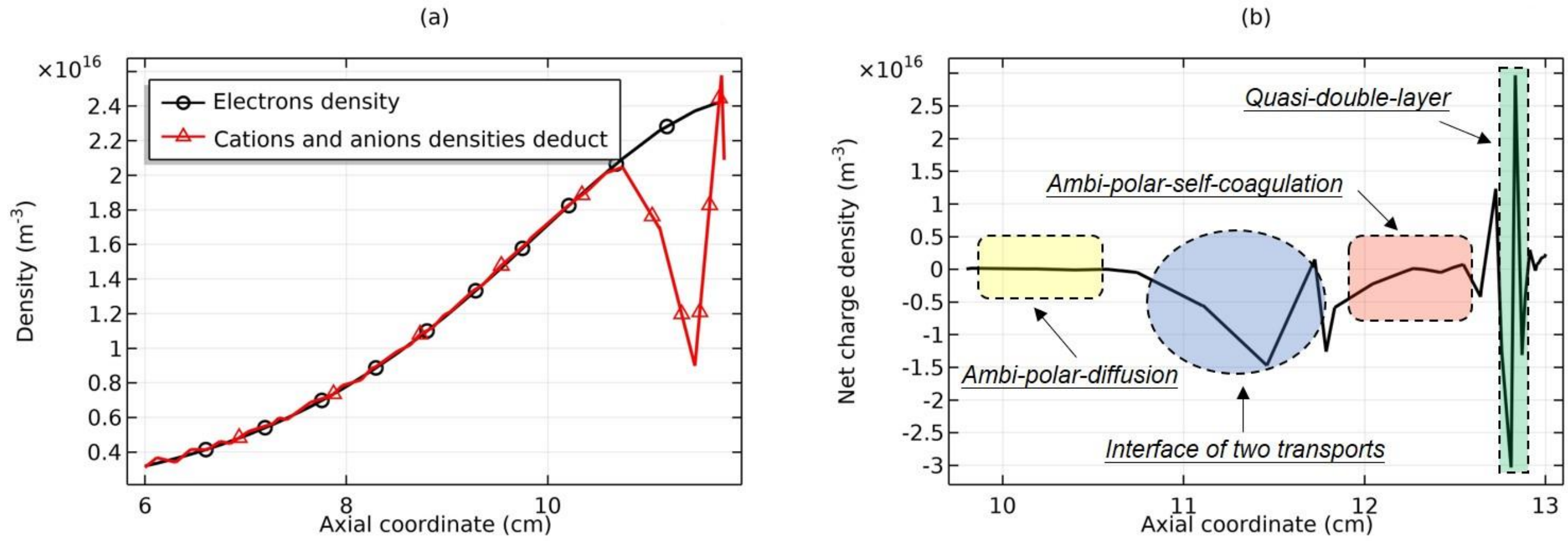

Figure 49 (a) Axial profiles of electron density and deduct of cations and anions densities, and (b) net charge density axial profile, where the ambi-polar-diffusion, ambi-polar-self-coagulation and their interface, as well as the quasi-double-layer are shown. The discharge conditions are the same as in Fig. 48.

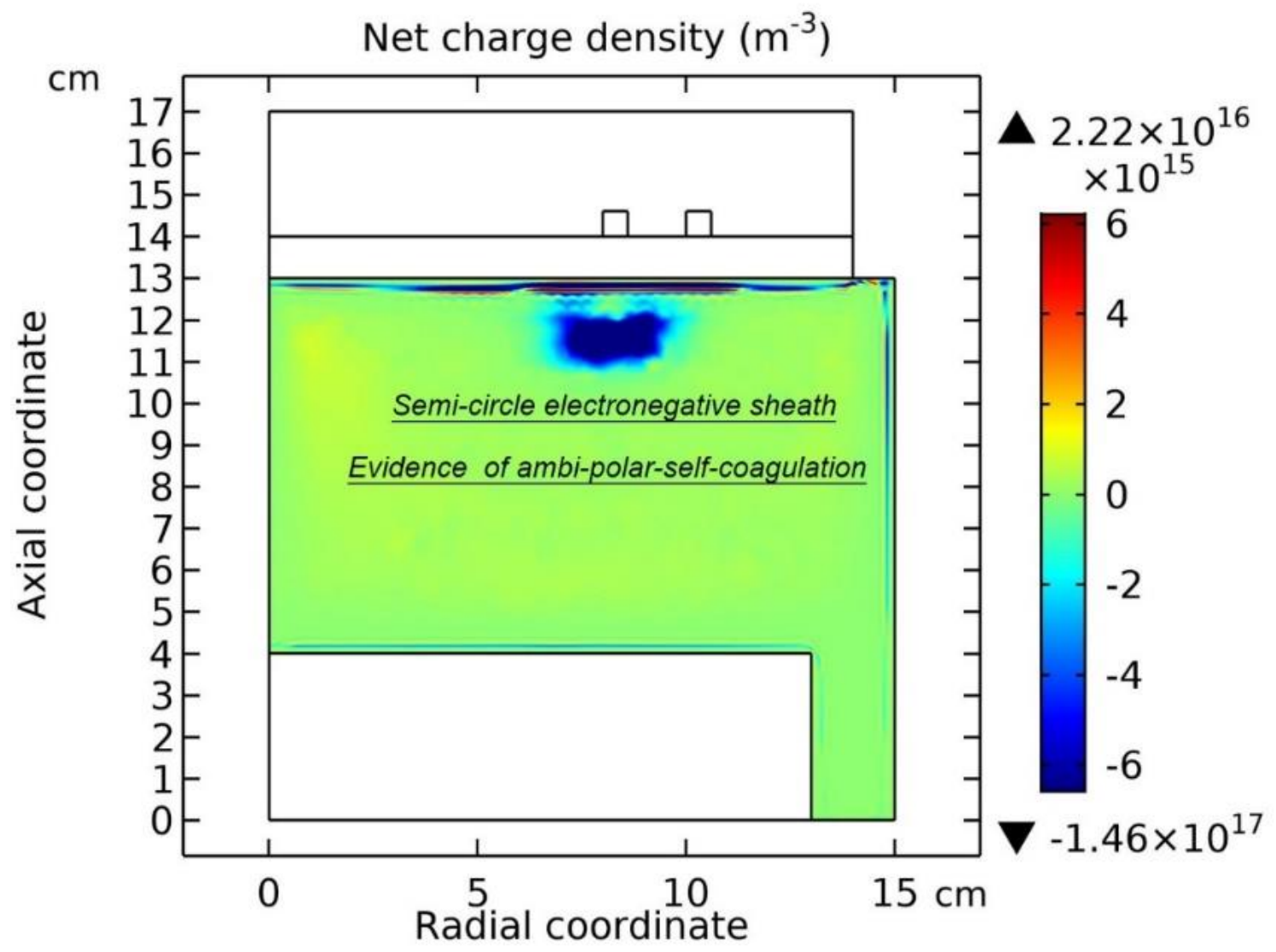

Figure 50 The net charge density of Ar/$SF_6$ inductive plasma, at the discharge conditions of 90mTorr, 300W and 10% $SF_6$ content, given by fluid model simulation.

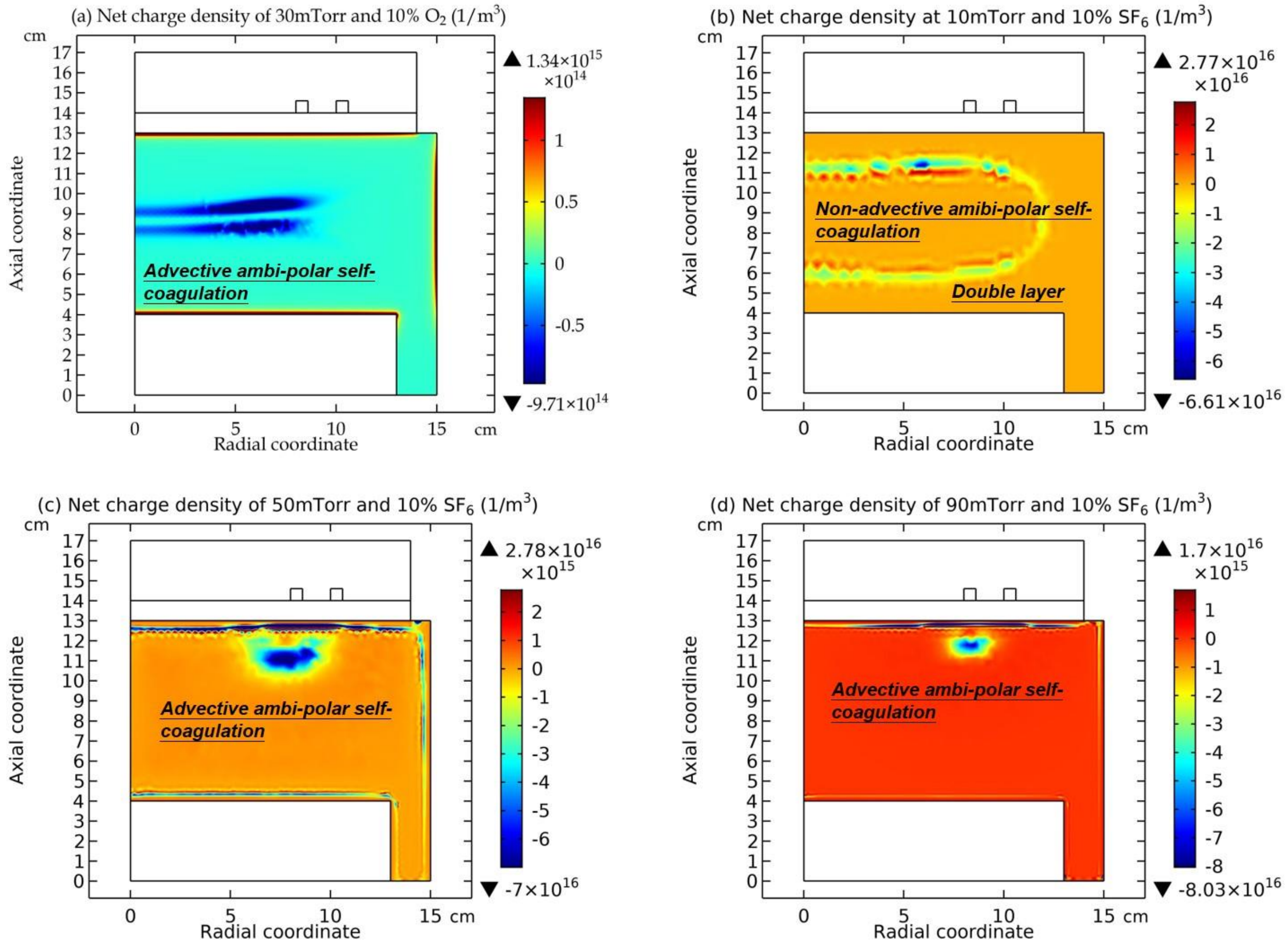

Figure 51 The net charge densities of (a) Ar/$O_2$ inductive plasma at 30mTorr and 10% $O_2$ content, and the Ar/$SF_6$ inductive plasma at 10% $SF_6$ content and different pressures, (b) 10mTorr, (c) 50mTorr and (d) 90mTorr, respectively. The figure data are given by fluid model simulation and the input power is 300W.

## (III.2.C) Pressure effect at high electronegativity

### (a) Evolution of ions density profile beside for the self-coagulation

In Figs. 52 and 53, the summed cations density profile in the core experiences the parabola, ellipse and flat-top, besides for the self-coagulation. These three analytic profiles are given by Lichtenberg *et al* at the condition that the positive chemical source is always larger than negative source; hence their works failed at predicting the self-coagulation behavior.

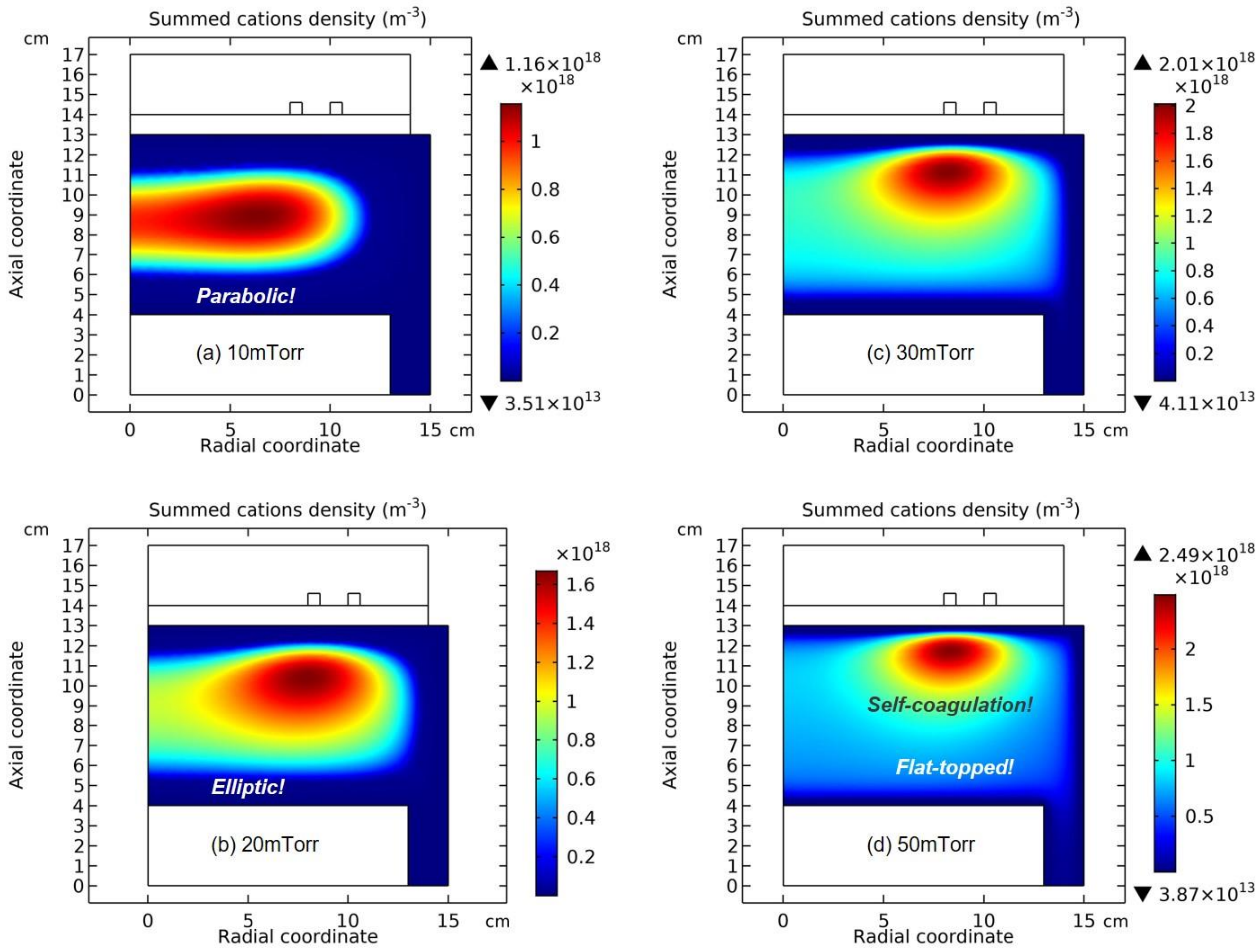

Figure 52 Evolution of cations density profile in Ar/$SF_6$ discharge with pressure, at 300W and 10% $SF_6$ content of gases mixture, given by fluid model simulation.

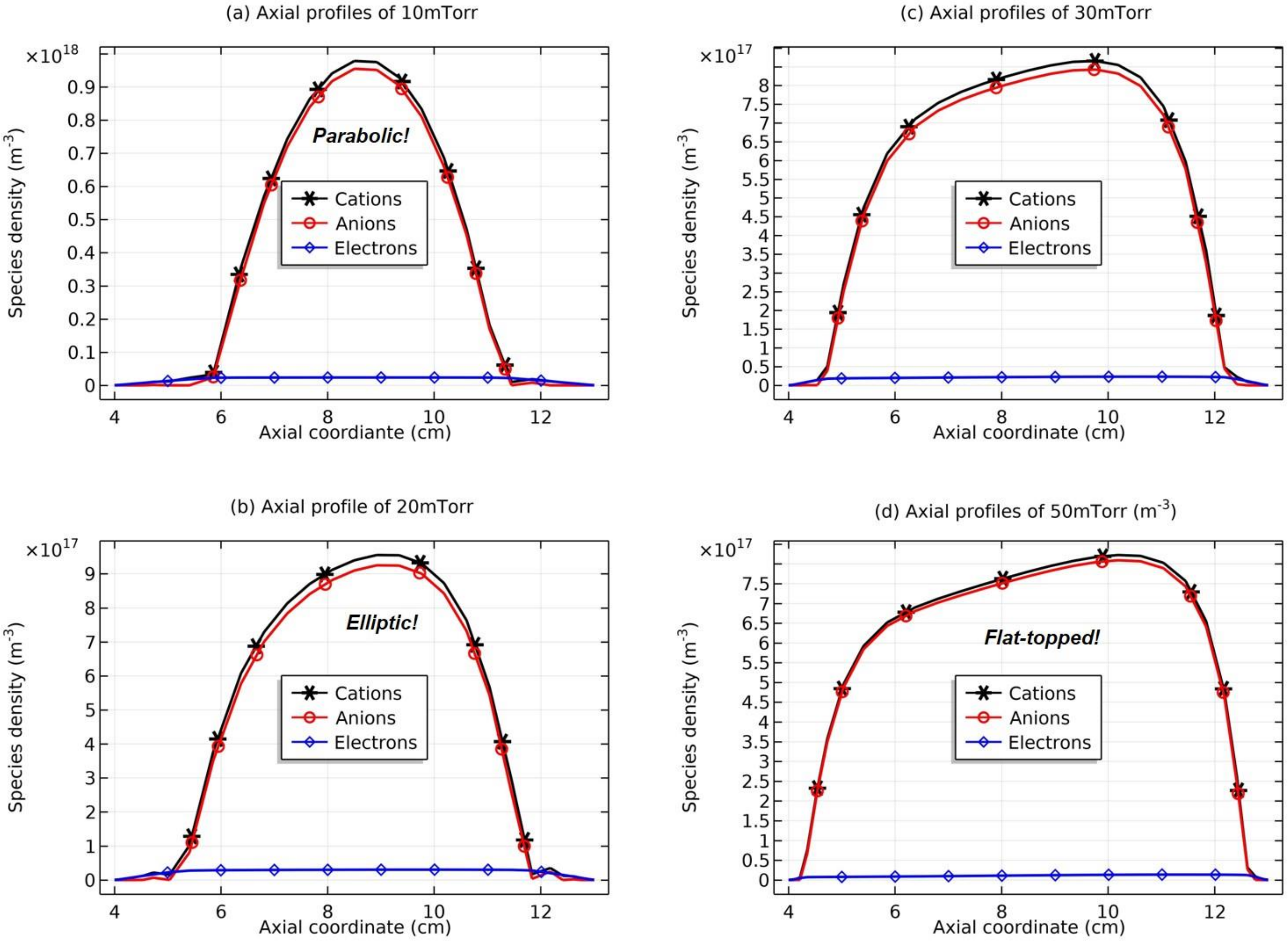

Figure 53 Axial profiles of plasma species, *i.e.*, cations, anions and electrons, against the pressure in the Ar/$SF_6$ discharge, at 300W and 10% $SF_6$ content of gases mixture, given by the fluid model simulation.

### (b) Ellipse and flat-top theories

The parabola theory has been illustrated in Section III.2(A). Herein, the ellipse and flat-top theories are described. First, the ellipse theory inequality of Lichtenberg *et al* is given in Fig. 54, at the condition of substantial recombination loss and the approximation that electron density variation is negligible. Then, the ellipse theory of them is displayed in Fig. 55, where the analytic solution is given as a standard elliptic integral. The transition from ellipse to flat-top model (defined by the Lichtenberg *et al*) is occurred when further increasing the electronegativity until the electron density variation cannot be neglected anymore. There is no analytic solution for flat-top model due to the interfere of electron density quantity. So, it is hard to understand the essence of flat-top discharge model. We rebuilt the flat-top model based on the fluid model simulation details in Fig. 56, which is easy to understand for readers. Its essence is that the recombination loss counteracts the generation rate in the inner bulk, and at infinitesimal positive source, the originally parabola profile part is flattened. At the bulk border, the electronegativity and anion density swiftly decrease, and the normal positive source and parabola is rebuilt. The combination of two parts of profile comprises the flat-top model. It is noted that the ambi-polar diffusion coefficient of *two-species system* is used in the flat-top model. In Fig. 57, the ellipse theory inequality analysis shows that the ambi-polar diffusion of two-species (represented by the exponential factor) dominates over the triple-species system (determined by the square root factor) when the anions drift fast inward because of the *recombination loss flux* defined by Lichtenberg *et al* (represented by the fact that anion Boltzmann relation is broken at recombination), opposite to the outward direction of cation. The mutual movement of anion with the combination of electron and cation destroys the triple-species ambi-polar diffusion, and two-species transport dominates. This produces predominant potential variation and accordingly the electron density begins to vary at the Boltzmann relation. In Fig. 58, the net sources of cations of $Ar/SF_6$ inductive plasma at 10mTorr and 90mTorr given by fluid simulations are shown. As seen, at 90mTorr where flat-top model prevails, its positive source is indeed almost counteracted by recombination loss. While at 10mTorr where the parabola profile is given, the positive source of cation (ionizations) is less influenced by the recombination, which validates the assumption in the parabola theory of Fig. 15 in Sec. III.2(A) that recombination loss can be neglected. The flat-top model of high pressure and high electronegativity also interprets the disappearance of stratification. It is difficult for the ambi-polar potential to push anion at a flat density profile plotted at rather high density, from the point of steady state.

Cation and anion continuity equations at drift-diffusion approximation

$$\frac{\mathrm{d}}{\mathrm{d}x}\left(-D_+\frac{\mathrm{d}n_+}{\mathrm{d}x}+n_+\mu_+E\right)=K_{iz}n_0n_e-K_{rec}n_+n_-$$

$$\frac{\mathrm{d}}{\mathrm{d}x}\left(-D_-\frac{\mathrm{d}n_-}{\mathrm{d}x}-n_-\mu_-E\right)=K_{att}n_0n_e-K_{rec}n_+n_-$$

Bring in the electron Boltzmann relation and neutrality condition

$$\frac{\mathrm{d}}{\mathrm{d}x}\left(-D_+\frac{\mathrm{d}}{\mathrm{d}x}(n_-+n_e)-\mu_+(n_-+n_e)\frac{D_e}{\mu_e}\frac{1}{n_e}\frac{\mathrm{d}n_e}{\mathrm{d}x}\right)$$
$$=K_{iz}n_0n_e-K_{rec}(n_-+n_e)n_-$$

$$\frac{\mathrm{d}}{\mathrm{d}x}\left(-D_+\frac{\mathrm{d}n_-}{\mathrm{d}x}+\mu_+n_-\frac{D_e}{\mu_e}\frac{1}{n_e}\frac{\mathrm{d}n_e}{\mathrm{d}x}\right)$$
$$=K_{att}n_0n_e-K_{rec}(n_-+n_e)n_-.$$

Add the cation and anion continuity equations and drop small terms

$$\frac{\mathrm{d}}{\mathrm{d}x}\left(-2D_+\frac{\mathrm{d}n_-}{\mathrm{d}x}-\gamma D_+\frac{\mathrm{d}n_e}{\mathrm{d}x}\right)=(K_{iz}+K_{att})n_0n_e-2K_{rec}n_+^2$$

(a)

$$\left|\frac{\gamma\,\mathrm{d}^2n_e/\mathrm{d}x^2}{2\mathrm{d}^2n_+/\mathrm{d}x^2}\right|\ll 1$$

$$\kappa\equiv\frac{1}{11.55\alpha_0}\left(\frac{D_+}{K_{rec}n_{e0}\alpha_0\ell_p^2}\right)^{1/2}$$
$$\times\exp\left(\frac{2K_{rec}n_{e0}\alpha_0\ell_p^2}{D_+}\right)^{1/2}<1.$$

(c)

$$\Gamma_-=-D_-\frac{\mathrm{d}n_-}{\mathrm{d}x}-n_-\mu_-E\quad\text{and}\quad\Gamma_-=\int_0^x K_{att}n_0n_e\mathrm{d}x-\int_0^x K_{rec}n_+n_-\mathrm{d}x.$$

$$n_-\mu_-E(x)=-\left(\int_0^x K_{att}n_0n_{e0}\mathrm{d}x-\int_0^x K_{rec}n_+n_-\mathrm{d}x+D_-\frac{\mathrm{d}n_-}{\mathrm{d}x}\right).$$

$$\frac{\mathrm{d}}{\mathrm{d}x}\left(-D_+\frac{\mathrm{d}n_+}{\mathrm{d}x}+n_+\mu_+E\right)=K_{iz}n_0n_e-K_{rec}n_+n_-$$

$r_\mu=\mu_+/\mu_-.$

$$\frac{\mathrm{d}}{\mathrm{d}x}\left[-\left(1+\frac{T_-}{T_+}\right)D_+\frac{\mathrm{d}n_+}{\mathrm{d}x}+r_\mu\int_0^x K_{rec}n_+^2\mathrm{d}x\right.$$
$$\left.-r_\mu\int_0^x K_{att}n_0n_e\mathrm{d}x\right]\simeq K_{iz}n_0n_e-K_{rec}n_+^2$$

$$\left(1+\frac{T_-}{T_+}\right)D_+\frac{\mathrm{d}^2n_+}{\mathrm{d}x^2}+(K_{iz}+r_\mu K_{att})n_0n_e$$
$$-(1+r_\mu)K_{rec}n_+^2=0.$$

(b)

Figure 54 Ellipse theory inequality of Lichtenberg *et al.*

$$\left(1+\frac{T_-}{T_+}\right)D_+\frac{\mathrm{d}^2 n_+}{\mathrm{d}x^2}+(K_{iz}+r_\mu K_{att})n_0 n_e -(1+r_\mu)K_{rec}n_+^2=0.$$

$n_+ n_- = n_{e0}^2\alpha(\alpha+1)$

$$-\frac{1}{2(1+T_-/T_+)D_+ n_{e0}}\frac{\mathrm{d}}{\mathrm{d}\alpha}\beta^2 =(K_{iz}+r_\mu K_{att})n_0 n_{e0}-(1+r_\mu)K_{rec}n_{e0}^2\alpha(\alpha+1)$$

$$\beta(\alpha)=-(1+T_-/T_+)D_+ n_{e0}\frac{\mathrm{d}\alpha}{\mathrm{d}x}.$$

Integrate at $\beta(\alpha_0)=0$

$$\beta(\alpha)=\beta_0\left[(\alpha_0-\alpha)-\frac{(1+r_\mu)K_{rec}n_{e0}}{(K_{iz}+r_\mu K_{att})n_0}\times\left(\frac{\alpha_0^3-\alpha^3}{3}+\frac{\alpha_0^2-\alpha^2}{2}\right)\right]^{1/2}$$

$$\beta_0=[2(1+T_-/T_+)D_+(K_{iz}+r_\mu K_{att})n_0 n_{eo}^2]^{1/2}.$$

$$\beta(\alpha)=-(1+T_-/T_+)D_+ n_{e0}\frac{\mathrm{d}\alpha}{\mathrm{d}x}.$$

Integration

$$x(\alpha)=\int_{\alpha(x)}^{\alpha_0}\frac{(1+T_-/T_+)D_+ n_{e0}\mathrm{d}\alpha}{\beta(\alpha)}$$

$\eta < 1,$

Standard elliptical integral

$$x=\frac{(1+T_-/T_+)D_+ n_{e0}}{\beta_0}\frac{\alpha_0}{(\eta/3)^{1/2}}\times\int_{\alpha(x)}^{\alpha_0}\frac{\mathrm{d}\alpha}{[(b\alpha_0-\alpha)(\alpha_0-\alpha)(a\alpha_0+\alpha)]^{1/2}}$$

$$\eta=\frac{(1+r_\mu)K_{rec}n_{e0}\alpha_0^2}{(K_{iz}+r_\mu K_{att})n_0}\qquad a=\frac{1}{2}+\frac{1}{2}\left(\frac{12}{\eta}-3\right)^{1/2}\qquad b=-\frac{1}{2}+\frac{1}{2}\left(\frac{12}{\eta}-3\right)^{1/2}.$$

Figure 55 Ellipse theory of Lichtenberg *et al*.

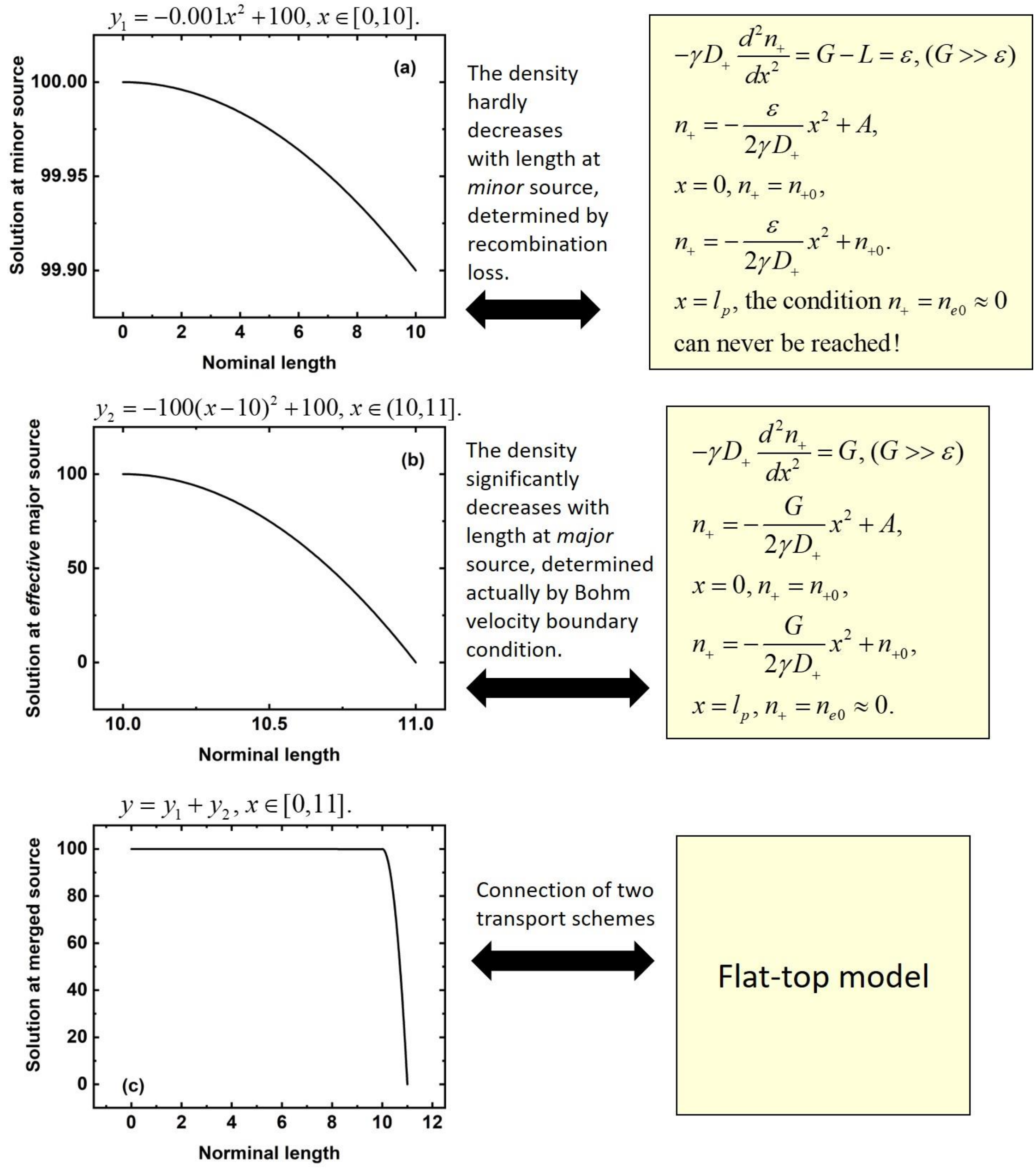

Figure 56 Flat-top theory rebuild

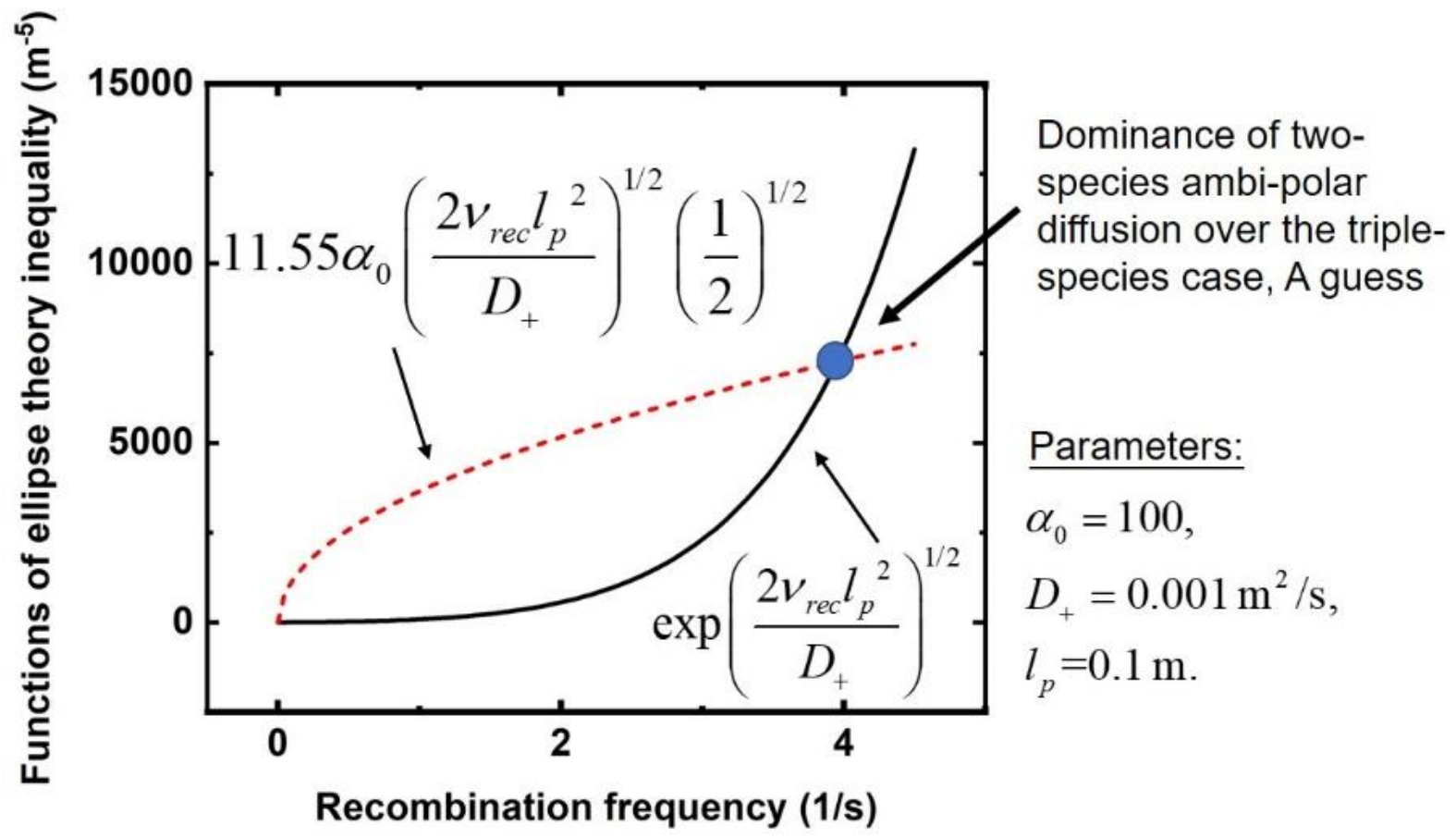

Figure 57 Ellipse theory inequality analysis

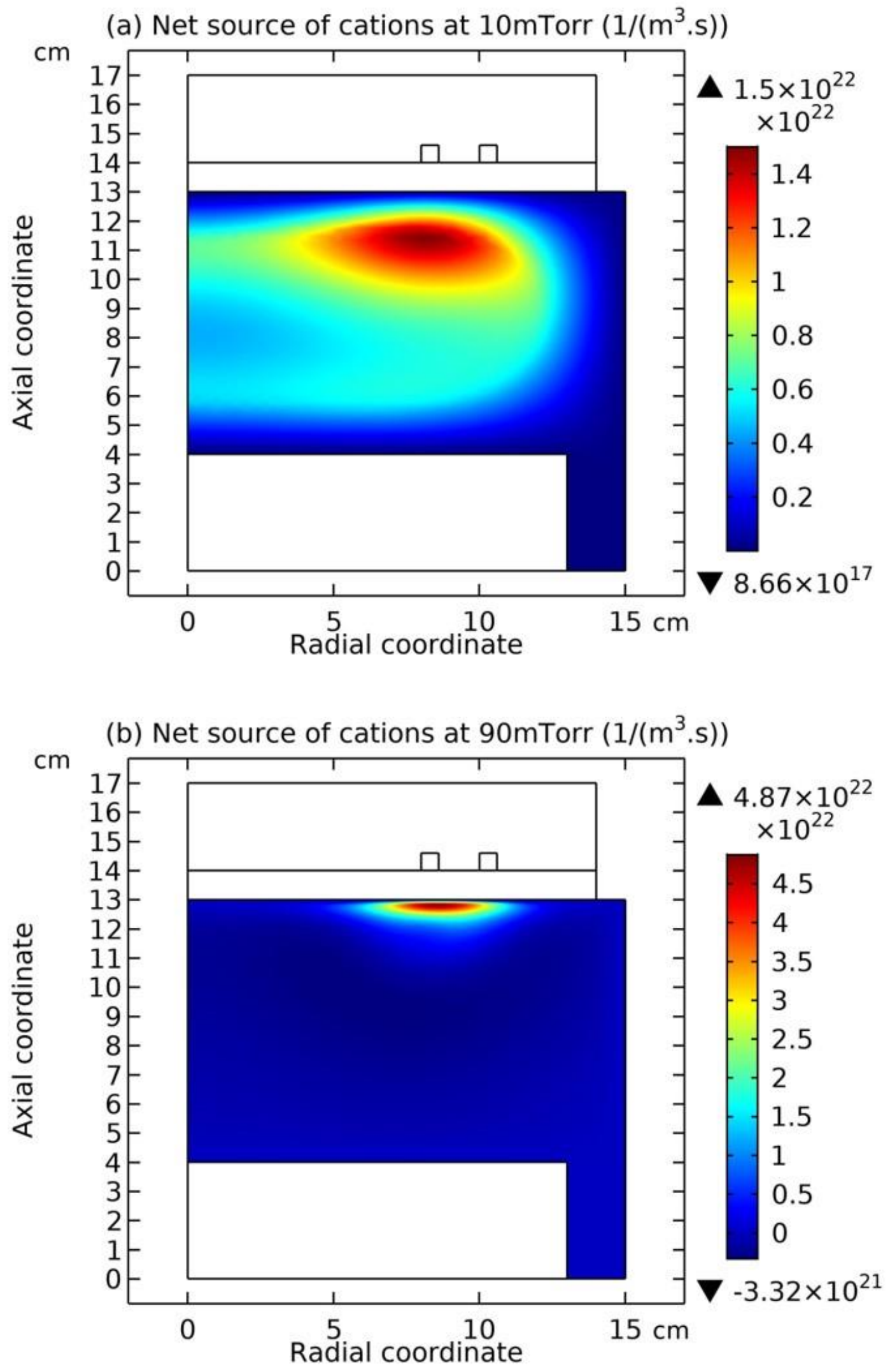

Figure 58 Net sources of cations of Ar/$SF_6$ inductive plasma at (a) 10mTorr and (b) 90mTorr, given by fluid model simulation at other discharge conditions of 300W and 10% $SF_6$ content.

### (c) Electron density profile at anion (heart) and its own (peripherical) self-coagulations

The electron density profile evolution of $Ar/SF_6$ inductive plasma with pressure is shown in Fig. 59. The density profile at 90mTorr is determined by both the heart anion self-coagulation and its own peripherical self-coagulation (shown in Section III.2(B).c). In Fig. 60, the electron density contours of pure argon plasma at two different electron transport coefficients are plotted. As increasing the pressure, at small transport coefficient, the argon plasma density is coagulated, but this coagulation is quite different with the case of $Ar/SF_6$ plasma. At large transport coefficient of electron, the argon plasma profile is almost kept, rather smooth without coagulation. In Fig. 61, the pressure dependences of electron density profile in $Ar/SF_6$ plasma at these two electron transport coefficient values are given. It is discovered that at large coefficient, the electron density profile still self-coagulates, but at a relatively low speed, which reveals the role of above *two* self-coagulation mechanisms on the electron density profile. In Fig. 62, the net sources of anions and electron at different pressures are plotted, which exhibits the locations of anions and electron self-coagulations, respectively.

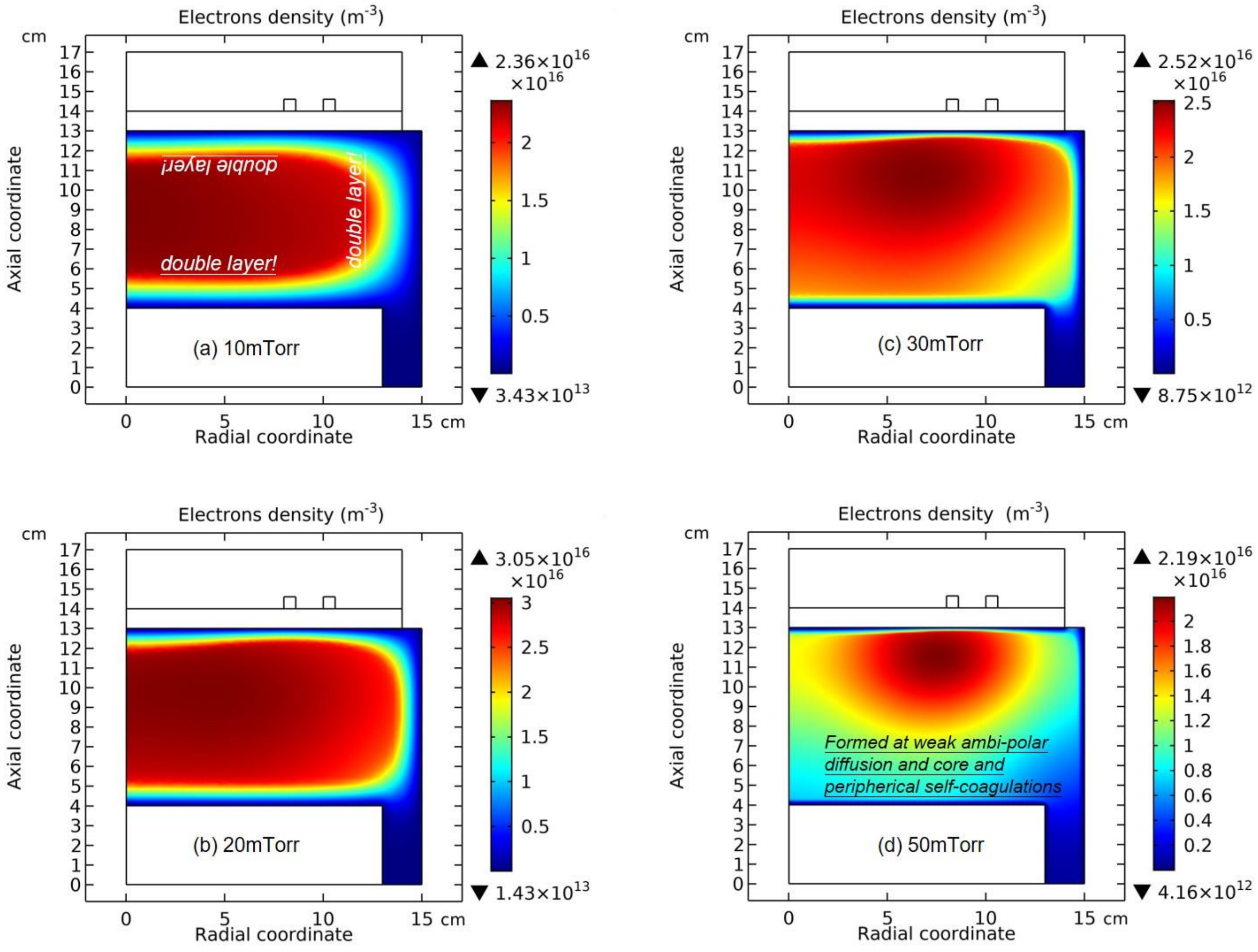

Figure 59 Evolution of electrons density profile in $Ar/SF_6$ discharge with pressure, at 300W and 10% $SF_6$ content of gases mixture, given by the fluid model simulation.

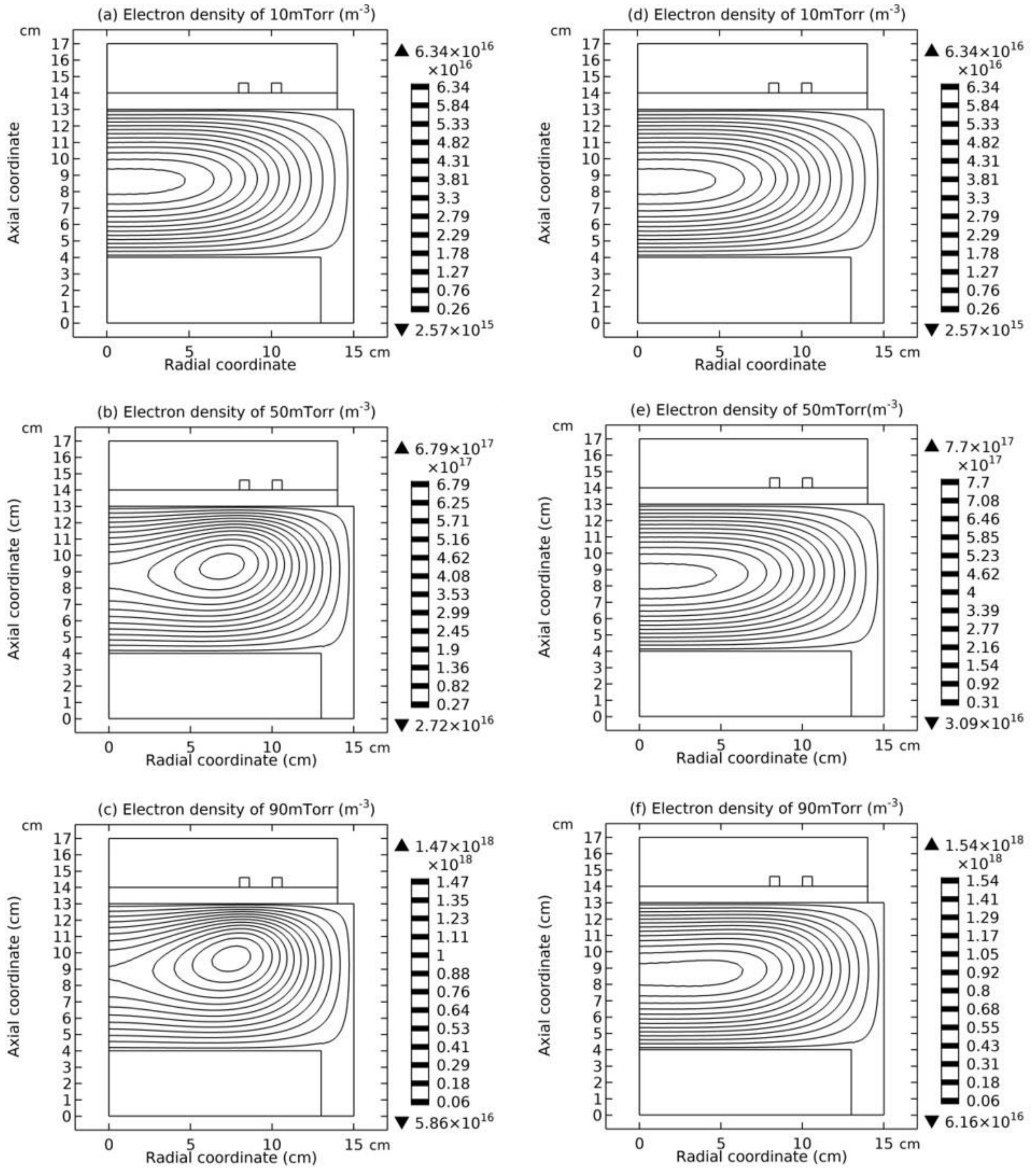

Figure 60 Electron density contours of pure argon plasma at (a) 10mTorr, (b) 50mTorr, and (c) 90mTorr, given by the fluid simulation utilizing a *small* reduced electron mobility, $\mu_e N_n = 8.2 \times 10^{23}\ (1/(V \cdot m \cdot s))$. Electron density contours of pure argon plasma at (d) 10mTorr, (e) 50mTorr, and (c) 90mTorr, given by the fluid simulation utilizing a *large* reduced electron mobility, $\mu_e N_n = 4.0 \times 10^{24}\ (1/(V \cdot m \cdot s))$. The input inducive power is 300W. Under the small electron transport coefficient, electron coagulates on increasing the pressure, and at large coefficient it does not.

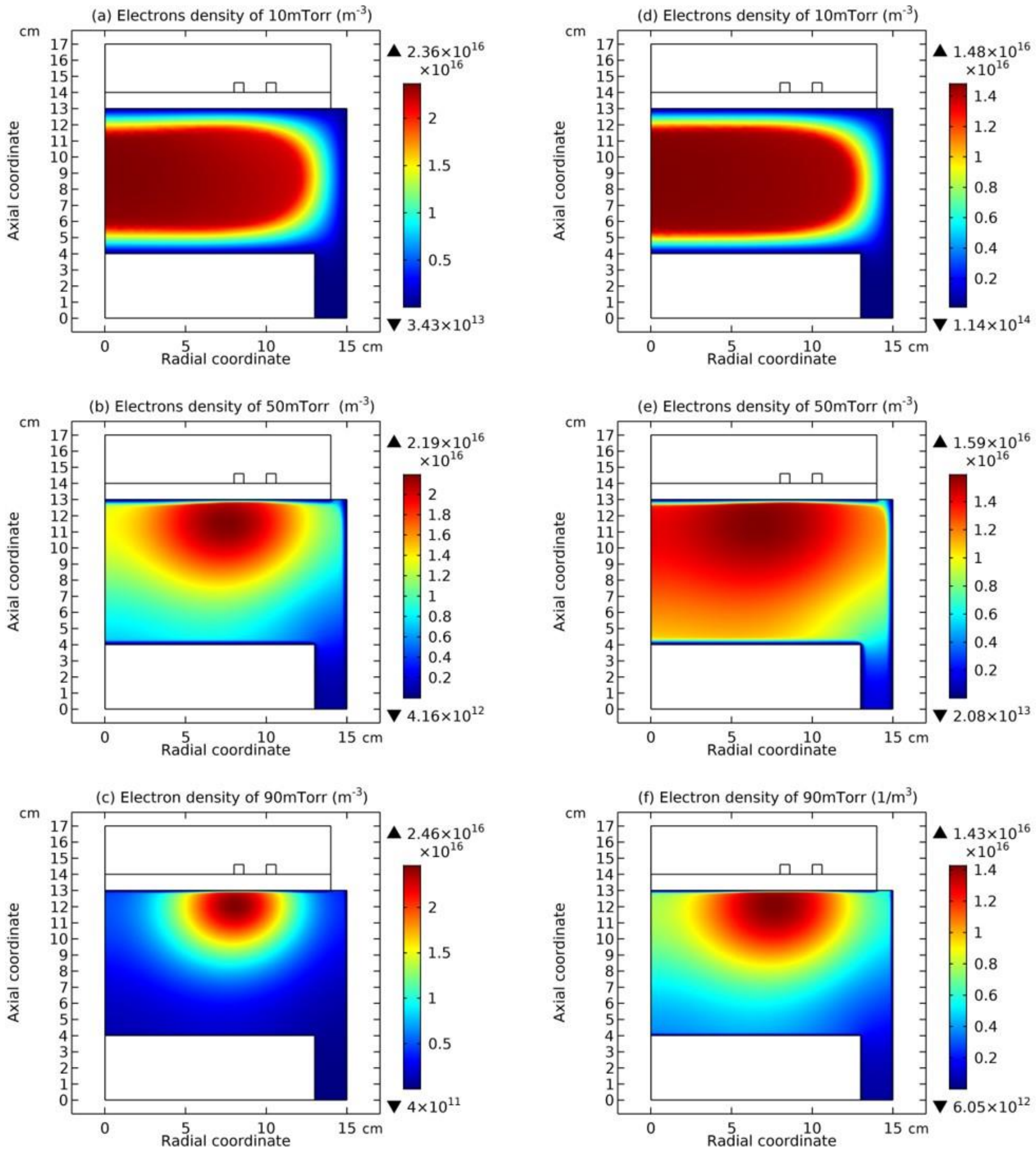

Figure 61 Electron density profiles of mixed Ar/$SF_6$ plasma at (a) 10mTorr, (b) 50mTorr, and (c) 90mTorr, given by the fluid simulation utilizing a *small* reduced electron mobility, $\mu_e N_n = 8.2 \times 10^{23}\ (1/(V \cdot m \cdot s))$. Electron density profiles of mixed Ar/$SF_6$ plasma at (d) 10mTorr, (e) 50mTorr, and (c) 90mTorr, given by the fluid simulation utilizing a *large* reduced electron mobility, $\mu_e N_n = 4.0 \times 10^{24}\ (1/(V \cdot m \cdot s))$. The input inducive power is 300W, and the $SF_6$ content is 10%. Electrons in Ar/$SF_6$ inductive plasma coagulate (with a slower speed) on increasing pressure even at large electron transport coefficient (when no coagulation is occurred in pure argon plasma; see Fig. 60(d-f)). The plot questions the saying that the phenomenon of coagulation-to-coil of high-density electronegative plasmas at high pressure (usually given by self-consistent simulations) is because of short mean free path, but implying new mechanism, heart and peripherical self-coagulations.

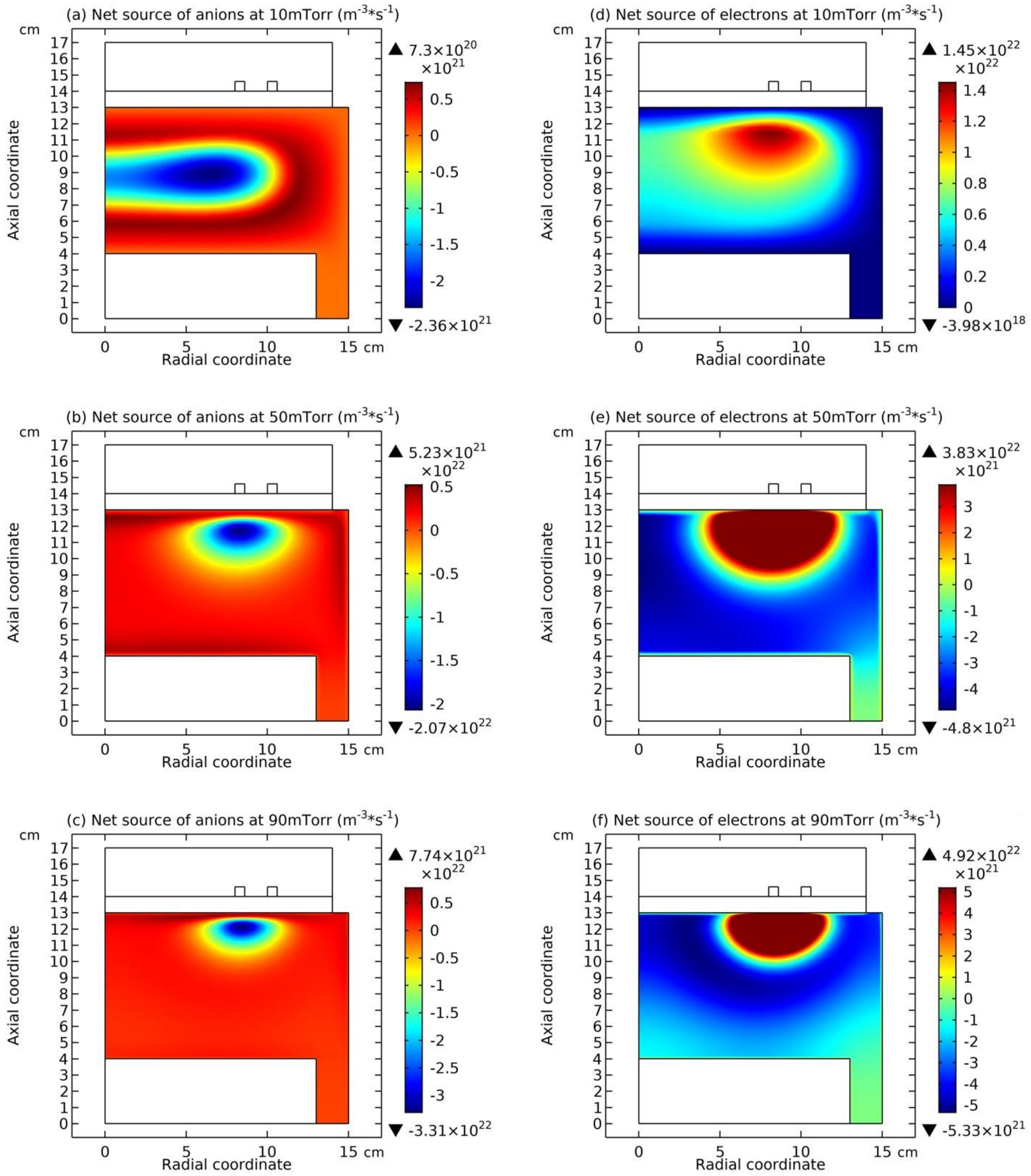

Figure 62 Pressure dependence of net sources of anions (a-c) and electron (d-f) of Ar/$SF_6$ inductive plasma, given by the fluid model simulation at the discharge conditions of 300W and 10% $SF_6$ content.

## (d) Electron deviates the Boltzmann relation

In this section, the fact that the electron deviates the Boltzmann relation at its own peripherical self-coagulation is focused and more detail about this fact is given in the supplementary material. As shown in Fig. 63, the electron deviates the Boltzmann relation at 50mTorr and 90mTorr, when the negative chemical source of electron cannot be neglected in Fig. 62(e,f). Accordingly, at 10mTorr, when the electron negative source is four orders less than the positive source in Fig. 62(d), *i.e.*, negligible, no self-coagulation happens on the electron, and its density is perfectly Boltzmann balanced, in Fig. 64.

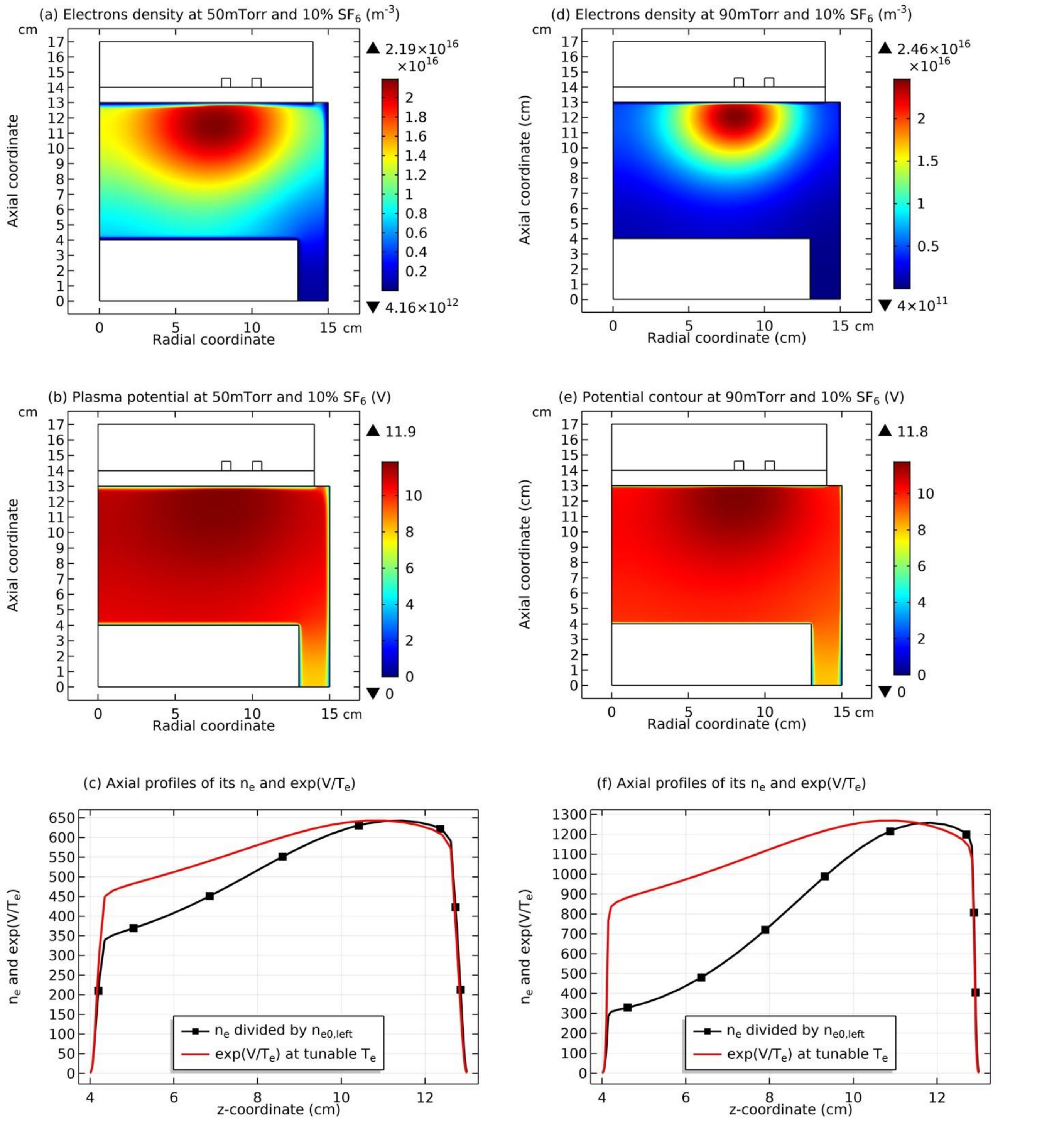

Figure 63 Electron density, plasma potential and their Boltzmann relation inspection at 50mTorr (a-c) and at 90mTorr (d-f), respectively, at the discharge conditions of 300W and 10% $SF_6$ content and fluid model simulation. Electrons deviate the Boltzmann relation at the high pressures, 50mTorr and 90mTorr.

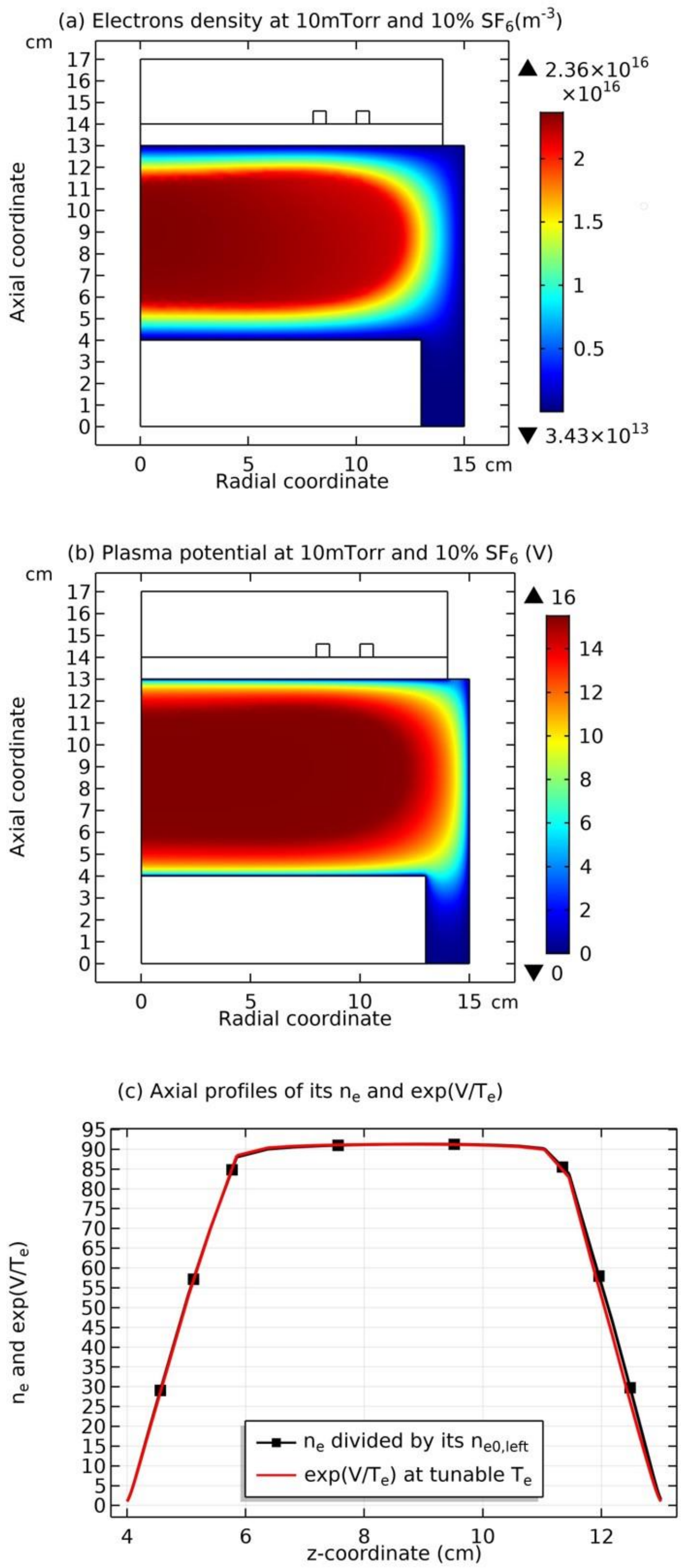

Figure 64 Electron density (a), plasma potential (b) and their Boltzmann relation inspection (c) at 10mTorr, at the discharge conditions of 300W and 10% $SF_6$ content and fluid model simulation. Electrons obey the Boltzmann relation at low pressure, 10mTorr.

## IV Conclusion

The discharge structure and transport process of $Ar/SF_6$ inductive plasma is interpreted in this work under a parametric study of pressure. The presentation of $Ar/O_2$ plasma previously published mainly writes the self-coagulation theory that is directly used to interpret the various coagulation behaviors of this $Ar/SF_6$ plasma. The stratification behavior, parabolic, elliptic and flat-topped features, double layer nature and self-coagulation property are exhibited. The highlighted is the limited dipole moment model (representing the double layer) that imports discontinuity into continuative profile and the discovery that electrons do not satisfy the Boltzmann balance at self-coagulation. Self-coagulation is novel transport phenomena. It constricts plasma by chemistry, not physics, foreseeing new means of generating *high- temperature and density* source (potential plasma application). It interprets the experimental observation of ions' density peak that is connected to flat-top profile. At the high pressures, the self-coagulation causes the mass point behavior in the collective plasma. For satisfying the neutrality, the ambi-polar self-coagulation concept is proposed and verified by the negative sheaths simulated. Besides, the re-self-coagulation dynamics of minor cation and its astronomic significance are discussed. The laboratory and natural plasmas are hence correlated.

The combination of self-consistent simulation and analytic theory is a good method in the studies of complex physics and chemical process, like the Ar/$SF_6$ inductively coupled plasma. On one hand, the simulation is difficult to understand since it is multi-physics-field tightly coupled and the behind mechanism is hard to refine, but exhibiting dynamic detail inside. On the other hand, the analytic solution cannot consider all aspects, but giving clear physics. It is the significance of methodology revealed from this electronegative plasma study.

## Acknowledgement

This work is financially supported by the foundation of DUT19LK59.

## Conflict of interest

The authors have no conflicts to disclose.

## Data available statement

The data that support the findings of this study are available within the article and its supplementary materials.